\documentclass[letterpaper,onecolumn,superscriptaddress,longbibliography,accepted=2023-10-20]{quantumarticle}
\pdfoutput=1
\usepackage[margin=0.8in]{geometry}
\usepackage{tikz}
\usepackage[utf8]{inputenc}
\usepackage{amssymb}
\usepackage[hidelinks]{hyperref}
\usepackage{mathtools}
\usepackage{tcolorbox}
\usepackage{colortbl}
\usepackage[braket, qm]{qcircuit}
\usepackage[version=4]{mhchem}

\usepackage{appendix}
\usepackage{amsmath,amssymb,braket}
\usepackage[ruled]{algorithm2e}
\usepackage{rotating}
\usepackage{booktabs}
\usepackage{multirow}

\newcommand{\SEVEplus}{QSP-EVE}
\newcommand{\SEVE}{std-EVE}
\makeatletter

    \def\CT@@do@color{%
      \global\let\CT@do@color\relax
            \@tempdima\wd\z@
            \advance\@tempdima\@tempdimb
            \advance\@tempdima\@tempdimc
    \advance\@tempdimb\tabcolsep
    \advance\@tempdimc\tabcolsep
    \advance\@tempdima2\tabcolsep
            \kern-\@tempdimb
            \leaders\vrule
                    \hskip\@tempdima\@plus 1fill
            \kern-\@tempdimc
            \hskip-\wd\z@ \@plus -1fill }
    \makeatother
\makeatletter
\newcommand{\subalign}[1]{%
  \vcenter{%
    \Let@ \restore@math@cr \default@tag
    \baselineskip\fontdimen10 \scriptfont\tw@
    \advance\baselineskip\fontdimen12 \scriptfont\tw@
    \lineskip\thr@@\fontdimen8 \scriptfont\thr@@
    \lineskiplimit\lineskip
    \ialign{\hfil$\m@th\scriptstyle##$&$\m@th\scriptstyle{}##$\hfil\crcr
      #1\crcr
    }%
  }%
}
\makeatother

\begin{document}
\title{Fault-tolerant quantum computation of molecular observables}

\newcommand{\BI}{\affiliation{%
Quantum Lab, Boehringer Ingelheim, 55218 Ingelheim am Rhein, Germany}}

\newcommand{\QCW}{\affiliation{QC Ware Corp, Palo Alto, CA 94306, USA}}

\newcommand{\PSIQ}{\affiliation{PsiQuantum, 700 Hansen Way, Palo Alto, CA 94304, USA}}

\author{Mark Steudtner}
\email{msteudtner@psiquantum.com}
\orcid{0000-0002-6419-302X}
\PSIQ

\author{Sam Morley-Short}
\orcid{0000-0002-4445-734X}
\PSIQ

\author{William Pol}
\orcid{0000-0002-1235-9200}
\PSIQ

\author{Sukin Sim}
\orcid{0000-0001-5324-5165}
\PSIQ

\author{Cristian L. Cortes}
\orcid{0000-0002-1163-2981}
\QCW
\author{Matthias Loipersberger}
\orcid{0000-0002-3648-0101}
\QCW
\author{Robert M. Parrish}
\orcid{0000-0002-2406-4741}
\QCW

\author{Matthias Degroote}
\email{matthias.degroote@boehringer-ingelheim.com}
\orcid{0000-0002-8850-7708}
\BI
\author{Nikolaj Moll}
\orcid{0000-0001-5645-4667}
\BI
\author{Raffaele Santagati}
\orcid{0000-0001-9645-0580}
\BI
\author{Michael Streif}
\orcid{0000-0002-7509-4748}
\BI

\begin{abstract}
Over the past three decades significant reductions have been made to the cost of estimating ground-state energies of molecular Hamiltonians with quantum computers. However, comparatively little attention has been paid to estimating the expectation values of other observables with respect to said ground states, which is important for many industrial applications. In this work we present a novel expectation value estimation (EVE) quantum algorithm which can be applied to estimate the expectation values of arbitrary observables with respect to any of the system's eigenstates. In particular, we consider two variants of EVE: std-EVE, based on standard quantum phase estimation, and QSP-EVE, which utilizes quantum signal processing (QSP) techniques. We provide rigorous error analysis for both both variants and minimize the number of individual phase factors for QSP-EVE. These error analyses enable us to produce constant-factor quantum resource estimates for both std-EVE and QSP-EVE across a variety of molecular systems and observables. For the systems considered, we show that QSP-EVE reduces (Toffoli) gate counts by up to three orders of magnitude and reduces qubit width by up to 25\% compared to std-EVE. While estimated resource counts remain far too high for the first generations of fault-tolerant quantum computers (between roughly $10^{14}$ and $10^{19}$ Toffoli gates for the examples considered), our estimates mark a first of their kind for both the application of expectation value estimation and modern QSP-based techniques.
\end{abstract}
\maketitle

\section{Introduction}
Calculating molecular energies is the most popular potential application of fault-tolerant quantum computers in quantum chemistry. Several studies have estimated and optimized the computational resources required to compute the ground state energy of classically challenging systems, improving the required number of non-Clifford gates (e{.}g{.} Toffoli) from $\sim\!10^{16}$ to $\sim\!10^{10}$~\cite{Poulin2014, Reiher2017, babbush2018encoding, Berry_2019, Google_THC, first_quant, PsiQ_battery, Xanadu_battery,VonBurg2021,goings2022}.
However, calculating ground state energies alone will have limited practical applicability. For example, in drug design, the calculations of molecular forces with respect to the ground state is required to simulate molecular dynamics~\cite{o2021efficient, cramer2013essentials, santagati2023drug} and electric multipole moments can be used to determine the permeability of drugs~\cite{fong2015permeability}.

Calculating the expectation values of general observables on a quantum computer requires more complex circuits than those for energy calculations~\cite{knill2007, brassard2002quantum}.
Energies are eigenvalues of the Hamiltonian and can be evaluated with quantum phase estimation (QPE) directly~\cite{Kitaev1995, Poulin2009}. For operators which do not commute with the Hamiltonian, the ground state is not an eigenstate of the observable.
One could prepare the ground state and then determine the observable from measurements. 
However, this approach is not optimal for the ground state preparation---usually achieved by post-selecting the ground state from multiple runs of QPE or with more refined quantum state preparation algorithms~\cite{poulin2018quantum, ge2019faster,lin2020near}, including filtering methods \cite{zhang2022computing}---which must be repeated for every sample. 
An alternative possibility recasts the expectation value calculation as an eigenvalue problem of a related operator, which enables to readout the expectation value with quantum phase estimation. 
A quadratic improvement in the sampling complexity, as well as an additive rather than multiplicative dependence on the complexity of state preparation, can be achieved with overlap estimation~\cite{ brassard2002quantum,knill2007optimal}. Further improvements can be expected by exploiting quantum singular value transformation (QSVT) \cite{QSVT,o2021efficient, rall2020quantum} or when measuring multiple expectation values simultaneously~\cite{huggins2021nearly,cornelissen2022near}.

So far, the practicality of overlap estimation algorithms remains an open question, as no explicit resource estimates have been presented, and most studies have focused on evaluating asymptotic scalings. This paper fills this gap by providing complete resource estimates of an expectation value estimation algorithm based on phase estimation~\cite{knill2007} and block encodings of the Hamiltonian and the observable, which we name standard expectation value estimation (\SEVE{}). We introduce a second version of the algorithm, which takes advantage of state-of-the-art quantum signal processing (QSP) techniques~\cite{Low2017, QSVT}, named \SEVEplus{}.
 The overlap estimation algorithm in \cite{o2021efficient,rall2020quantum} employs highly sophisticated QSP routines, which rely heavily on the ability to solve complex optimization problems~\cite{Low2017} to obtain the phase factors required for their compilation.
 To avoid large optimization problems for \SEVEplus{}, we choose to replace QSP-heavy subroutines with quantum phase estimation routines~\cite{rall2021faster,martyn2021grand}.
While this requires marginally more qubits, it makes finding the optimal phase factors much more accessible.
Our cost analysis is based on end-to-end compilation of the quantum algorithm. We explicitly perform the classical preprocessing step and obtain the QSP phase factors required. To our knowledge, this is the first time a QSP-based algorithm has been analyzed to such an extent. 
By explicitly stating all steps in the compilation and the corresponding resources, we provide a baseline for future endeavors to further optimize the cost of expectation value estimation on quantum computers.

The paper is structured as follows: the expectation value estimation algorithm and its complexity are presented in Section~\ref{sec:algor}. In Section~\ref{sec:correctness}, we verify that the algorithm outputs an expression of the expectation value, assuming all its components work flawlessly. This is, of course, not necessarily the case: the finite precision of certain algorithmic components manifests in systematic errors of the expectation value estimate and finite failure probabilities of the entire algorithm. We quantify these errors and failure probabilities in Section~\ref{sec:errors} while suggesting remedies, including QSP techniques. Section~\ref{sec:results} shows numerical evidence of these remedies working. We also present resource estimates, including all constant factors for a realistic problem: estimating the dipole operator, force components, and kinetic energies of p-benzyne and various smaller molecules. Finally, we discuss our findings in Section~\ref{sec:discussion}.

\section{Algorithm overview}\label{sec:algor}
Let us say we are given two hermitian operators, an observable $F$ and a Hamiltonian $H$. To be able to encode the target problem on a quantum computer, we represent $F$ and $H$ as factorized qubit operators $\hat{H}$ and $\hat{F}$ such that $\| \hat{F}\|, \| \hat{H}\| \leq 1$, and use $\lambda_F$ and $\lambda_H$ as their corresponding norm factors in the respective units of $F$ and $H$. The target problem is then formulated as follows: 
given an observable operator $\hat{F}$, we want to evaluate the expectation value $F_G$ of $\hat{F}$ with respect to the ground state $|\psi_G\rangle$ of the Hamiltonian $\hat{H}$,
\begin{align}
    F_G = \langle\psi_G|\hat{F}|\psi_G\rangle\, ,
\end{align}
knowing only the ground state energy $G = \langle\psi_G|\hat{H}|\psi_G\rangle$ and using block encodings of $\hat{F}$ and $\hat{H}$. In principle, the following algorithm can be trivially extended to expectation values concerning any excited state of the Hamiltonian.
 
\begin{figure*}[!tb]
    \centering
    \includegraphics[width=\linewidth]{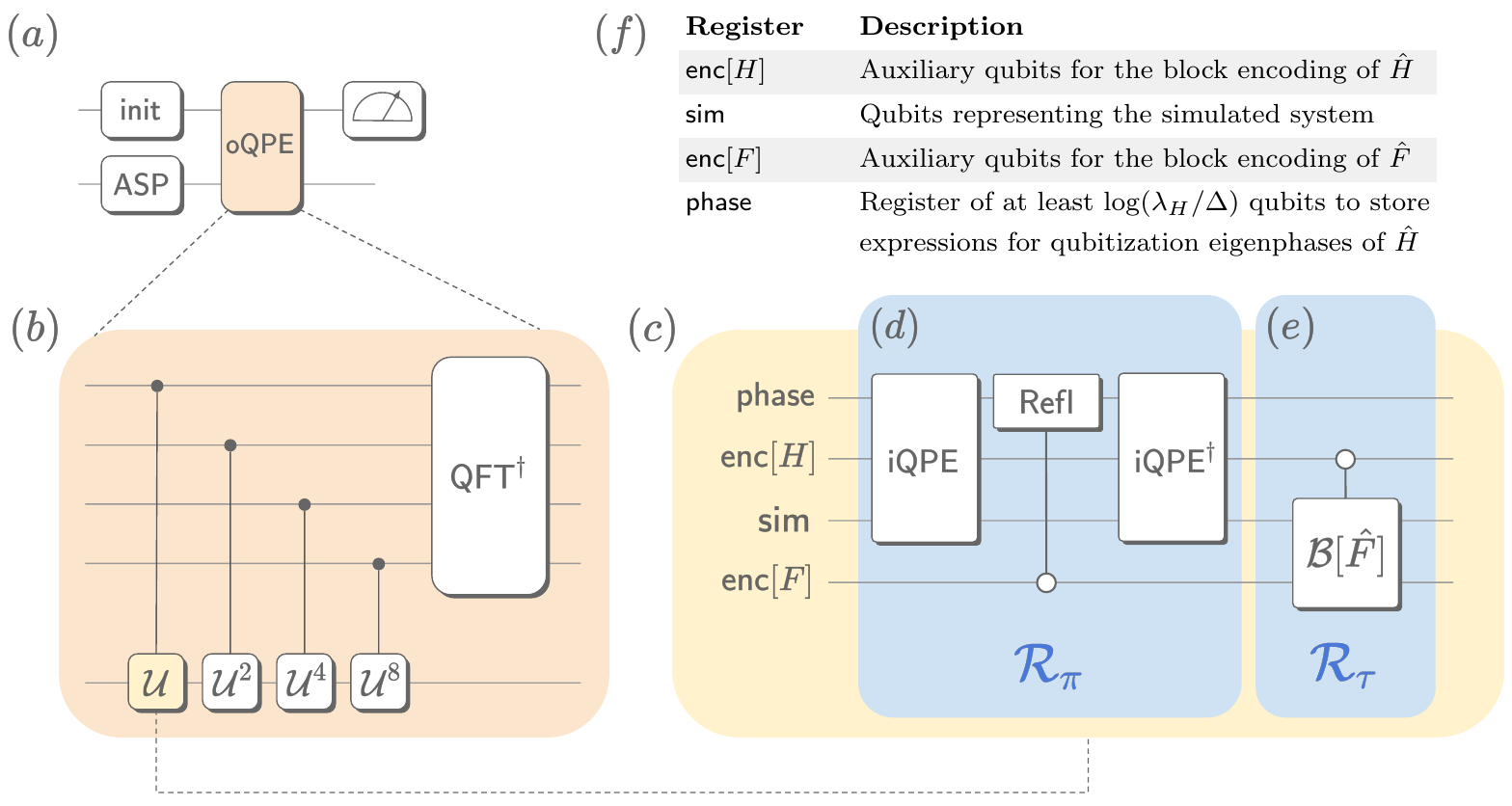}
    \caption{Overview of the quantum algorithm for expectation value estimation. $\boldsymbol{(a)}$ The first of two registers is initialized as a window function using the routine $\mathsf{init}$. In the standard formulation of quantum phase estimation, $\mathsf{init}$ would just apply Hadamard gates to every qubit, but more refined circuits can be used to achieve optimal phase readouts \cite{van2007optimal,rendon2022effects}. In the second register, we prepare a target state with an Ansatz state preparation ($\mathsf{ASP}$) routine, for instance a preparation of a Slater determinant state, followed by qubitization-based quantum phase estimation with which we postselect on the right energy. Both registers are fed into an outer quantum phase estimation ($\mathsf{oQPE}$) routine and an expression for the expectation value is obtained by measuring all qubits of the upper register in the computational basis. $\boldsymbol{(b)}$ Zooming in on the $\mathsf{oQPE}$ routine. Here the upper register has been split into single qubits. Note that the size of the upper register is arbitrary; it only has four qubits in this case. Here $\mathcal{U}$ are the phase oracle iterates applied to the lower register for phase kickback and $\mathsf{QFT}$ denotes the quantum Fourier transform. $\boldsymbol{(c)}$ Zooming in on the phase oracle $\mathcal{U}$, which consists of two reflections in panels ${(d)}$ and ${(e)}$ acting on the four registers $\mathsf{phase}$, $\mathsf{enc}[H]$, $\mathsf{sim}$, and $\mathsf{enc}[F]$. $\boldsymbol{(d)}$ Reflection featuring inner quantum phase estimation routines $(\mathsf{iQPE})$ using qubitization of the Hamiltonian $\hat{H}$. The $\mathsf{iQPE}$ routines write expressions of the eigenphases into the $\mathsf{phase}$ register, and the computational-state reflection $\mathsf{Refl}$, controlled on the all-zero state of the $\mathsf{enc}[F]$ register, tags eigenstates associated with $|\psi_G\rangle$ in the $\mathsf{sim}$ register. The reflection is identified with $\mathcal{R}_\pi$ in Section \ref{sec:correctness}. $\boldsymbol{(e)}$ Block encoding of $\hat{F}$, conditioned on the all-zero state in the $\mathsf{enc}[H]$ register. The reflection is identified with $\mathcal{R}_\tau$ in Section \ref{sec:correctness}. $\boldsymbol{(f)}$ Descriptions of the qubit registers that the iterate $\mathcal{U}$ acts on. $\lambda_H$ is the norm of the Hamiltonian $H$ and $\Delta$ is its spectral gap.}
    \label{fig:overview}
\end{figure*}

The general structure of the expectation value estimation algorithm is presented in Fig.~\ref{fig:overview}.
At the highest hierarchical level, we divide a number of qubits into two quantum registers. State preparation initializes the first register in a QPE \textit{filter} or \textit{window function} with the routine $\mathsf{init}$, while an ansatz state preparation ($\mathsf{ASP}$) initializes a subset of the lower register in one of the qubitization eigenstates associated with the Hamiltonian ground state $|\psi_G\rangle$. There is no need for the preparation to be perfect, but the squared overlap with the target state is a factor in the success probability of the entire algorithm. Note that in contrast to some prior art~\cite{mitarai2022perturbation, knill2007optimal}, this state preparation is not part of the main routine and thus appears only $O(1)$ times: we could therefore allow $\mathsf{ASP}$ to contain more costly state preparation routines such as in Refs.~\cite{Poulin2009, ge2019faster, lin2020near}. However, since the success probability of the $\mathsf{ASP}$ is decoupled from the success probability of the remaining algorithm, we can use (traditional) qubitized phase estimation on an initially-prepared state (using e{.}g{.} a Hartree-Fock state or a sum of Slater determinants) to yield the ground state on one register, and an estimate of the ground state energy which we can read out and store classically. The precision required for this estimate can be the same as the precision of the ground state reflections in later steps: it scales linearly with the energy gap $\Delta$ between the ground state and the first excited state.

These prepared states are then fed into an outer quantum phase estimation ($\mathsf{oQPE}$), which entangles eigenstates of the oracle $\mathcal{U}$ on the lower register with expressions for the eigenphase $\theta$ of their corresponding eigenvalues $e^{i 2 \pi \theta}$ on the registers above, see Fig.~\ref{fig:overview}$(b)$. By measurement of the upper register at the end of the circuit in Fig.~\ref{fig:overview}$(a)$ (and up to discretization errors), we project the lower register into an eigenstate of $\mathcal{U}$ and learn the eigenphase from the configuration of the upper register, which encodes $\theta$ as a binary fixed-point number in its computational basis. With high probability, we have projected into an eigenstate whose eigenphase $\theta = \theta_{\pm}$ contains information on the desired expectation value~\cite{knill2007}:
\begin{align}
\label{eq:return}
    \theta_{\pm} = \frac{1}{2}\left( 1 \pm \frac{1}{\pi}\arccos\frac{1+\langle\psi_G|\hat{F}|\psi_G \rangle}{2} \right)\, , 
\end{align}
which helps us to  estimate the value of $\langle \psi_G |\hat{F}|\psi_G\rangle$. 
Here $\pm$ indicates the existence of two solutions, $\theta_+$ and $\theta_{-}$, and it is assumed that $\hat{F}$ is the normalized version of the observable with $||\hat{F}||\leq 1$, where $||\cdot||$ is the spectral norm. 
Note that our expectation value estimation is unambiguous in the sign of the expectation value, meaning the sign $\pm$ in Eq.~\eqref{eq:return} does not obfuscate the sign of $\langle\psi_G|\hat{F}|\psi_G \rangle$. 
The iterate $\mathcal{U}$, depicted in Figure \ref{fig:overview}$(c)$, therefore does not require the additional calculations as necessary in the overlap estimation routine~\cite{knill2007}. $\mathcal{U}$ is divided into two subcircuits. The subcircuit in Figure~\ref{fig:overview}$(e)$ is a block encoding of the operator $\hat{F}$, and the subcircuit $(d)$ uses block encodings of the Hamiltonian $\hat{H}$ inside an inner QPE ($\mathsf{iQPE}$), as well as information about the ground state energy $G$, and the spectral gap $\Delta$ of $\hat{H}$. The 1-norms of $\hat{H}$ and $\hat{F}$ after factorization are $\lambda_H$ and $\lambda_F$, respectively. We can estimate the complexity of the $\mathsf{oQPE}$ by counting the repetitions of its main element, the block encoding of the Hamiltonian $\hat{H}$. 

As $\mathsf{iQPE}$ needs to be precise enough to distinguish the  eigenphase that qubitization makes with respect to the Hamiltonian ground state $|\psi_G\rangle$ from the qubitization eigenphase of the first excited state, it makes at least $O(\lambda_H/\Delta)$ queries to the block encoding of $\hat{H}$ (following arguments in Ref.~\cite{berry2018improved}).
However, due to QPE discretization errors, the accuracy of $\mathsf{iQPE}$ has to be improved to satisfy the target precision of the observable. 
To achieve a target error $\widetilde{\varepsilon}$ in the inner phase estimation, it is necessary to increase its complexity by a factor of $\varkappa(\widetilde{\varepsilon})$. Note that the $\mathsf{ASP}$ is not affected by this bit discretization error as much as the $\mathsf{iQPE}$, as the $\mathsf{ASP}$ only needs to prepare $|\psi_G\rangle$ with a fidelity close to $1$, but not necessarily as close as $1-\widetilde{\varepsilon}$. In later sections, we will introduce two versions of the expectation value estimation algorithm that allow for $\varkappa(\widetilde{\varepsilon})=1/\widetilde{\varepsilon}$ and $\varkappa(\widetilde{\varepsilon})=\log(1/\widetilde{\varepsilon})$, respectively. The target error of the observable also shapes the number of qubits in the $\mathsf{oQPE}$. 
To estimate the expectation value up to an error $\varepsilon$, we have to repeat the iterate $\mathcal{U}$ a total of $O(\lambda_F/\varepsilon)$ times. Setting $\widetilde{\varepsilon} = O(\varepsilon / \lambda_F$) also for the target precision of $\mathsf{iQPE}$, we find that $\mathsf{oQPE}$ has a query complexity of
\begin{align}
\label{eq:bigcomplexity}
    O\!\left(\frac{\lambda_F\lambda_H}{\Delta\, \varepsilon} \varkappa\!\left(\frac{\varepsilon}{\lambda_F}\right)\right)\, .
\end{align}
Considering the use of asymptotically optimal algorithms for the $\mathsf{ASP}$ \cite{lin2020near}, we present the complexity of the algorithm in Table~\ref{tab:complexities}, where $\gamma$ is the overlap amplitude of the initial state feeding into the $\mathsf{ASP}$ before $\mathsf{oQPE}$. 
\begin{table*}[tb]
    \centering
        \caption{Complexity of sampling (ideal and early fault-tolerant), the original overlap estimation algorithm, together with the improved versions (including \SEVEplus{}), where we include the complexity of state preparation routines. Here, $\varepsilon$ is the target accuracy of the observable value, $\Delta$ is the spectral gap of the Hamiltonian, $\lambda_H$ and $\lambda_F$ are norms of the Hamiltonian and observable, respectively, and $\gamma$ is the overlap amplitude between the Hamiltonian ground state $|\psi_G\rangle$ and its initial approximation. For instance, if we initialize the system in a Hartree-Fock state $| \mathrm{HF} \rangle$, we find $\gamma = |\langle \psi_G | \mathrm{HF} \rangle|$. The state preparation itself can be a state-of-the-art optimal routine such as \cite{lin2020near}. The original overlap estimation algorithm assumes that a reflection about the state $|\psi_G\rangle$ is implemented constructively, meaning one has to resort to the aforementioned Ansatz state preparation. The dots in the second line indicate that no logarithmic dependence has been given in the source material.}
    \vspace{-10pt}
    \begin{align*}
    \arraycolsep=2.4pt
    \def\arraystretch{2.2}
    \begin{array}{lll}
    \hline\hline
    \text{Method} & \;&\text{Complexity $O(\cdot)$}\\
    \hline
        \text{Sampling}&& \frac{ \lambda_F^2\lambda_H}{\gamma\Delta\varepsilon^2}\log^2\frac{\lambda_F}{\varepsilon} \\
        \text{Early fault-tolerant computation \cite{zhang2022computing}}&& \frac{\lambda_F^2\lambda_H}{\gamma^2\Delta\varepsilon^2}\, \mathrm{poly log}(\dots) \\
       \text{Original overlap estimation/Rall~ \cite{knill2007,rall2020quantum}} && \frac{\lambda_F \lambda_H}{\gamma\Delta\varepsilon}\log^2\frac{\lambda_F}{\varepsilon} \\
       \text{This algorithm/Improved overlap estimation~\cite{o2021efficient}} && \left(\frac{1}{\gamma}\log\frac{\lambda_F}{\varepsilon} +\frac{\lambda_F}{\varepsilon}\right)\frac{\lambda_H}{\Delta}\log\frac{\lambda_F}{\varepsilon}\\\hline\hline
    \end{array}
    \end{align*}
    \label{tab:complexities}
    \vspace{-10pt}
\end{table*}
The table shows a two-fold advantage of the expectation value estimation over sampling and related approaches---a quadratic improvement of the sampling complexity and an additive, rather than a multiplicative dependence on the overlap amplitude $\gamma$ between the Hamiltonian ground state and its initial approximation.

The upper register in Fig.~\ref{fig:overview}$(a)$ features $O(\log(\lambda_F/\varepsilon))$ qubits. The lower register is split into four sub-registers $\mathsf{phase}$, $\mathsf{enc}[H]$, $\mathsf{enc[F]}$ and $\mathsf{sim}$ that we describe in Figure~\ref{fig:overview}$(f)$. When necessary for clarity, we will decorate states with the registers they are supported on, e{.}g{.}~$|\psi_G\rangle_{\mathsf{sim}}$.

\section{Algorithm details}
\label{sec:correctness}
We show that the algorithm of Fig.~\ref{fig:overview}$(a)$ computes the desired expectation value within the chosen error. The Hamiltonian $\hat{H}$ and the operator $\hat{F}$ have a spectral decomposition 
\begin{align}
\label{eq:ham}
    \hat{H} &= \sum_E E\, |\psi_E\rangle\!\langle\psi_E|\, ,\\
        \label{eq:obs}
    \hat{F} &= \sum_{\eta}\eta \, |\phi_\eta\rangle\!\langle\phi_\eta|\, ,
\end{align}
where $|\psi_E\rangle$ and $|\phi_\eta\rangle $ are eigenstates corresponding to the energies $E$ and eigenvalues $\eta$ of $\hat{H}$ and $\hat{F}$, respectively, with both bounded between $-1$ and $+1$. Knowing the complete spectrum of $\hat{H}$ and $\hat{F}$---the exact values of $E$ and $\eta$ in the sums of Eq.~\eqref{eq:ham} and Eq.~\eqref{eq:obs}---is not required. We only require knowledge of the ground state energy $E=G$ and a lower bound on the spectral gap $\Delta$.

Since the expectation value estimation algorithm is a form of phase estimation, we can verify its function by showing that the iterate in Fig.~\ref{fig:overview}$(c)$ has the eigenphases corresponding to Eq.~\eqref{eq:return}. The iterate $\mathcal{U}$ consists of two subcircuits: each subcircuit (Figure~\ref{fig:overview}$(d)$ and $(e)$) is a reflection---a self-inverse unitary operator. We verify the eigenphases of the $\mathcal{U}$ in three steps:
first, we consider the eigenphases of iterates consisting of any two reflections in Section~\ref{subsec:tworefls}. Second, we detail the first reflection of Fig.~\ref{fig:overview}$(d)$ and its use of quantum phase estimation in Section~\ref{subsec:qubitization}. Finally we identify the first and second reflection with the general reflections of Section~\ref{subsec:tworefls} in Section~\ref{subsec:expectation}, showing that $\theta_{\pm}$ are among the eigenphases of $\mathcal{U}$.

\subsection{General iterates of two reflections}
\label{subsec:tworefls}
We define an iterate as the product of two reflection operators $\mathcal{R}_\pi$ and $\mathcal{R}_\tau$, which are given through their respective projectors $\hat{\pi}$ and $\hat{\tau}$:
\begin{align}
\label{seve3}
\mathcal{R}_\pi &= 1 - 2 \hat{\pi} \, ,\notag \\
\mathcal{R}_\tau &= 1 - 2 \hat{\tau}\, .
\end{align}
The projectors are not necessarily known; we only need to know how to construct the resulting reflections. Furthermore, the projectors can be of arbitrary rank. We are interested in the singular values of their product:
\begin{align}
\label{seve4}
\hat{\tau} \cdot \hat{\pi} \; = \; \sum_{k} w_k\, |t_k\rangle\!\langle p_k | \, .
\end{align}

The left singular vectors $|t_k\rangle$ and right singular vectors $|p_k\rangle$ of the same singular value $w_k > 0$ can now be used to construct two eigenstates $|v_{k+}\rangle$ and $|v_{k-}\rangle$ of the iterate $\mathcal{R}_\tau \mathcal{R}_\pi$. The technical details are in Appendix \ref{sec:singvalest}. The eigenstates of the iterate are
\begin{align}
\label{eq:eigenvectors}
    |v_{k\pm}\rangle & =  \frac{1}{\sqrt{2}}\left(\vphantom{\sum_1} |p_k\rangle \pm i |p_k^\perp\rangle \right) \\
                    &=  \frac{\exp(\pm i\arccos w_k)}{\sqrt{2}} \left(\vphantom{\sum_1} |t_k\rangle \mp i|t_k^\perp\rangle\right) \, 
\end{align}
where $|p_k^\perp\rangle$ is the component of $|t_k\rangle$ orthogonal to $|p_k\rangle$ and vice-versa for $|t_k^\perp\rangle$: 
\begin{align}
\label{eq:perpstates}
    |p_k^\perp\rangle = \frac{|t_k\rangle - w_k|p_k\rangle}{\sqrt{1-w_k^2}}\, ,\qquad |t_k^\perp\rangle = \frac{|p_k\rangle - w_k|t_k\rangle}{\sqrt{1-w_k^2}}\, .
\end{align}
The corresponding eigenvalues of $|v_{k\pm}\rangle$ are expressed in terms of $w_k$ as 
\begin{align}
\label{eq:eigenvalues}
\mathcal{R}_\tau \mathcal{R}_\pi |v_{k\pm}\rangle = \exp(\pm i2\arccos w_k) |v_{k\pm}\rangle.
\end{align}
Using the iterate $\mathcal{U} = \mathcal{R}_\tau \mathcal{R}_\pi$ within phase estimation would thus estimate the eigenphases $\pm i2\arccos w_k$. The expectation value estimation ensures that at least one of the singular values $w_k$ is a function of the expectation value $\langle\psi_G|\hat{F}|\psi_G \rangle$. Therefore, by estimating the singular values $w_k$, we can obtain the expectation value.

\subsection{Reflections with quantum phase estimation inside}
\label{subsec:qubitization}
\begin{figure*}[tb]
    \centering
    \includegraphics[scale=0.4]{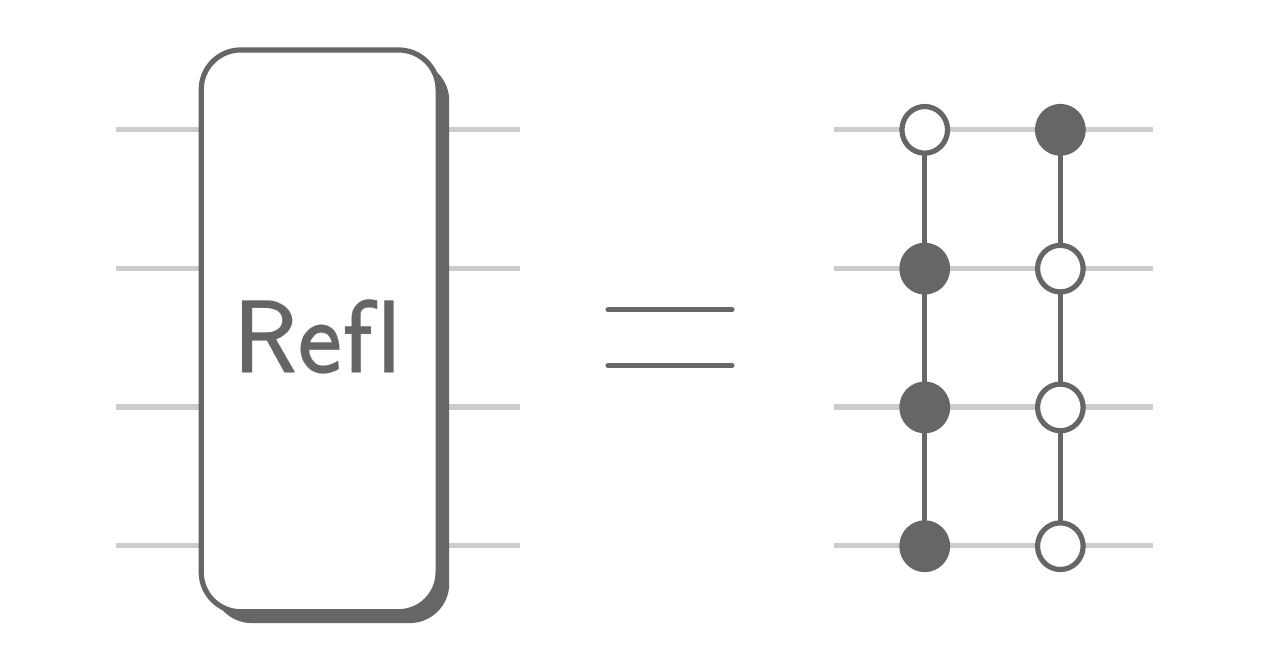}
    \caption{Example of a computational state reflection $\mathsf{Refl}$. The reflection is the equivalent of a number of multi-qubit Toffoli gates, where each provides a phase flip $(-1)$ on a specified computational subspace of qubits they act on. As per convention, empty control symbols indicate that the specified subspace for that qubit is  $|0\rangle$, and filled control symbols indicate that the subspace is $|1\rangle$. This figure shows a reflection on the computational states $|0111\rangle$ and $|1000\rangle$, corresponding to integers $m=7$ and $m=8$, respectively (see Eq.~\eqref{eq:refl}). These gates can be cheaply realized using the elbow circuits shown in Figures~4 and~5 of \cite{babbush2018encoding}.}
    \label{fig:refl}
\end{figure*}

The subcircuit of Fig.~\ref{fig:overview}$(d)$ contains phase estimation routines on the registers $\mathsf{phase}$, $\mathsf{sim}$ and $\mathsf{enc}[H]$. Within the $\mathsf{iQPE}$ routines, we are performing qubitization \cite{low2019hamiltonian}. Just like EVE, qubitization is a special case of the estimation procedure in Section~\ref{subsec:tworefls}. In qubitization, $\mathcal{R}_\tau$ is a reflection on the all-zero state $|\boldsymbol{0}\rangle$ in the $\mathsf{enc}[H]$ register, while $\mathcal{R}_\pi = \mathcal{B}[\hat{H}]$ is a block encoding of $\hat{H}$ on the $\mathsf{sim}$ and $\mathsf{enc}[H]$ registers. Block encodings are defined to be self-inverse, and therefore they are reflections. With the singular values $\sqrt{(1-E)/2}$ and the energies $E$ as given in Eq.~\eqref{eq:ham}, the qubitization iterate has eigenphases $\pm \arccos E$ associated with eigenstates
\begin{align}
\label{eq:qubitizationeigenstate}
    |Q_{E,\pm}\rangle_{\subalign{\\&\mathsf{sim}   \\ &\mathsf{enc}[H]}} &= \frac{1}{\sqrt{2}}\left( 1 \pm i \frac{E - \mathcal{B}[\hat{H}]}{\sqrt{1-E^2}}\right) |\psi_E\rangle_{\mathsf{sim}} \otimes |\boldsymbol{0}\rangle_{\mathsf{enc}[H]} \, .
\end{align}
A good reference for the above are equations 10 -- 13 of \cite{berry2018improved}.
With $\mathsf{iQPE}$ operating on the $\mathsf{phase}$ register, it would \textit{ideally} output computational basis states $|\Theta_{E,\pm}\rangle$ relating to integers $\Theta_{E,\pm} \in [0, 2^n-1]$, such that $\Theta_{E,\pm} = \pm 2^{n-1} (\arccos E)/\pi\, \mathrm{mod}\, 2^n$, where $n$ is the number of qubits in the $\mathsf{phase}$ register:
\begin{align}
\label{eq:idealQPE}
    \mathsf{iQPE} |Q_{E,\pm}\rangle_{\subalign{\\&\mathsf{sim}   \\ &\mathsf{enc}[H]}} \otimes |\boldsymbol{0}\rangle_{\mathsf{phase}} &= |Q_{E,\pm}\rangle_{\subalign{&\mathsf{sim}   \\ &\mathsf{enc}[H]}} \otimes |\Theta_{E,\pm}\rangle_{\mathsf{phase}}\, .
\end{align}
We use this property to implement reflections on certain qubitization eigenstates $|Q_{E,\sigma}\rangle$ indirectly by dressing a reflection on the $\mathsf{phase}$ register with $\mathsf{iQPE}$ circuits. To that end, we introduce $\mathsf{Refl}$, an arithmetic reflection that tags a set $\mathcal{M}$ of integers $m\in [0, 2^n-1]$ in the computational basis of the $\mathsf{phase}$ register by means of Toffoli gates and data-loaders, such that
\begin{align}
\label{eq:refl}
    \mathsf{Refl} = 1 - 2 \sum_{m \in \mathcal{M}}|m \rangle\!\langle m |_{\mathsf{phase}}\, ,
\end{align}
where $|m\rangle$ is a computational state encoding the integer $m$ just like $|\Theta_{E,\pm}\rangle$ encodes $\Theta_{E,\pm}$ (see the circuit in Fig.~\ref{fig:refl}). 
Knowing the integers $\Theta_{E,\sigma}$ for energies $E$ and signs $\sigma=\pm$ allows us to implement a reflection on the corresponding qubitization eigenstates,
\begin{align}
\label{eq:refl0}
    \mathsf{iQPE}^\dagger\, \mathsf{Refl}\; \mathsf{iQPE} &= 1 - 2 \sum_{E,\sigma} |Q_{E,\sigma}\rangle\!\langle Q_{E,\sigma}|_{\subalign{\\&\mathsf{sim}   \\ &\mathsf{enc}[H]}} \otimes \varrho(E,\sigma)_{\mathsf{phase}}\, .
\end{align}
Combining Eq.~\eqref{eq:refl0} with Eq.~\eqref{eq:refl} we see that $\varrho(E,\sigma)$ is a projector into the all-zero state only in certain cases:
\begin{align}
\label{eq:refl1}
    \varrho(E^\prime, \sigma^\prime) = |\boldsymbol{0}\rangle\!\langle \boldsymbol{0}| \quad \text{only if} \quad \Theta_{E^\prime, \sigma^\prime} \in \mathcal{M} \, .
\end{align}
 We will use that property in the next section, where $\mathcal{M}$ will only contain expressions for the ground state energy $G$. This approach is less costly than implementing $1 - 2 |\psi_G\rangle\!\langle\psi_G|$ starting from an initial reflection $1 - 2 |\mathrm{HF}\rangle\!\langle\mathrm{HF}|$ by acting with $\mathsf{ASP}$ routines on the Hartree-Fock state $|\mathrm{HF}\rangle$. We will also see that this approach uses fewer individual QSP phase factors than when implementing $1 - 2 |\psi_G\rangle\!\langle\psi_G|$ directly as in Ref.~\cite{o2021efficient}.  

\subsection{Expectation value estimation}
\label{subsec:expectation}
We now formulate the expectation value estimation routine as an instance of the singular value estimation of Section \ref{subsec:tworefls}. In Fig.~\ref{fig:overview}$(c)$ the subcircuit in panel $(d)$ shall now correspond to the reflection $\mathcal{R}_\pi$ and the subcircuit in panel $(e)$ shall correspond to $\mathcal{R}_\tau$ in Eq.~\eqref{seve3}. The latter reflection is a block encoding $\mathcal{B}[\hat{F}]$ of the observable $\hat{F}$ on the $\mathsf{sim}$ and $\mathsf{enc}[F]$ registers. Being self-inverse, the block encoding $\mathcal{B}[\hat{F}]$ is a reflection on the subspace spanned by the states $|\omega_\eta\rangle$, associated with the eigenvalues $\eta$, such that
\begin{align}
    \mathcal{B}[\hat{F}] = 1 - 2 \sum_\eta |\omega_\eta\rangle\!\langle\omega_\eta|_{\subalign{\\&\mathsf{sim}   \\ &\mathsf{enc}[F]}}\, .
\end{align}
From the fact that $\langle\boldsymbol{0}| \mathcal{B}[\hat{F}]|\boldsymbol{0}\rangle_{\mathsf{enc[F]}} = \hat{F}$, we find that $|\omega_\eta\rangle$ has the form
\begin{align}
    |\omega_\eta\rangle_{\subalign{\\&\mathsf{sim}   \\ &\mathsf{enc}[F]}} & = \sqrt{\frac{1-\eta}{2}} |\phi_\eta;\boldsymbol{0}\rangle_{\subalign{\\&\mathsf{sim}   \\ &\mathsf{enc}[F]}}
    - \sqrt{\frac{1+\eta}{2}} |\phi_\eta;\boldsymbol{0}^\perp\rangle_{\subalign{\\&\mathsf{sim}   \\ &\mathsf{enc}[F]}}\, ,
\end{align}
with 
\begin{align}
\label{eq:relation}
    |\phi_{\eta}; \boldsymbol{0}\rangle_{\subalign{\\&\mathsf{sim}   \\ &\mathsf{enc}[F]}} = |\phi_\eta\rangle_\mathsf{sim} \otimes | \boldsymbol{0}\rangle_{\mathsf{enc}[F]}\, ,
\end{align}
where $|\phi_\eta\rangle$ are the eigenstates of the observable, see Eq.~\eqref{eq:obs}, and$|\phi_\eta;\boldsymbol{0}^\perp\rangle$ is the  component orthogonal to $|\phi_\eta;\boldsymbol{0}\rangle$ within $|\omega_\eta\rangle$. Since the projection of $|\phi_\eta;\boldsymbol{0}^\perp\rangle$ onto the all-zero state of the $\mathsf{enc}[F]$ register vanishes, we find that 
\begin{align}
\label{eq:select_olap}
    \langle \omega_{\eta^\prime} |\phi_{\eta}; \boldsymbol{0}\rangle = \delta_{\eta\eta^\prime}\, \sqrt{\frac{1-\eta}{2}}\, .
\end{align}
 Keeping in mind that $\mathcal{B}[\hat{F}]$ and $|w_\eta\rangle$
are supported on the $\mathsf{sim}$ and $\mathsf{enc}[F]$ registers we find that the reflection is
\begin{align}
    \mathcal{R}_\tau =& |\boldsymbol{0}\rangle\!\langle \boldsymbol{0}|_{\mathsf{enc}[H]}\otimes\mathcal{B}[\hat{F}] + \left( 1 - |\boldsymbol{0}\rangle\!\langle \boldsymbol{0}|_{\mathsf{enc}[H]} \right) \, .
\end{align}
Therefore, its projector is
\begin{align}
    \hat{\tau} = |\boldsymbol{0}\rangle\!\langle\boldsymbol{0}|_{\mathsf{enc}[H]} \otimes \sum_\eta |\omega_\eta\rangle\!\langle\omega_\eta|_{\subalign{\\&\mathsf{sim}   \\ &\mathsf{enc}[F]}} \, .
\end{align}
Using Eq.~\eqref{eq:refl0}, the projector of $\mathcal{R}_\pi$ is
\begin{align}
\label{eq:unity0}
    \hat{\pi} &= |\boldsymbol{0}\rangle\!\langle\boldsymbol{0}|_{\mathsf{enc}[F]} \otimes \sum_{E, \sigma} |Q_{E, \sigma}\rangle\!\langle Q_{E, \sigma}|_{\subalign{\\&\mathsf{sim}   \\ &\mathsf{enc}[H]}} \otimes \varrho(E,\sigma)_{\mathsf{phase}}\,\\
     & = 
     \label{eq:unity}\sum_\eta|\phi_\eta;\boldsymbol{0}\rangle\!\langle\phi_\eta;\boldsymbol{0}|_{\subalign{\\&\mathsf{sim}   \\ &\mathsf{enc}[F]}} \cdot \sum_{E, \sigma} |Q_{E, \sigma}\rangle\!\langle Q_{E, \sigma}|_{\subalign{\\&\mathsf{sim}   \\ &\mathsf{enc}[H]}} \otimes \varrho(E,\sigma)_{\mathsf{phase}} \, .
\end{align}
Going from Eq.~\eqref{eq:unity0} to Eq.~\eqref{eq:unity}, the projector $|\boldsymbol{0}\rangle\!\langle\boldsymbol{0}|$ has been factored out of the tensor and is subsequently expanded with $1 = \sum_\eta |\phi_\eta\rangle\!\langle\phi_\eta|$ on the $\mathsf{sim}$ register. Using Eq.~\eqref{eq:relation}, \eqref{eq:qubitizationeigenstate}, and \eqref{eq:select_olap}, the product of the two projectors is
\begin{align}
    \hat{\tau} \cdot \hat{\pi} &= \sum_\eta \sum_{E,\sigma} \langle\phi_\eta|\psi_E\rangle \sqrt{\frac{1-\eta}{4}}   \left(| \boldsymbol{0}\rangle_{\mathsf{enc}[H]} \otimes |\omega_\eta \rangle_{\subalign{\\&\mathsf{sim}   \\ &\mathsf{enc}[F]}}\vphantom{\frac{1}{1}} \right)\notag\\&\qquad\times \left(\vphantom{\frac{1}{1}}\langle Q_{E,\sigma}|_{\subalign{\\&\mathsf{sim}   \\ &\mathsf{enc}[H]}} \otimes \langle \boldsymbol{0}|_{\mathsf{enc}[F]}\right) \otimes \varrho(E, \sigma)_{\mathsf{phase}}\, .
\end{align}
We can obtain the square of the singular values by solving the eigenvalue problem of 
\begin{align}
\label{eq:square}
    \hat{\pi} \cdot \hat{\tau} \cdot \hat{\pi} &= \sum_k w_k^2 |p_k\rangle \!\langle p_k| \\ &= \sum_{E,\widetilde{E}, \sigma, \widetilde{\sigma}} |\boldsymbol{0}\rangle\!\langle\boldsymbol{0}|_{\mathsf{enc}[F]} \otimes |Q_{\vphantom{\widetilde{E}}E,\sigma}\rangle\!\langle Q_{\widetilde{E},\widetilde{\sigma}}|_{\subalign{&\mathsf{sim}   \\ &\mathsf{enc}[H]}}  \otimes \varrho(E,\sigma)_{\mathsf{phase}} \cdot \varrho(\widetilde{E},\widetilde{\sigma})_{\mathsf{phase}}\notag \\ 
    & \qquad\qquad \quad \times \underbrace{\sum_\eta \frac{1-\eta}{4} \langle \psi_E|\phi_\eta\rangle\!\langle \phi_\eta|\psi_{\widetilde{E}}\rangle }_{\left\langle \psi_{\vphantom{\widetilde{E}}E} \right|\frac{1 - \hat{F}}{4} \left| \psi_{\widetilde{E}}\right\rangle}\, ,
\end{align}
where we have used the spectral decomposition of the observable from Eq.~\eqref{eq:obs}. When $\mathcal{M}$ in Eq.~\eqref{eq:refl0} only contains $\Theta_{G,s}$ for a fixed $\sigma=s$ (determined during the $\mathsf{ASP}$), then, by Eq.~\ref{eq:refl1}, we have $\varrho(G,s) \cdot \varrho(E,\sigma) = \delta_{E,G}\, \delta_{\sigma, s}|\boldsymbol{0}\rangle\!\langle \boldsymbol{0}|$. One solution to Eq.~\eqref{eq:square} is therefore
\begin{align}
    \label{eq:squaresolution}
    |Q_{G,s}\rangle_{\subalign{&\mathsf{sim}   \\ &\mathsf{enc}[H]}}\otimes |\boldsymbol{0}\rangle_{\mathsf{enc}[F]} \otimes |\boldsymbol{0}\rangle_{\mathsf{phase}}\, 
\end{align}
with its eigenvalue equal to $(1-\langle\psi_G|\hat{F}|\psi_G\rangle)/4$. Using Eq.~\eqref{eq:eigenvalues} we find that a phase estimation with the iterate $\mathcal{R}_\tau \mathcal{R}_\pi$ allows us to estimate the phase angles \eqref{eq:return}
when the input is the state of Eq.~\eqref{eq:squaresolution}.
\begin{figure*}[tb]
    \centering
    \includegraphics[width=\linewidth]{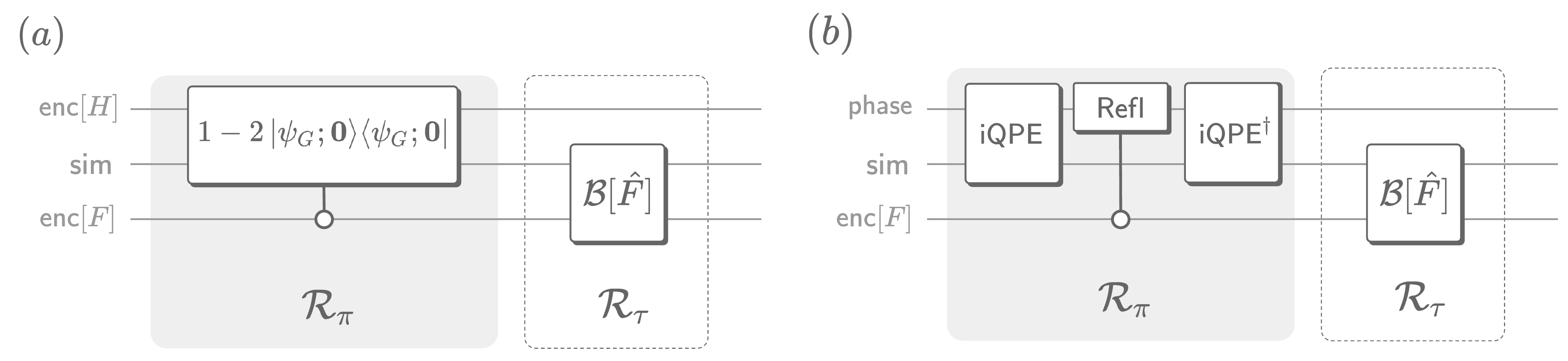}
    \caption{Variations of the expectation value estimation iterate in Figure \ref{fig:overview}$(c)$. $\boldsymbol{(a)}$ A more general version of the iterate, in which the reflection $\mathcal{R}_\pi$ is implemented with quantum singular value transformation (QSVT), quantum signal processing (QSP) or quantum eigenvalue transformation (QET) techniques   \cite{dong2022ground} without the need for an $\mathsf{iQPE}$. When block encodings $\mathcal{B}[\hat{H}]$ are used, this version still requires a $\mathsf{enc}[H]$ register, but without the need to condition the block encoding $\mathcal{B}[\hat{F}]$ on the all-zero state of the $\mathsf{enc}[H]$ qubits within $\mathcal{R}_\tau$. No $\mathsf{phase}$ register is required. $\boldsymbol{(b)}$ An implementation of the iterate in which the $\mathsf{iQPE}$ routines feature a Hamiltonian simulation technique that does not use block encodings. We, therefore, do not require a $\mathsf{enc}[H]$ register or control of the block encoding $\mathcal{B}[\hat{F}]$.}
    \label{fig:variations}
\end{figure*}

There are other variations of this iterate. Here we use Eq.~\eqref{eq:refl0} to replace a reflection $1-2|\psi_G\rangle\!\langle\psi_G|$ on the $\mathsf{sim}$ register. Such a reflection could be implemented with QSP/QSVT techniques directly, resulting in a circuit depicted in Fig.~\ref{fig:variations}$(a)$. However, this would require us to solve a large-scale optimization problem to obtain the phase factors, which we want to avoid. One could also conceive a version of the circuit where $\mathsf{iQPE}$ is used within $\mathcal{R}_{\pi}$, but with a Hamiltonian simulation algorithm different from qubitization as in Fig.~\ref{fig:variations}$(b)$. While qubitization has an asymptotically optimal scaling for normalized Hamiltonians, there are instances for which other methods are more efficient \cite{campbell2021early}. A last variation of the iterate is depicted in Figure \ref{fig:overview}$(c)$, but the set $\mathcal{M}$ in $\mathsf{Refl}$ is made to include $\Theta_{G,+}$ and $\Theta_{G,-}$ at the same time. All three variations of the iterate would allow us to estimate the angles $2 \pi \theta_{\pm} = \pm \arccos \langle\psi_G|\hat{F}|\psi_G\rangle$, which is twice as much signal as in Eq.~\eqref{eq:return}. So what would disqualify the third variation when it offers a better sensitivity? As shown in the next section, we have to consider the discretization error of the $\mathsf{iQPE}$ routines as a limiting factor for the accuracy of the estimated expectation value. We have deliberately chosen to dismiss the third variation of the iterate, as we would, in this case, have to contend with $\varrho(G, + ) \neq \varrho(G, -)$, introducing an additional source of error on top of the errors already considered. 

\section{Errors and failure probability}
\label{sec:errors}

Unfortunately, unlike the ideal formulation in Eq.~\eqref{eq:idealQPE}, an actual implementation of a quantum phase estimation will create errors propagating through the entire circuit. Every quantum phase estimation suffers from errors due to the discretization of the eigenphase expressions: a quantum phase estimation circuit with $n$-qubits outputs the eigenphases $\sigma \arccos E $ of the eigenstate $|Q_{E,\sigma}\rangle$ without error as long as $\Theta_{E,\sigma}$ is an integer. Discrete eigenphases are generally unlikely, and for non-integer $\Theta_{E,\sigma}$ quantum phase estimation outputs a distribution $p_n(\cdot)$ of integer expressions $k \in [0,2^n-1]$ for pseudo eigenphases $2\pi k/2^n$ in $n$ qubits:
\begin{align}
\label{eq:realQPE}
    &\mathsf{iQPE} |Q_{E,\sigma}\rangle_{\subalign{\\&\mathsf{sim}   \\ &\mathsf{enc}[H]}} \otimes |\boldsymbol{0}\rangle_{\mathsf{phase}} = \sum_{k=0}^{2^n-1} p_n\!\left(\Theta_{E,\sigma}-k\right)|Q_{E,\sigma}\rangle_{\subalign{\\ & \mathsf{sim}   \\ &\mathsf{enc}[H]}} \otimes |k\rangle_{\mathsf{phase}}\, ,
\end{align}
where $|k\rangle$ is the computational basis state encoding the integer $k$. In general, the pseudo-eigenphase distribution causes the projectors $\varrho(E, \sigma)$ on the $\mathsf{phase}$ register to ``smear out" such that we can no longer expect Eq.~\eqref{eq:refl1} to hold. Since $\Theta_{G,s}$ is not guaranteed to be an integer, we have to choose an integer number $\hat{m}$ such that $\hat{m}$ and $\hat{m}+1$ make up the set $\mathcal{M}$ in Eq.~\eqref{eq:refl}, in order to lower- and upper-bound the target eigenphase with $\hat{m} \leq \Theta_{G,s} \leq \hat{m}+1$. Here, $\mathcal{M}$ has only two elements, but it could generally include any $r$ configurations, and in consequence, the projectors $\varrho(E,\sigma)$ would have rank $r$. As pseudo-eigenphases   of all $|Q_{E,\sigma}\rangle$ could overlap with configurations in $\mathcal{M}$, the states that we utilize to estimate $\langle\psi_G|\hat{F}|\psi_G\rangle$ can be contaminated with excited states of $\hat{H}$. While it is challenging to find expressions for the errors in estimating the expectation value without knowledge of the complete spectrum of $\hat{H}$, we can estimate the largest error contribution caused by the contamination of the ground state $|\psi_G\rangle$ with the first excited state $|\psi_\mathcal{E}\rangle$ of energy $\mathcal{E}$. Let us quantify this contamination: we can write the singular value decomposition of the product as 
\begin{align}
    \label{eq:second_singvals}
    \varrho(G,s) \cdot \varrho(\mathcal{E},s)= \sum_{j=1}^r \Omega_j|G^s_j\rangle\!\langle \mathcal{E}^s_j|
\end{align}
where $\Omega_j$ are the singular values, and $|G^s_j\rangle$, $|\mathcal{E}^s_j\rangle$ are left and right singular vectors, respectively. Contamination of $|Q_{G,s}\rangle$ with $|Q_{\mathcal{E},s}\rangle$ happens when there are (non-zero) singular values $\Omega_j>0$. Without knowing the spectrum of $\hat{F}$, we can bound the difference between $F_{\mathrm{est}}$, the estimated value of $\hat{F}$ with respect to the contaminated state, and the actual expectation value as
\begin{align}
\label{eq:error}
    \left| F_{\mathrm{est}} - \langle\psi_G|\hat{F}|\psi_G\rangle \right| \;\leq\; 2 \max_{j=1\dots r} \Omega_j\, .
\end{align}
 The contamination additionally causes the expectation value algorithm to fail with a non-zero probability. This is because the state in Eq.~\eqref{eq:squaresolution} does not completely overlap with the solutions that allow us to estimate the expectation value up to the error in Eq.~\eqref{eq:error}. We find that having prepared the state in Eq.~\eqref{eq:squaresolution}, the success probability of the expectation value estimation algorithm is between $\langle \boldsymbol{0}|\varrho(G,s) |\boldsymbol{0}\rangle / 2$ and $\langle \boldsymbol{0}|\varrho(G,s) |\boldsymbol{0}\rangle$. The success probability depends on the observable and approaches its maximum value fast if 
\begin{align}
|\langle\psi_G|\hat{F}|\psi_G\rangle - \langle\psi_{\mathcal{E}}|\hat{F}|\psi_{\mathcal{E}}\rangle| \gg \max_j\Omega_j|\langle\psi_G|\hat{F}|\psi_{\mathcal{E}}\rangle|\, .
\end{align}
The proof for the observable error and the success probability can be found in Appendix \ref{sec:errppendix}. The error and failure probability of the expectation value algorithm can be decreased by improving the accuracy of the $\mathsf{iQPE}$ routines by either increasing the number of qubits in the $\mathsf{phase}$ register or by using of QSP routines. The two approaches span different versions of the same algorithm, which we shall call \SEVE{} (standard expectation value estimation) and \SEVEplus{} (QSP expectation value estimation). Both versions are discussed in the following:\\

\noindent\textbf{Standard version (\SEVE{})} -- The number of $\mathsf{phase}$ qubits must initially be chosen according to the norm and the spectral gap of $\hat{H}$. However, one might choose to add more qubits to the $\mathsf{phase}$ register in order to deal with the discretization error. The idea is that the discretization error only really affects the lesser significant bits of the QPE, and adding qubits raises the significance of the bits distinguishing $\Theta_{G,s}$ from $\Theta_{\mathcal{E},s}$. Concerning the success probability, it has been demonstrated in Appendix C of Ref.~\cite{cleve1998quantum} how increasing the number of qubits increases the success probability of QPE. In the next section we will demonstrate how it decreases the expectation value error of our algorithm. To increase the success probability of \SEVE{}, we could add more elements in $\mathcal{M}$, but this would typically increase the expectation value error again, which in turn calls for a further increase of the $\mathsf{phase}$ register. Adding a single qubit of precision to a phase estimation routine doubles its complexity, so the cost of the phase estimation increases exponentially with the number of additional qubits.\\

\begin{figure*}[tb]
    \centering
    \includegraphics[width=0.9\linewidth]{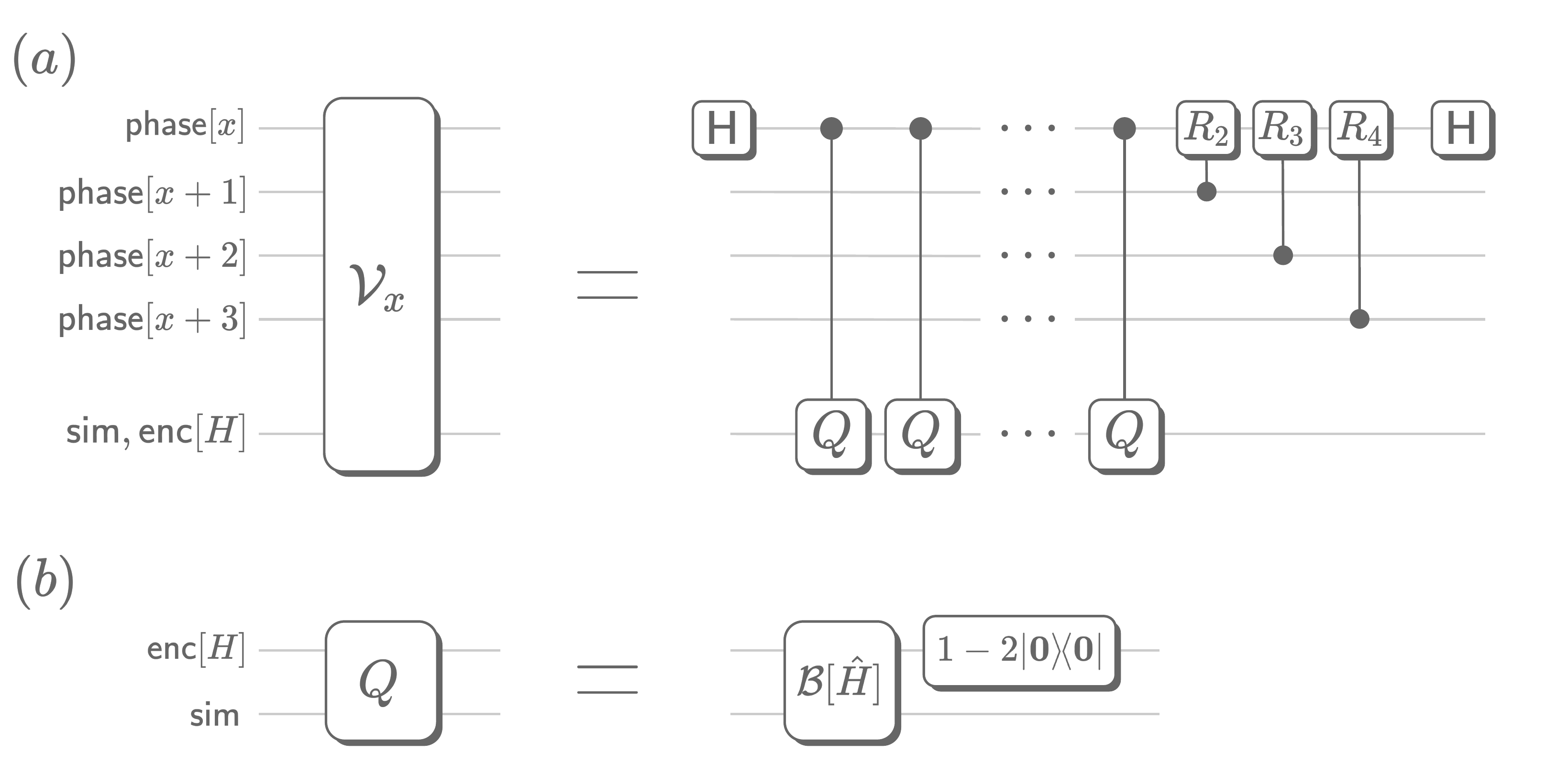}
    \caption{Building blocks of a QPE for qubitization.  $\boldsymbol{(a)}$ Quantum phase estimation building block that rotates the qubit $\mathsf{phase}[x]$ according to the phase kickback of the oracles $Q$ on the last register. Here, $Q$ is the qubitization iterate, and the last register is the combination of the registers $\mathsf{sim}$ and $\mathsf{enc}[H]$. One can apply a double-kickback trick discussed in \cite{babbush2018encoding}, Eq.~(16). We denote the Hadamard gates with $\mathsf{H}$, and $R_k$ are phase rotations $R_k = | 0 \rangle\langle 0 | + \exp(-i\pi/2^{k-1})| 1 \rangle\langle 1 |$ controlled on the less significant $\mathsf{phase}$ qubits for phase feedback. An inexpensive way to implement the latter is to feed the qubits $\mathsf{phase}[y]$ for $y>x$ into an adder circuit with a phase-gradient state (see Ref.~\cite{gidney2018halving}, page 4), flipping the value of all qubits $\mathsf{phase}[y]$ controlled on the value of qubit $\mathsf{phase}[x]$ before and after the addition. A circuit $\mathcal{V}_x$ with an $n$-qubit $\mathsf{phase}$ register makes an equivalent of $2^{x-1}$ queries to $Q$ for all $x=1\dots n$. $\boldsymbol{(b)}$ Qubitization iterate $Q$, featuring the block encoding of the Hamiltonian $\hat{H}$ and a reflection on the all-zero state of the $\mathsf{enc}[H]$ register.}
    \label{fig:primitive}
\end{figure*}

\begin{figure*}[tb]
    \centering
    \includegraphics[width=\linewidth]{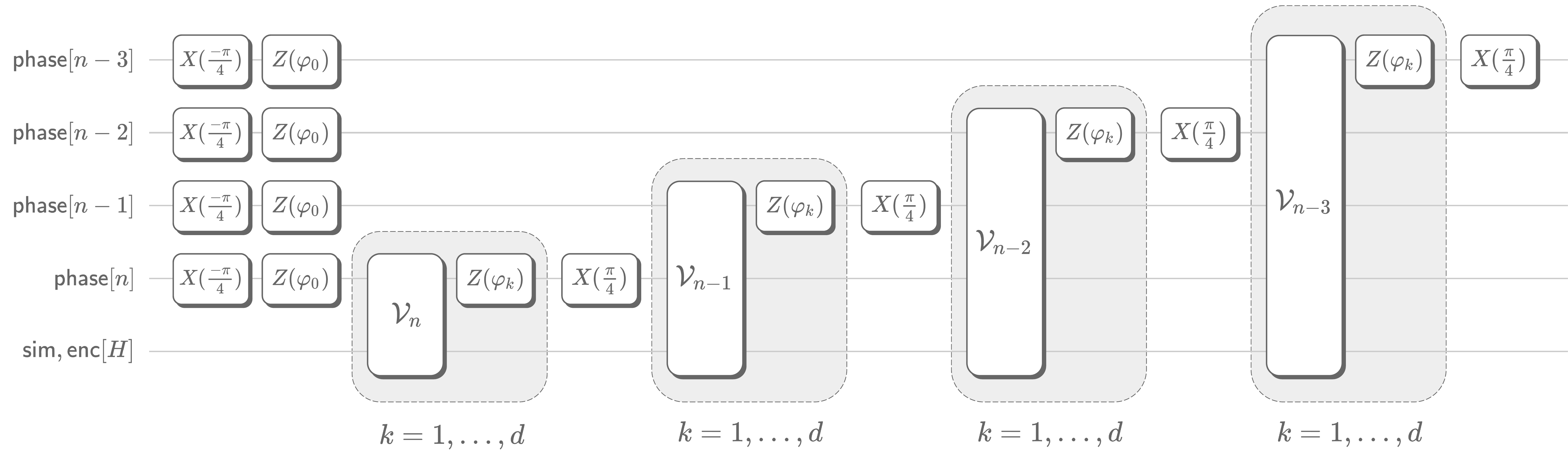}
    \caption{QSP-version of the inner QPE in \SEVEplus{}. $X(\alpha)$ and $Z(\beta)$ are $X$ and $Z$ rotations about the angles $\alpha$ and $\beta$, respectively, where $X(\alpha)=\cos\alpha I + i \sin\alpha X$ and  $Z(\beta) = \cos\beta I + i \sin\beta Z$. The shaded areas are repeated $d$ times with individual angles $\varphi_k$, where $d$ is the degree of a suitable polynomial approximation to the step function $S(\cdot)$, see Eq.~\eqref{eq:step_function}. The presented circuits are a symmetric QSP implementation \cite{wang2022energy} according to the Chebychev polynomial in Eq.~\eqref{eq:polyapproxi} such that $P_x(\cdot) = P(\cdot)$ for all $\mathsf{phase}$ qubits $\mathsf{phase}[x]$. As symmetric QSP only allows fixing the real part of $P_x(\cdot)$ in Eq.~\eqref{eq:qsvt_unitary} but never yields an imaginary part in the off-diagonal function $Q_x(\cdot)$, we can effectively transfer the imaginary part of $P_x(\cdot)$ onto $Q_x(\cdot)$ by a basis transform of the qubit $\mathsf{phase}[x]$, which is done by the Clifford rotations $X(\pm \pi/4)$. The subroutines $\mathcal{V}_x$ are defined in Figure \ref{fig:primitive}.}
    \label{fig:qsppqe}
\end{figure*}

\noindent \textbf{QSP-version (\SEVEplus{})} -- The singular qubits in the $\mathsf{phase}$ register are labeled by $\mathsf{phase}[x]$ for $x=1 \dots n$, where $\mathsf{phase}[1]$ is the most significant bit and $\mathsf{phase}[n]$ is the least significant bit in the phase estimation. The textbook version of quantum phase estimation used in \SEVE{} then consists of the sequence
\begin{align}
   \mathsf{iQPE}(\text{\SEVE{}})\; = \; \mathcal{V}_1 \mathcal{V}_2 \cdots \mathcal{V}_n
\end{align}
where the routine $\mathcal{V}_x$ sets the value of qubit $\mathsf{phase}[x]$ by calling the controlled iterate $\mathcal{U}$ a total of $2^{x-1}$ times (or an equivalent double kickback series of two reflections). The routine $\mathcal{V}_x$ uses the feedback from the less significant $\mathsf{phase}$ qubits $\mathsf{phase}[y]$ for $y>x$, as depicted in Fig.~\ref{fig:primitive}. 
The reason for pseudo eigenphases is that the routines $\mathcal{V}_x$ are not setting the qubit $\mathsf{phase}[x]$ to either $|0\rangle$ or $|1\rangle$ but to a superposition of the two. Refs.~\cite{martyn2021grand, rall2021faster} attempt to round the qubit $\mathsf{phase}[x]$ to the most-likely computational basis state using QSVT techniques. This is also the idea behind the version of $\mathsf{iQPE}$ used in the \SEVEplus{} algorithm, but there we specifically use QSP techniques rather than the more general QSVT. In this QSP version of QPE, the rounding is achieved with a symmetric QSP \cite{wang2022energy} circuit, depicted in Fig.~\ref{fig:qsppqe}.
Using $d_x$ queries to $\mathcal{V}_x$ and $d_x+1$ $Z$-rotations on qubit $\mathsf{phase}[x]$, we define a sequence
\begin{widetext}
\begin{align}
    \mathcal{W}_x = \exp\!\left(i\varphi_{x,0}Z_{\mathsf{phase}[x]}\right)\mathcal{V}_x \exp\!\left(i\varphi_{x,1}Z_{\mathsf{phase}[x]}\right) \mathcal{V}_x \cdots \mathcal{V}_x \exp\!\left(i\varphi_{x,d_x}Z_{\mathsf{phase}[x]}\right)\, .
\end{align}
\end{widetext}
The angles $\varphi_{x,k}$ are carefully chosen such that a similarly-structured sequence, where the $\mathcal{V}_x$ routines are replaced with $X$-rotations about the angle $\vartheta$ yields
\begin{widetext}
\begin{align}
\label{eq:qsvt_unitary}
\exp\!\left(i\varphi_{x,0}Z\right)\, \exp\!\left(i\vartheta X\right)\, \exp\!\left(i\varphi_{x,1}Z\right)\, \exp\!\left(i\vartheta X\right)\, \cdots\, \exp\!\left(i\vartheta X\right)\, \exp\!\left(i\varphi_{x,d_x}Z\right) \, = 
    \left[\begin{matrix}
    P_x(\cos \vartheta) & i Q^*_x(\cos \vartheta) \\
    i Q_x(\cos\vartheta) & P_x(\cos\vartheta)
    \end{matrix} \right]
\end{align}
\end{widetext}
  with some function $Q_x(z)$ and the degree-$d_x$ polynomial function $P_x(z)$ approximating a step function $S(z)$ defined as
\begin{align}
\label{eq:step_function}
    |S(z)| = \left\lbrace \begin{array}{ll} 1 & \text{for} \; |z| > \frac{1}{\sqrt{2}} \\ \\ 0 & \text{else}\, .\end{array}\right.
\end{align}
The deviation of $P_x(z)$ from $S(z)$ is largest in a region of $\kappa_x$ around the discontinuities at $z = \pm 1/\sqrt{2}$. The maximum value of $|P_x(z) - S(z)|$ outside these regions is defined as $\Delta_x$. This maximum difference can be decreased exponentially fast. The degree of the polynomial that matches the required characteristics can be found as $d_x = O(\kappa_x^{-1} \log(\Delta_x^{-1}))$ \cite{QSVT, Low2017, Low2017QuantumSP}. The $\mathsf{iQPE}$ routine in \SEVEplus{} is then defined as the sequential application of the  $\mathcal{W}_x$ subroutines, starting with the subroutine setting the least significant bit:
\begin{align}
    \mathsf{iQPE}(\text{\SEVEplus}) \;=\; \mathcal{W}_1 \mathcal{W}_2 \cdots \mathcal{W}_n \, .
\end{align}
Different approximations to $S(z)$ will give different $d_x$. With any choices of $P_x$, we can compute the singular values $\Omega_j$ and expectation values $\langle \boldsymbol{0}|\varrho(G,s) |\boldsymbol{0}\rangle$ with an efficient classical routine defined in Appendix~\ref{sec:iQPE}, allowing for an analysis of errors and success probabilities.

\section{Results}
\label{sec:results}
\subsection{Comparison of \SEVE{} and \SEVEplus{}}
\label{sec:comparison}

For a meaningful comparison between \SEVE{} and \SEVEplus{}, we will consider a baseline QPE querying the qubitization iterate with $n_0$ qubits, thereby querying the Hamiltonian block encoding $O(2^{n_0})$ times. Here, $n_0$ is the number of qubits high enough to separate the ground and first excited state eigenphases, $\pm\arccos G$ and $\pm\arccos \mathcal{E}$, by at least $2^{-n_0+1}\pi$. This separation is chosen such that consecutive numbers $\lbrace \hat{m}, \hat{m} + 1\rbrace \in \mathcal{M}$ can be used to tag states in the $\mathsf{Refl}$ routine, where
\begin{align}
\label{eq:bin0}
2^{-n_0+1} \pi \hat{m} \, \leq \,
&s \arccos G \; \mathrm{mod}\, 2\pi \, \leq \,
2^{-n_0+1} \pi (\hat{m}+1), \\
\label{eq:bin1}
&s \arccos \mathcal{E} \; \mathrm{mod}\, 2\pi \, \geq \, 2^{-n_0+1} \pi (\hat{m}+2)  
\end{align} 
for a given sign $s\in \lbrace 1, -1 \rbrace$. Counts for \SEVE{} and \SEVEplus{} are then expressed in units of the baseline QPE complexities, $c_{\mathsf{bQPE}}$. 

In \SEVE{}, we would add $n_{\mathsf{x}}$ additional qubits to separate the eigenphases of the Hamiltonian ground state from the eigenphase of the first excited state. Every qubit added to the $\mathsf{iQPE}$ circuit with respect to the baseline QPE roughly doubles its complexity, and so we find its complexity to be upper-bounded by $2^{n_{\mathsf{x}}}c_{\mathsf{bQPE}}$.

In \SEVEplus{}, we would use a symmetric QSP sequence \cite{wang2022energy} to round every qubit to its closest value \cite{rall2021faster} with QSP. We choose the same target polynomial $P(\cdot)$ for all QSP routines $\mathcal{W}_x$ associated with the rounding of the qubit $\mathsf{phase}[x]$, i{.}e{.} we pick $P_x(\cdot)=P(\cdot)$ for all $x$. The polynomial $P(\cdot)$ is a degree $d$ approximation of 

\begin{align}
\label{eq:polyapproxi}
P(z) \approx 1 - \frac{1}{2} \textsf{erf} \bigg( k \bigg(\frac{1}{\sqrt{2}} - z \bigg) \bigg) - \frac{1}{2} \textsf{erf} \bigg( k \bigg( \frac{1}{\sqrt{2}} + z \bigg) \bigg),
\end{align}
 where $\mathsf{erf}$ is the Gaussian error function. The variable $k$ is inversely related to $\kappa_x$-region (Eq.~73 in Ref.~\cite{Low2017}) where we have chosen $\kappa_x=0.25$. Note that it is conceivable to relax the degree requirements for polynomials $P_y(\cdot)$ of more significant qubits $\mathsf{phase}[y]$, but this would, at most, improve the overall complexity by a factor of $2$. With a uniform $P(\cdot)$, the complexity of the $\mathsf{iQPE}$ can be upper bounded by $2d\cdot c_{\mathsf{bQPE}}$. The factor of $2$ is due to one extra qubit being added to $\mathsf{iQPE}$ in \SEVEplus{}---a modification we have found to be necessary even when the polynomial is the perfect step function, i{.}e{.} $P(\cdot)=S(\cdot)$. 

\begin{figure*}[tb]
    \centering
\begin{tikzpicture}
    \node at (0,0){\includegraphics[scale=0.5]{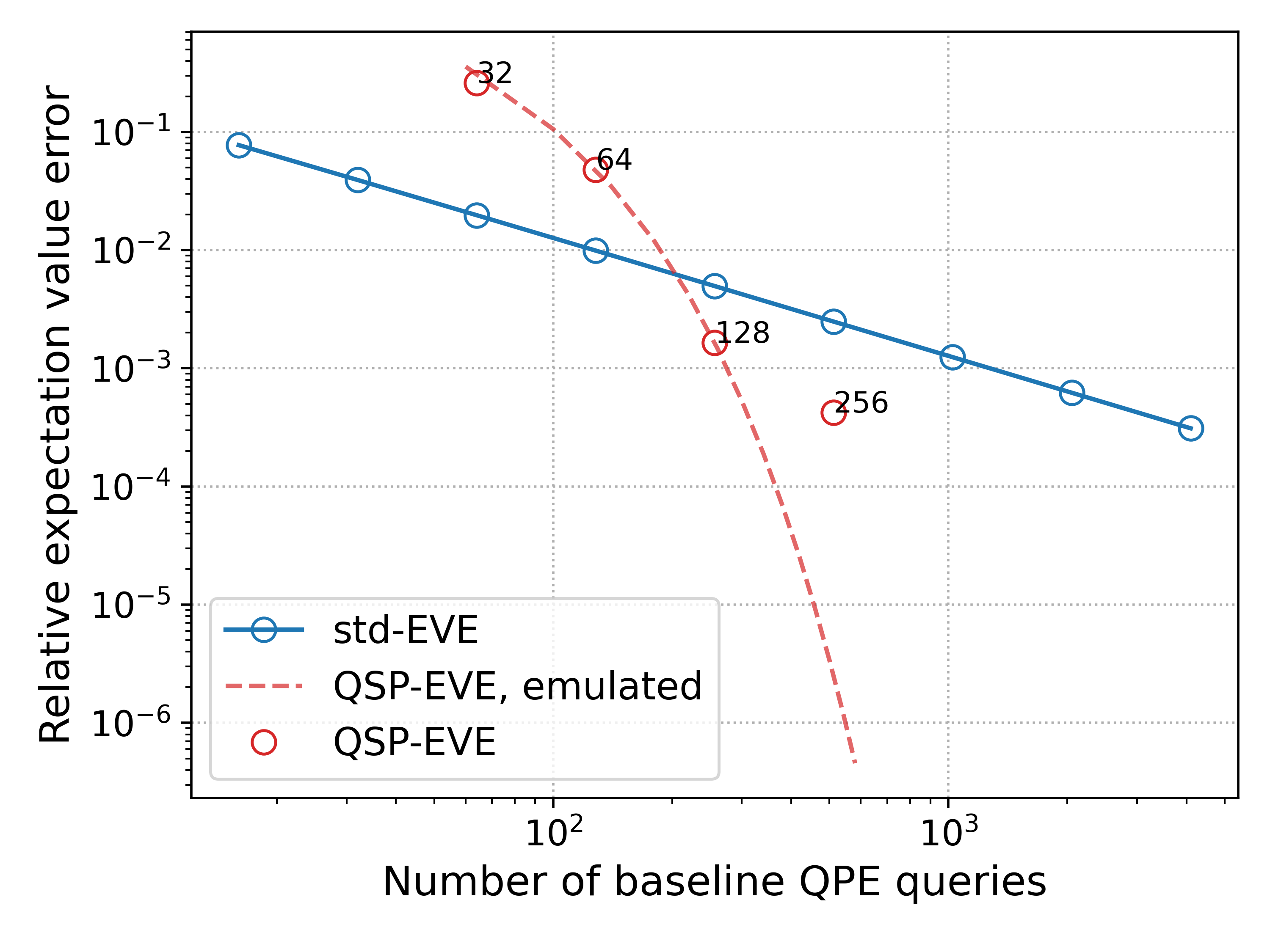}};
    \node at (8,.076) {\includegraphics[scale=0.5]{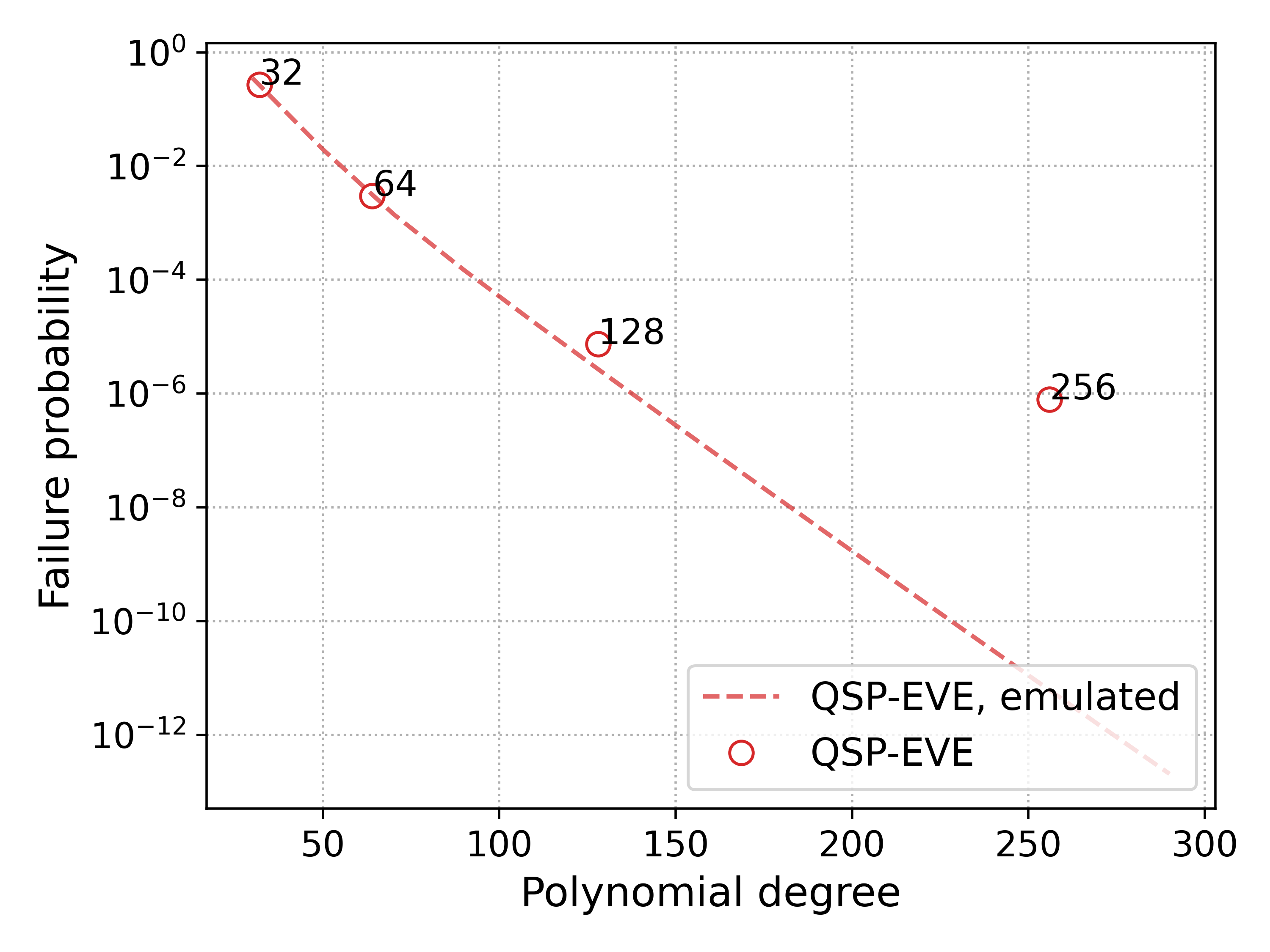}};
    \node at (-3.4,3) {$(a)$};
    \node at (4.5,3) {$(b)$};
\end{tikzpicture}    
    \caption{Performance of \SEVE{} and \SEVEplus{} algorithms in general. $\boldsymbol{(a)}$ Comparing the expectation value errors in \SEVE{} and \SEVEplus{} with respect to the complexity of the $\mathsf{iQPE}$ in terms of $c_{\mathsf{bQPE}}$, the complexity of a baseline QPE with $n_0=20$ qubits. These results are relatively robust against change in the number of qubits $n_0$ and are relative to the cost of a Hamiltonian block encoding. In \SEVE{}, the expectation value errors decrease roughly linearly with the complexity of $\mathsf{iQPE}$, which is set by the number of extra qubits $n_{\mathsf{x}}$: $2^{n_{\mathsf{x}}} c_{\mathsf{bQPE}}$. In \SEVEplus{}, the expectation value error decreases exponentially with the degree $d$ of the target polynomial $P(\cdot)$. The complexity of $\mathsf{iQPE}$ in \SEVEplus{} is given by individual degrees as $2 d \cdot c_{\mathsf{bQPE}}$. The dashed line presents the ``emulated" \SEVEplus{} curve, which is the curve for an ideal implementation of the target polynomial $P(\cdot)$. The singular data points represent the errors obtained with pre-computed phase factors. Every such data point is decorated with its corresponding degree $d$. The QSP angles used in the data points have a precision of 21 bits. We can see that the data points follow the theoretical prediction but diverge after $d = 128$ due to the suboptimality of the computed phase factors. $\boldsymbol{(b)}$ Minimum failure probability $1 - \langle \boldsymbol{0} | \varrho(G,s) | \boldsymbol{0} \rangle$  of \SEVEplus{} different from unity, as a function of the polynomial degree $d$. The dashed curve outlines the theoretically achievable curve, whereas the data points denote results based on numerically attained phase factors. The success probability climbs quickly from roughly $8/\pi^2 \approx 0.81$ at $d=1$ towards $1$ as $d$ increases. At $d=128$, the failure probability is $7\cdot 10^{-6}$. The underlying $\mathsf{iQPE}$ routine has a total of $20$ $\mathsf{phase}$ qubits, and the bit precision for QSP angles is 21. The minimum failure probability of \SEVE{} is constant at 19\%.}
    \label{fig:seve_comparison}
\end{figure*}

 The achievable precisions for \SEVE{} and \SEVEplus{} in terms of the target error of the observable (relative to $\lambda_F$) are depicted in Fig.~\ref{fig:seve_comparison}$(a)$ as a function of their algorithmic complexities in units of multiples of the baseline QPE. The dashed curve of \SEVEplus{} in the graph denotes the expectation value errors with respect to target polynomial $P(\cdot)$, while the data points denote expectation value errors obtained from QSP circuits using pre-computed phase factors with finite bit precision. The phase angles were obtained through optimization using the LBFGS solver in QSPPACK \cite{dong2021efficient}. While the first expectation value errors match the theoretical predictions of the dashed curve, we see a clear difference after $d = 128$. We attribute this behavior to a suboptimality of the phase factors due to the increasing hardness of the underlying optimization problem for larger $d$. {\em Infinite QSP}, a different optimization method \cite{dong2022infinite}, which performs well for target functions such as sine and cosine, is ineffective here, as the Chebychev coefficients of $P(\cdot)$ do not decay rapidly enough. 
  For our results in Fig.~\ref{fig:seve_comparison}$(a)$, we have chosen a baseline QPE with $n_0=20$ qubits. The presented data is relatively robust against varying numbers of baseline qubits $n_0$---as long as $n_0$ is sufficiently large---since the discretization error only affects the first few insignificant qubits.

The data shows a rapid decay of the expectation value error by increasing the complexity of \SEVEplus{} -- a decay that turns out to be exponential. We attribute this behavior to our ability to exponentially suppress the deviation of $P(\cdot)$ from $S(\cdot)$ at their  plateaus. 
 At the same time, the expectation value error decays roughly linear with the complexity of $\mathsf{iQPE}$ in \SEVE{}. It follows that the overall complexities of both algorithms, see Eq.~\eqref{eq:bigcomplexity}, become 
\begin{align}
\label{eq:final_complex}
O\!\left(\frac{\lambda_F^2\lambda_H}{\Delta\, \varepsilon^2}\right)\, \quad \text{and} \quad O\!\left(\frac{\lambda_F\lambda_H}{\Delta\, \varepsilon} \log\!\left(\frac{\lambda_F}{\varepsilon}\right)\right)\,
\end{align}
for \SEVE{} and \SEVEplus{}, respectively. It should be noted that for very large choices of target error (up until roughly $6\cdot10^{-3}$) that \SEVE{} makes fewer queries to the baseline QPE than \SEVEplus{}.  This target error regime is insufficient for our application, but in principle, if one could relax the target error sufficiently, the overhead introduced by the use of QSP makes \SEVEplus{} less competitive; this may be relevant in other applications where QPE is used on its own rather than as part of a larger routine such as in this algorithm, where the $\mathsf{iQPE}$ is used in a reflection about the ground state.

Note also that the maximal success probability of \SEVE{} is always fixed to roughly $8/\pi^2\approx 81\%$ while it quickly approaches $1$ in \SEVEplus{} as depicted in Fig.~\ref{fig:seve_comparison}$(b)$. Through the use of \SEVEplus{} we reduce the number of individual phase factors from $\Delta^{-1} \log \Delta^{-1} \sim 10^{5}$ in  \cite{mitarai2022perturbation} to $d\sim 10^{2}$.

\subsection{Resource estimates for molecular systems}
\label{sec:numerical_study}
In the following, we calculate resource requirements for computing nuclear forces \cite{o2021efficient}, electric dipole moments \cite{hait2018accurate}, and kinetic energies using \SEVE{} and \SEVEplus{}, relying on the theoretical curve for the latter. The choice of observables covers both one- and two-body operators and thus allows us to cost our algorithms for different tasks. We note that the forces and dipole moments are energy derivatives and could also be estimated using ground state energy calculations together with finite-difference formulas \cite{o2021efficient}. As molecular test systems, we use \ce{H2}, \ce{Be}, a single water molecule, ammonia, and p-benzyne with molecular geometries defined in Tab.~\ref{tab:molecular_geometries} in Appendix~\ref{sec:geometries}. All calculations are performed in the {cc-pVDZ} basis set. The Hamiltonian matrix elements are obtained with the $\mathsf{PySCF}$ code~\cite{sun2020pyscf,sun2018pyscf}. For $\ce{Be}$ and $\ce{H2}$ we calculate the spectral gap using the FCI routine of PySCF~\cite{sun2020pyscf,sun2018pyscf}, for ammonia and water we use the $\mathsf{Block2}$ DMRG code~\cite{block2} with a bond dimension $M=1000$ on the full space, and for p-benzyne in a (30e, 30o) active space. The data is processed and visualized with an in-house software package.

The number of logical qubits required for the execution of each algorithm, as well as the required number of Toffoli gates, is then estimated by matching the precision $O(\lambda_F/\varepsilon)$ with requirements for the complexity of $\mathsf{iQPE}$ in the curves of Fig.~\ref{fig:seve_comparison}$(a)$. A detailed description of the procedure can be found in Appendix~\ref{sec:resource_details}. We summarize the results of our calculations in the following paragraphs and report tables with detailed resource estimates in Appendix~\ref{sec:data_code}.

Accurate nuclear forces are required for many applications, such as optimizing molecular geometries or simulating molecular dynamics \cite{marx2000ab}. For a molecule with $N_\mathrm{a}$ atoms, there are 3$N_\mathrm{a}$ nuclear forces. The $i$-th force component of the $A$-th nuclei can be represented by a two-body force operator $F_A^{(i)}$,  
\begin{align}
F_A^{(i)}= \sum_{pq}f^{(1)}_{pq} a^{\dag}_pa_q+\sum_{pqrs}f^{(2)}_{pqrs} a^{\dag}_pa^{\dag}_r a_qa_s\,,
\end{align}
with the one- and two-body integrals $f^{(1)}_{pq}$ and $f^{(2)}_{pqrs}$ being defined as in Eq.~(22) of Ref.~\cite{o2021efficient}. There is no universally defined notion of chemical accuracy as target precision for the force operator. Instead we follow \cite{o2021efficient} and use $\varepsilon_{\mathrm{force}} = $~5~mHa/\r{A} as target precision. We calculate the resource requirements for all $3N_\mathrm{a}$ operators and report the results in Tab.~\ref{tab:force}. 
\begin{table*}[tb]
    \centering
    \caption{Quantum resources for computing molecular forces using EVE. For the water, ammonia, and p-benzyne system, we present the number of spin orbitals $N$, the norm $\lambda_H$ of the Hamiltonian in Ha, the gap of the Hamiltonian $\Delta_H$ in Ha, and the norm of force operator $\lambda_{F_A^{(i)}}$ in Ha/\r{A} together with the resulting Toffoli and logical qubit counts for \SEVE{} and \SEVEplus{} instances estimating the expectation values of force operators. The force operators of different atoms and different components have different norms. Therefore, we report only the components associated with minimum and maximum gate costs here. For complete datasets, including subroutine cost breakdowns, see Appendix~\ref{sec:data_code}.}
    \vspace{10pt}
\begin{tabular}{l@{\hskip 5mm}rrrcr@{\hskip 5mm}rr@{\hskip 5mm}rr}
\hline\hline
\multicolumn{1}{c}{} & \multicolumn{5}{c}{{Property}} & \multicolumn{2}{c}{{Toffoli gates}} & \multicolumn{2}{c}{{Logical qubits}} \\
System & N & $\lambda_H$ & $\Delta_H$ & & $\lambda_{F_A^{(i)}}$ & \SEVE{} & \SEVEplus{} & \SEVE{} & \SEVEplus{} \\
\hline
\multirow[c]{2}{*}{Water} & \multirow[c]{2}{*}{24} & \multirow[c]{2}{*}{328} & \multirow[c]{2}{*}{0.302} & min & 32 & 7.72e+17 & 4.42e+15 & 1790 & 1360 \\
 & & & & max & 82 & 3.15e+18 & 8.84e+15 & 1826 & 1363 \\
\hline 
\multirow[c]{2}{*}{Ammonia} & \multirow[c]{2}{*}{29} & \multirow[c]{2}{*}{434} & \multirow[c]{2}{*}{0.280} & min & 39 & 1.92e+18 & 5.49e+15 & 2130 & 1597 \\
 & & & & max & 113 & 1.56e+19 & 2.20e+16 & 2174 & 1603 \\
\hline 
\multirow[c]{2}{*}{p-Benzyne} & \multirow[c]{2}{*}{104} & \multirow[c]{2}{*}{3833} & \multirow[c]{2}{*}{0.114} & min & 129 & 2.03e+21 & 3.49e+18 & 7577 & 5881 \\
 & & & & max & 1093 & 1.36e+23 & 2.79e+19 & 7926 & 5891 \\
\hline\hline
\end{tabular}
    \label{tab:force}
\end{table*}
We find that QSP-EVE requires $\sim\!10^{15}$ to $\sim\!10^{19}$ Toffoli gates and thousands of logical qubits to reach the target precision, compared to $\sim\!10^{17}$ to $\sim\!10^{23}$ for std-EVE. For all force components, QSP-EVE significantly outperforms std-EVE in terms of gate counts and number of logical qubits, showing at most a $\sim\!4875\times$ reduction in gate count and $\sim\!26\%$ reduction in qubit count (for p-benzyne).

As a second example, we calculate the resource requirements for estimating the expectation value of the dipole moment operator. The $i^\textrm{th}$ component of dipole moment can be represented as a one-body operator
\begin{align}
D^{(i)}= \sum_{pq} d^{(i)}_{pq} a_p^\dagger a_q\quad
\mathrm{with}\quad d_{pq}^{(i)}
=
\int \mathrm{d} \vec x_1 \
\psi_{p}^{*} (\vec x_1)
\left [
\vec x_{1}^{(i)}
\right ]
\psi_{q} (\vec x_1)
\end{align}
where the index $i$ denotes the Cartesian coordinate. For typical applications, one aims to estimate the dipole moment up to a relative error of a few percent \cite{hait2018accurate}. Assuming $1$~Debye as the typical strength of the dipole moment and a relative error of 1\% motivates our choice of $\varepsilon_{\mathrm{dip.}} = $10~mDebye for this observable. We calculate the resource requirements for all components of the dipole operator. We note that for diatomic molecules, we only calculate the resource requirements for the non-vanishing dipole moment along the internal coordinate. The results for the dipole moment are shown in Tab.~\ref{tab:dipole}. Notably, we find that at most QSP-EVE provides an $\sim\!2376\times$ reduction in gate count and $\sim\!25\%$ reduction in qubit count when compared to std-EVE (for p-benzyne).
\begin{table*}[tb]
    \centering
    \caption{Quantum resources for computing dipole moments using EVE. For the water, ammonia, and p-benzyne system, we present the number of spin orbitals $N$, the norm $\lambda_H$ of the Hamiltonian in Ha, the gap of the Hamiltonian $\Delta_H$ in Ha, and the norm of the dipole operator $\lambda_{D^{(i)}}$ in Debye, along with the resulting numbers of Toffoli gates and logical qubits for \SEVE{} and \SEVEplus{} instances estimating the expectation values of dipole operators along different axes. Note that the z component is omitted for Water due to the system's geometrical symmetry. For complete datasets, including subroutine cost breakdowns, see Appendix~\ref{sec:data_code}.}
    \vspace{10pt}
\begin{tabular}{l@{\hskip 5mm}rrrcr@{\hskip 5mm}rr@{\hskip 5mm}rr}
\hline\hline
\multicolumn{2}{c}{{}} & \multicolumn{4}{c}{{Property}} & \multicolumn{2}{c}{{Toffoli gates}} & \multicolumn{2}{c}{{Logical qubits}} \\
System & N & $\lambda_H$ & $\Delta_H$ & Axis &$\lambda_{D^{(i)}}$ & \SEVE{} & \SEVEplus{} & \SEVE{} & \SEVEplus{} \\
\hline
\multirow[c]{2}{*}{Water} & \multirow[c]{2}{*}{24} & \multirow[c]{2}{*}{328} & \multirow[c]{2}{*}{0.302} & x & 65 & 3.15e+18 & 8.98e+15 & 1818 & 1355 \\
 & & & & y & 17 & 1.90e+17 & 2.25e+15 & 1746 & 1349 \\
\hline 
\multirow[c]{3}{*}{Ammonia} & \multirow[c]{3}{*}{29} & \multirow[c]{3}{*}{434} & \multirow[c]{3}{*}{0.280} & x & 91 & 1.56e+19 & 2.23e+16 & 2165 & 1594 \\
 & & & & y & 20 & 4.71e+17 & 5.58e+15 & 2082 & 1588 \\
 & & & & z & 26 & 9.41e+17 & 5.58e+15 & 2083 & 1588 \\
\hline 
\multirow[c]{3}{*}{p-Benzyne} & \multirow[c]{3}{*}{104} & \multirow[c]{3}{*}{3833} & \multirow[c]{3}{*}{0.114} & x & 239 & 8.25e+21 & 7.07e+18 & 7681 & 5872 \\
 & & & & y & 205 & 8.25e+21 & 7.07e+18 & 7681 & 5872 \\
 & & & & z & 481 & 3.35e+22 & 1.41e+19 & 7797 & 5875 \\
\hline\hline
\end{tabular}
    \label{tab:dipole}
\end{table*}

As second one-body operator, we study the kinetic energy operator of the electronic structure Hamiltonian
\begin{align}
K = \sum_{pq} k_{pq} a_p^\dagger a_q\quad
\mathrm{with}\quad k_{pq}
=
\int \mathrm{d} \vec x_1 \
\psi_{p}^{*} (\vec x_1)
\left [
-
\frac{1}{2}
\nabla_i^2
\right ]
\psi_{q} (\vec x_1)
\,
.
\end{align}
The expectation value of the kinetic energy operator can serve as a check of the correctness of the wave function through the virial theorem \cite{slater1933virial}. As the kinetic energy operator is an energy operator, we choose chemical accuracy as target precision, i.e., $\varepsilon_{\mathrm{kin.}} =$~1.6~mHa. The results for the kinetic energy are shown in Tab.~\ref{tab:kinetic}. Similarly to the previous observables, we find that at best QSP-EVE provides a $\sim\!4564\times$ reduction in gate count and $\sim\!26\%$ reduction in qubit count when compared to std-EVE (for p-benzyne).
\begin{table*}[b]
    \centering
    \caption{Quantum resources for computing kinetic energies using EVE. For the \ce{H2}, \ce{Be}, water, ammonia, and p-benzyne system, we present the number of spin orbitals $N$, norm of the Hamiltonian $\lambda_H$ in Ha, the gap of the Hamiltonian $\Delta_H$ in Ha, and the norm of kinetic energy operator $\lambda_K$ in Ha, along with the resulting number of Toffoli gates and logical qubits for \SEVE{} and \SEVEplus{} instances estimating the expectation values of kinetic energy operators. For complete datasets, including subroutine cost breakdowns, see Appendix~\ref{sec:data_code}.}
    \vspace{10pt}
\begin{tabular}{l@{\hskip 5mm}rrrr@{\hskip 5mm}rr@{\hskip 5mm}rr}
\hline\hline
{} & \multicolumn{4}{c}{{Property}} & \multicolumn{2}{c}{{Toffoli gates}} & \multicolumn{2}{c}{{Logical qubits}} \\
System & N & $\lambda_H$ & $\Delta_H$ & $\lambda_K$ & \SEVE{} & \SEVEplus{} & \SEVE{} & \SEVEplus{} \\
\hline
\ce{H2} & 10 & 71 & 0.472 & 17 & 1.56e+17 & 4.08e+14 & 932 & 665 \\
Be & 14 & 65 & 0.207 & 15 & 2.15e+17 & 5.65e+14 & 1152 & 829 \\
Water & 24 & 328 & 0.302 & 99 & 5.21e+19 & 3.82e+16 & 1890 & 1361 \\
Ammonia & 29 & 434 & 0.280 & 91 & 1.29e+20 & 4.74e+16 & 2244 & 1597 \\
p-Benzyne & 104 & 3833 & 0.114 & 316 & 1.36e+23 & 2.98e+19 & 7913 & 5878 \\
\hline\hline
\end{tabular}
    \label{tab:kinetic}
\end{table*}

For all three observables, we find that QSP-EVE results in around 2-4 orders of magnitude lower Toffoli gate counts and logical qubits numbers over std-EVE, as well as up to an approx.\ 25\% reduction in qubit counts. In combination, such improvements lead QSP-EVE to provide up to five orders of magnitude to reduce circuit volume for the systems studied.

\begin{figure*}[tb]
    \centering
    \begin{tikzpicture}
        \node at (0,0) {\includegraphics[width=0.48\linewidth]{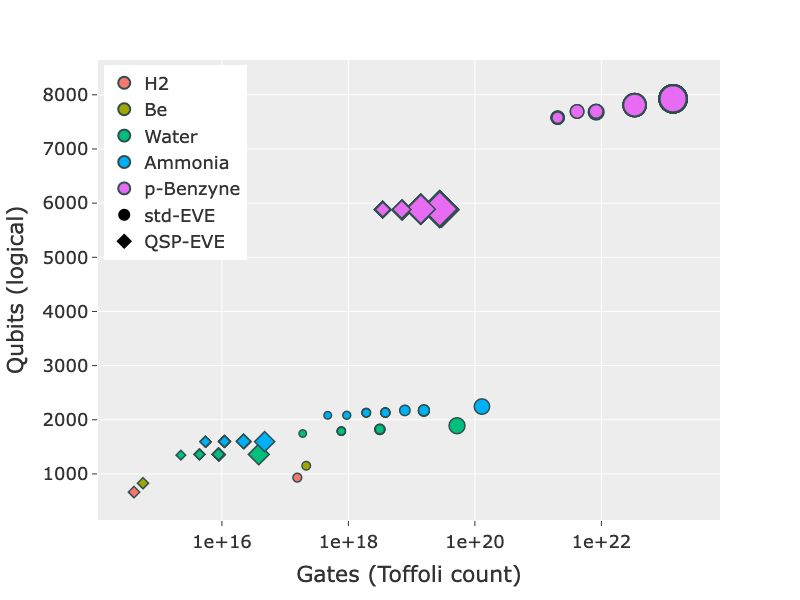}};
        \node at (8,0) {\includegraphics[width=0.48\linewidth]{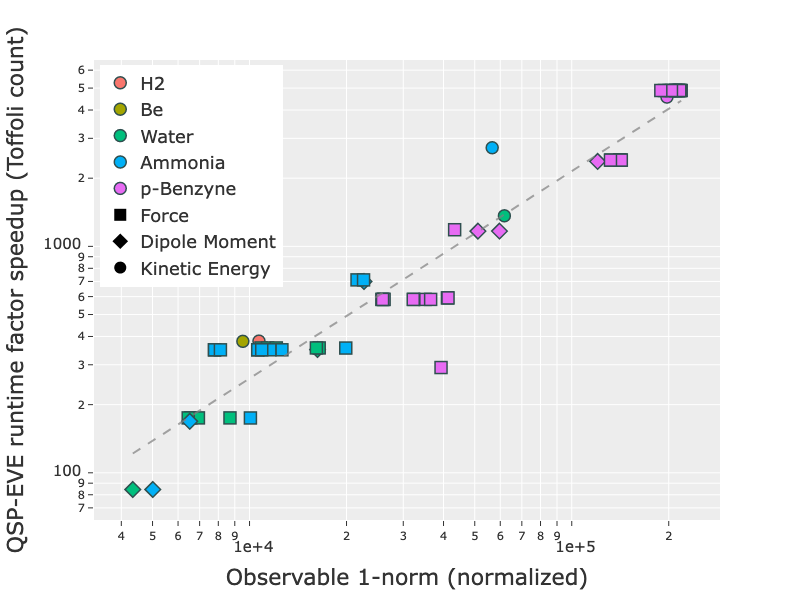}};
        \node at (-3.7, 3.0) {$(a)$};
        \node at (4.5, 3.0) {$(b)$};
    \end{tikzpicture}
    \caption{Resource analysis for all observables considered (3 force components, kinetic energies and dipole moments) of various molecules in cc-pVDZ basis. $\boldsymbol{(a)}$ Logical qubit counts plotted against Toffoli gate counts for the estimation for \SEVE{} and \SEVEplus{} algorithms. Marker sizes are proportional to observable 1-norms, and each data point represents one of the observables. $\boldsymbol{(b)}$ Advantage of \SEVEplus{} over \SEVE{}. Toffoli gate count of the \SEVE{} algorithm relative to a run of \SEVEplus{} with respect to different normalized observable 1-norms $\Lambda_F$. That is, the absolute 1-norm of the observable $\lambda_F$, relative to the target error $\varepsilon$, which is $1.6\,$mHa for kinetic energies, $10\,\text{mDebye}$ for dipole moments and $5\,\text{mHa/\AA}$ for the forces, such that $\Lambda_F = \lambda_F / \varepsilon$. This quantity has been chosen as it indicates the complexity of the outer QPE. The linear fit (dashed line) in the plot roughly confirms the scaling of $O(\Lambda_F/\log \Lambda_F)$ expected from Eq.~\eqref{eq:final_complex}, with a gradient of roughly $0.91\Lambda_F$.}
    \label{fig:resource_plots}
\end{figure*}

We summarize all our resource estimates in Fig.~\ref{fig:resource_plots}$(a)$, displaying the logical-qubit and magic-state requirements. 
The complete data is also listed in Appendix \ref{sec:data_code}. We confirm that the resource requirements for \SEVEplus{} are lower than for \SEVE{} in all cases considered. The relative advantage of \SEVEplus{} over \SEVE{} is separately depicted in Fig.~\ref{fig:resource_plots}$(b)$ in the number of Toffoli gates for the listed observables with respect to their relative 1-norms $\Lambda_F=\lambda_F/\varepsilon$. Eq.~\eqref{eq:final_complex} indicates that the advantage should scale as $O(\Lambda_F \log \Lambda_F)$.
A fit through the data points roughly confirms this behavior.

\begin{figure*}[ht!]
    \centering
 \includegraphics[scale=0.6]{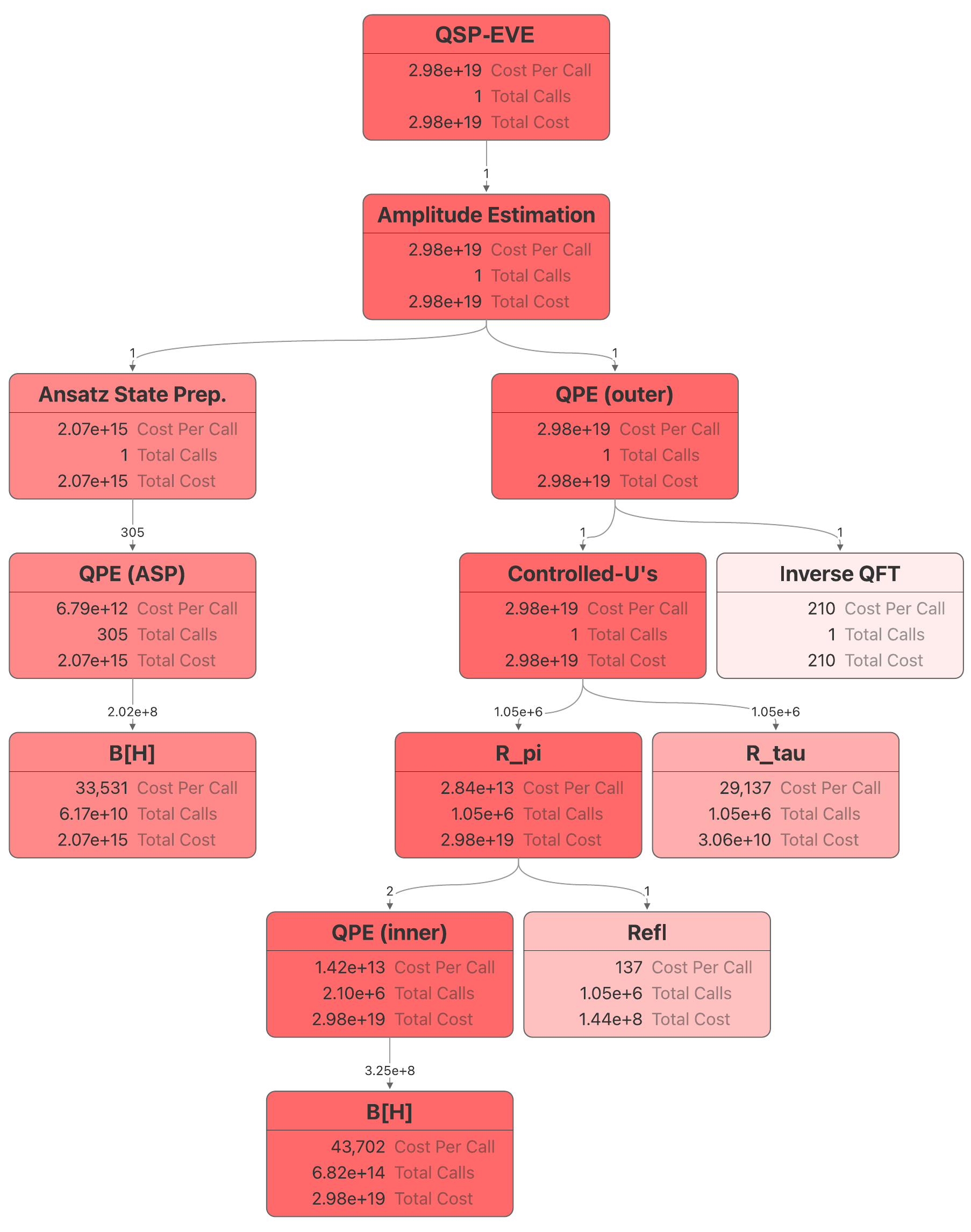}
    \caption{Callgraph depicting the estimated quantum resources required for computing the kinetic energy of a p-benzyne molecule using \SEVEplus{}. The graph displays the distribution of the costs among various subroutines, named as in Fig.~\ref{fig:overview}, and where $\mathsf{B[H]}$ is the block encoding of the Hamiltonian and $R_\tau$ is the block encoding of the observable (i.e. $\mathsf{B[F]}$). All costs are given in terms of Toffoli gate counts, with each subroutine node depicting the per-call cost, the total number of calls and cost when taken over the full algorithm (i.e. over all parent calls). Note that some routines are deliberately omitted in the count, as they contribute very little. Edge numbers define the number of calls of the target routine within a single call of its parent routine. Darker shading indicates a greater total gate cost for the subroutine. The $\mathsf{ASP}$ comprises several low-precision phase estimation routines in the repeat-until-success circuit. The initial state overlap of the Hartree-Fock state with the Hamiltonian ground state is assumed to be 1\% and is subsequently lifted to above 91\% (via a simple repeat-until-success scheme). Details on the resource estimation of all subroutines can be found in Appendix \ref{sec:resource_details}. The callgraph diagram for the same problem instance in \SEVE{} can be found in Fig{.}~\ref{fig:callgraph} in Appendix~\ref{sec:data_code}}
    \label{fig:callgraph_plus}
\end{figure*}

An overview of the cost distribution of \SEVEplus{} can be found in Figure \ref{fig:callgraph_plus}, where the gate counts for estimating the kinetic energy of a p-benzyne system are split into the different subroutines, down to the block encodings of $\hat{H}$ and $\hat{F}$. 
The majority of the cost (not just for \SEVEplus{} but also for \SEVE{} as can be seen in Appendix \ref{sec:resource_details}) comes from the block encodings of the Hamiltonians $\mathcal{B}[H]$, not because a single call is expensive but due to the large number of calls throughout the algorithm. The cost for a repeated phase estimation within the Ansatz state preparation (the $\mathsf{ASP}$ routine in Fig.~\ref{fig:overview}) is included in the resource counts as well as the callgraph in Fig.~\ref{fig:callgraph_plus}. We assume a pessimistic overlap of 1\% between the Hamiltonian ground state $|\psi_G\rangle$ and the Hartree-Fock state that serves as input on the $\mathsf{sim}$ register.

\section{Conclusion}
\label{sec:discussion}

In this work, we have presented two quantum algorithms for expectation value estimation, \SEVE{} and \SEVEplus{}, and calculated their computational costs. Comparing the two algorithms we have shown that by exploiting the latest developments in quantum signal processing (QSP) it is possible to improve the asymptotic quantum resource requirements. Additionally, we have provided explicit resource estimates for calculating the expectation values of molecular forces, dipole moments, and kinetic energies for different exemplary molecules, requiring between $\sim\!10^3$ and $\sim\!10^4$ logical qubits and between $\sim\!10^{15}$ and $\sim\!10^{19}$ Toffoli gates. Furthermore, we have presented a breakdown of the contributions of the different algorithm's subroutines to the total computational cost in Figure~\ref{fig:callgraph}. We found that the inner phase estimation within the reflections in the EVE algorithms represents the most significant source of computational cost, driven by the direct dependence on the Hamiltonian norm and the spectral gap.

Complementary to the resource estimates, we have obtained the QSP phase angles that would be required at the time of the algorithm's compilation. Although the results of the numerical optimization are currently unsatisfactory for larger problem instances, we have lowered the required numbers of individual phase factors sufficiently for us to believe that the QSP phase angles can be obtained even for larger instances with more computing power or through further research into optimization methods. Our research makes clear that, although QSP techniques are the current state of the art in fault-tolerant algorithms, they do imply a heavy classical precomputation. It is important to balance this increase in classical computational cost versus the reduction of quantum resources requirements when making algorithm design choices.

We acknowledge that the current gate counts are still unreasonably high to run on mid-term fault-tolerant hardware. Once algorithmic improvements yield more feasible resource estimates, architecture-based optimization might drive the cost down further. Most importantly, our results show that adopting ideas from QSP while limiting the complexity of its optimization within the EVE algorithm can bring up to three orders of magnitude improvement when it comes to the number of Toffoli gates.

Consequently, future algorithmic improvements together with more efficient Hamiltonian compression schemes \cite{cohn2021quantum, Google_THC} could yield orders of magnitude improvements, as already seen for estimating the Hamiltonian's ground-state energy~\cite{Poulin2014, Reiher2017, babbush2018encoding, Berry_2019, Google_THC, first_quant, PsiQ_battery, Xanadu_battery,VonBurg2021,goings2022}. We identified the Hamiltonian simulation subroutine as the most cost determining part of our algorithm and any progress in this area would directly reduce our resource requirements as well.

A final observation concerns the use of QSP for QPE as a subroutine, related to the comment made in Section~\ref{sec:comparison} where we compare the resource costs of \SEVEplus{} and \SEVE{}. For large target errors, the number of extra phase bits to use in \SEVE{} is small enough such that it outperforms \SEVEplus{} (in terms of queries to the baseline QPE). While for all examples considered here the precision requirements favor the use of QSP-EVE, there may be contexts with less stringent target error requirements in which std-EVE may be a better choice. For instance, in ground-state energy estimation where one makes a single query to QPE, an upper bound on the probability of success for estimating an eigenenergy to some precision is already $81\%$. In this scenario, one may not need to account for bit discretization error, and can instead opt to run standard QPE a handful of times rather than a potentially more expensive QSP QPE. This work highlights the importance of selecting appropriate variants of QPE within different coherent and non-coherent settings or for different required precisions, and the significant resultant impact upon resource requirements.

\section{Acknowledgments}

MS, WP, SS, and SMS thank all our colleagues at PsiQuantum for useful discussions. In particular, we thank Owen Williams for implementing the callgraph visualization engine, and Harriet Apel for her comments on the manuscript. MD, NM, RS and MS thank Clemens Utschig-Utschig for his comments on the manuscript and the support during the project. We also thank Sophia Simon for her feedback.
\bibliography{biblio}
\bibliographystyle{quantum}
\newpage

\appendix
\section{Singular value estimation details}
\label{sec:singvalest}
In this Section, we provide details about the eigenvalues in Eq.~\eqref{eq:eigenvalues} and eigenstates in Eq.~\eqref{eq:eigenvectors} of the iterate featuring the product of two general reflections. We start by recalling Eq.~\eqref{seve3} and Eq.~\eqref{seve4}, and conclude that the projectors have the form
\begin{align}
\label{seve5}
\hat{\pi} = \sum_k |p_k\rangle\!\langle p_k | + \pi^\perp\quad \text{and}\quad 
\hat{\tau} = \sum_k |t_k\rangle\!\langle t_k | + \tau^\perp\, ,
\end{align}
such that $\langle t_j | p_k \rangle = \langle t_j |\hat{\tau}\cdot \hat{\pi} |p_k \rangle =\delta_{jk} w_k$ and
where $\pi^\perp$ and $\tau^\perp$ are projectors orthogonal to all $|p_k\rangle$ and $|t_k\rangle$, as well as $\pi^\perp\cdot\tau^\perp = 0$.
In that notation, we can make an Ansatz vector for an eigenstate of $\mathcal{R}_\tau \mathcal{R}_\pi$:
\begin{align}
\label{seve7}
 |v_k\rangle = \; a_k | p_k \rangle + b_k | t_k \rangle \, ,
\end{align}
with coefficients $a_k$ and $b_k$. For $\mathcal{R}_\tau \mathcal{R}_\pi|v_k\rangle$ we find
\begin{align}
\label{seve8}
&(1-2 \hat{\tau}) (1-2 \hat{\pi}) (a_k | p_k \rangle + b_k | t_k \rangle) \\
\notag
& = \; (1-2 \hat{\tau}) ((-a_k - 2 b_k w_k) | p_k \rangle + b_k | t_k \rangle) \\ \notag
& = \; (-a_k - 2 b_k w_k) | p_k \rangle + (- b_k + 2 a_k w_k+   4 b_k w_k^2 ) | t_k \rangle   \, .
\end{align}
If $|v_{k}\rangle$ is an eigenstate of $\mathcal{R}_\tau\mathcal{R}_\pi$, then the last line must be equal to $\lambda_{k} |v_{k}\rangle$, where $\lambda_{k}$ is an eigenvalue. Comparing coefficients, we retrieve the following system of equations:
\begin{align}
\label{seve9}
 -a_k - 2 b_k w_k &= \lambda_{k} a_k \, , \\
\label{seve10}
 - b_k + 2 a_k w_k+   4 b_k w_k^2 &= \lambda_{k} b_k \, .
\end{align}
Solving $\eqref{seve9}$ for $a_k$ yields
\begin{align}
\label{seve12} 
a_k = - \frac{2 w_k}{1 + \lambda_k} b_k \, ,
\end{align}
which can be used to eliminate $b_k$ from Eq.~\eqref{seve10}:
\begin{align}
\label{seve13} 
\lambda_{k} &= - 1 + 2 \left(- \frac{2 w_k}{1 + \lambda_{k}}\right) w_k+   4 w_k^2 \\
& = - \frac{4 w_k ^ 2}{1 + \lambda_{k}} +   4 w_k^2 - 1\, .
\notag
\end{align}
We then convert Eq.~\eqref{seve13} into a quadratic equation.
\begin{align}
\label{seve14}
 (1 + \lambda_{k})\lambda_{k} &= - 4 w_k ^ 2+ (1 + \lambda_{k}) (4 w_k^2 - 1)\\ \notag
  (1 + \lambda_{k}) \lambda_{k} - (4w_k - 1 )\lambda_{k} &= - 4 w_k ^ 2+   (4 w_k^2 - 1)\\
  \notag
  \lambda_{k}^2 + 2(1 - 2 w_k^2) \lambda_{k}   &= -1
\end{align}
The quadratic equation can be solved for $\lambda_k$ by
\begin{align}
\label{seve15}
  \lambda_{k} &= -(1 - 2 w_k ^ 2) \mp \sqrt{(1 - 2 w_k^2)^2 -1 } \\
  \notag
  &= (2 w_k^2 - 1) \pm i \sqrt{1 - (2 w_k^2 - 1)^2 }\\
  \notag
  &= \exp\left(\pm i2 \arccos w_k\right) ,
\end{align}
since $\cos (2\arccos w_k)= 2 w_k ^2 - 1$. Now we turn our attention to the eigenstate: making the eigenvalue in Eq.~\eqref{seve15} explicit in Eq.~\eqref{seve12}, we obtain 
\begin{align}
a_k &= - \frac{2 w_k}{1 + \exp (\pm i2 \arccos w_k)} \, b_k\notag\\
  &= - \underbrace{\exp (\pm i \arccos w_k)}_{w_k \pm i \sqrt{1 - w_k^2}} \, b_k\, ,
\end{align}
where we have used $(1 + e^{-2ix}) = (2 e^{-ix} \cos x)$ for arbitrary $x$.
Plugging the new expression for $a_k$ into $(7)$ gives us
\begin{align}
|v_{k\pm}\rangle = b_k \left( |t_k\rangle - w_k |p_k\rangle \mp i \sqrt{1- w_k^2} |p_k\rangle \right) \, .
\end{align}
Finally, substituting $|p_k\rangle$ ($|t_k\rangle$) with the definition of the orthogonal state $|t_k^\perp\rangle$ ($|p_k^\perp\rangle$) in Eq{.}~\eqref{eq:perpstates}, we prove Eq.~\eqref{eq:eigenvectors}.
\section{Details about errors and failure probabilities}
\label{sec:errppendix}
In this Section, we prove the following two statements. 
\begin{description}
\item[I. Expectation value error] The systematic error to the estimated expectation value in the approximation outlined in Section \ref{sec:errors} is $2 \max_j\Omega_j$.
\item[II. Success probability] The probability of the algorithm succeeding within the approximation outlined in Section \ref{sec:errors} is between $\langle\boldsymbol{0}|\varrho(G,s)|\boldsymbol{0}\rangle / 2$ and $\langle\boldsymbol{0}|\varrho(G,s)|\boldsymbol{0}\rangle $ . 
\end{description}
Let us start with the first statement by recalling the singular value decomposition in Eq.~\eqref{eq:second_singvals} while also noting that 
\begin{align}
\label{seve43}
\varrho(G,s) = \sum_j |G^s_j \rangle\! \langle G^s_j| \quad \text{and} \quad \varrho(\mathcal{E},s) = \sum_j |\mathcal{E}^s_j \rangle\! \langle \mathcal{E}^s_j|\, .
\end{align}
Using $ \langle G^s_j | \varrho(G,s) \cdot \varrho(\mathcal{E},s)|\mathcal{E}^s_k\rangle = \delta_{jk} \Omega_j$ and the spectral decomposition of $\hat{F} = \sum_\eta \eta\cdot |\phi_\eta \rangle\!\langle\phi_\eta|$, we find
\begin{widetext}
\begin{align}
\label{seve45}
\hat{\pi}\cdot \hat{\tau}\cdot \hat{\pi} = |\boldsymbol{0} \rangle\! \langle \boldsymbol{0} |_{\mathsf{enc}[F]} \otimes \sum_j
\frac{1}{4}\left[\begin{matrix} 1- \langle \psi_G |\hat{F}| \psi_G \rangle & - \Omega_j \langle \psi_G |\hat{F}| \psi_\mathcal{E} \rangle \\ - \Omega_j \langle \psi_\mathcal{E} |\hat{F}| \psi_G \rangle & 1 -\langle \psi_\mathcal{E} |\hat{F}| \psi_\mathcal{E} \rangle \end{matrix} \right]_j
\end{align}
\end{widetext}
where the matrices $[\;\cdot\;]_j$ are formulated with respect to the basis states
\begin{align}
&|Q_{G, s}\rangle_{\subalign{\\&\mathsf{sim}   \\ &\mathsf{enc}[H]}} \otimes |G^s_j\rangle_{\mathsf{phase}}\, , \quad |Q_{\mathcal{E}, s}\rangle_{\subalign{\\&\mathsf{sim}   \\ &\mathsf{enc}[H]}}\otimes|\mathcal{E}^s_j\rangle_{\mathsf{phase}} \, .
\end{align}
From here on we will use the notation $F_E = \langle \psi_E |\hat{F}| \psi_E \rangle$ for $E \in \lbrace G,\, \mathcal{E}\rbrace$ and $\langle \psi_\mathcal{E} |\hat{F}| \psi_G \rangle = z$. The matrix
\begin{align}
\label{seve46}
\frac{1}{4}\left(\begin{matrix} 1-F_G & -\Omega_j z^* \\ -\Omega_j z & 1-F_\mathcal{E} \end{matrix} \right)
\end{align}
has the eigenvalues 
\begin{align}
\label{seve47}
w^2_{j\pm} = \frac{1}{4}\left( 1 - \frac{F_G   + F_\mathcal{E}}{2} \right) \pm \frac{1}{4}\sqrt{\frac{\left( F_G - F_\mathcal{E} \right)^2}{4} + |\Omega_j z|^2}\, .
\end{align}
The eigenvalues $w^2_{j+}$ and $w^2_{j-}$ belong to different solutions to the problem. They are squares of the singular values $w_{j\pm}$, but only one of them will be used to estimate the value of $F_G$. Note that Eq.~\eqref{seve46} is related to a representation of $\hat{F}$ in the subspace of $|\psi_G\rangle$ and $|\psi_\mathcal{E}\rangle$:
\begin{align}
\label{seve48}
\hat{F} = \left(\begin{matrix} F_G & z^* \\ z & F_\mathcal{E} \end{matrix} \right)\, {,}
\end{align}
which has the eigenvalues $\lambda_{\pm}$:
\begin{align}
\label{seve49}
\lambda_{\pm} = \frac{F_G + F_\mathcal{E}}{2} \pm \sqrt{\frac{(F_G - F_\mathcal{E})^2}{4} + |z|^2}\, .
\end{align}
Inspecting both Eq.~\eqref{seve47} and Eq.~\eqref{seve49}, we find
\begin{align}
\label{seve50}
\left. w^2_{j\pm}\vphantom{\frac{1}{1}}\right|_{\Omega_j = 1} =\quad \frac{1- \lambda_{\pm}}{4} \, ,
\end{align}
which is smaller than $1$, considering that $|| \hat{F} || \leq 1$ due to its block encoding implementation. For a nonzero $\Omega_j$, we would estimate $F_G$ with the solution closest to $(1-F_G)/4$: by using $\sqrt{a^2 + b^2} \leq |a| + |b|$ for any real numbers $a$ and $b$ we find
\begin{align}
\label{seve51}
   \min_{\pm} \left| w^2_{j\pm} - \frac{1- F_G}{4} \right|   & = \; \frac{1}{4}\sqrt{\frac{1}{4}\left(F_G -F_\mathcal{E} \right)^2 + |\Omega_j z|^2}  - \frac{1}{8}\left|\vphantom{\frac{1}{1}}F_G - F_\mathcal{E} \right|   \\ 
  &\leq \;\frac{1}{4} \Omega_j |z|
\end{align}
which means that we spoil the estimate of $F_G$ by at most $\pm |\Omega_j z|$. By inspecting Eq.~\eqref{seve49} we can see that $|z|$ can, for some observables be as big as $2$, without that their eigenvalues would be unbound, $|\lambda_{\pm}|>1$. The accuracy of $F_G$ is therefore upper-bounded by
\begin{align}
\label{seve52}
 \max_j 2 \Omega_j\, ,
\end{align}
proving the first statement. 
We will now verify the statement about the success probability. Let us say that the matrix in Eq.~\eqref{seve46} is solved by a vector $(c_{G,j},\, c_{\mathcal{E},j})^\top$. The left singular vector of $\hat{\tau}\cdot \hat{\pi}$ is therefore
\begin{align}
    |p_j\rangle &= \left(\vphantom{\sum}c_{G,j} |Q_{G,s}\rangle \otimes |G^s_j\rangle_{\mathsf{phase}} + c_{\mathcal{E},j} |Q_{\mathcal{E},s}\rangle \otimes |\mathcal{E}^s_j\rangle_{\mathsf{phase}} \right) \otimes |\boldsymbol{0}\rangle_{\mathsf{enc}[F]}\, .
\end{align}
With the shorthand 
\begin{align}
    |Q_{G,s};\boldsymbol{0};\boldsymbol{0}\rangle = |Q_{G,s}\rangle_{\subalign{\\&\mathsf{sim}   \\ &\mathsf{enc}[H]}} \otimes |\boldsymbol{0}\rangle_{\mathsf{enc}[F]} \otimes | \boldsymbol{0}\rangle_{\mathsf{phase}}
\end{align}
we describe the success probability with the overlaps of the initial state $|Q_{G,s};\boldsymbol{0};\boldsymbol{0}\rangle$ with the two viable solutions of the expectation value estimation: 
\begin{widetext}
\begin{align}
\sum_{\sigma=\pm}\left|\langle Q_{G,s};\boldsymbol{0};\boldsymbol{0}| \cdot \frac{1}{\sqrt{2}}\left(\vphantom{\sum_1} |p_j\rangle + i\sigma |p^\perp_j\rangle \right)\right|^2 =   \left|\langle Q_{G,s};\boldsymbol{0};\boldsymbol{0}|p^\perp_j\rangle\vphantom{\sum}\right|^2 + \left|\langle Q_{G,s};\boldsymbol{0};\boldsymbol{0} |p_j\rangle\vphantom{\sum}\right|^2\, .
\end{align}
\end{widetext}
The overlap with $|p_k^\perp\rangle$ is small and can be neglected: to define $|p_j^\perp\rangle$ we find $|t_j\rangle$ as
\begin{widetext} 
\begin{align}
w_j |t_j\rangle &=   \hat{\tau} \cdot \hat{\pi} |p_j\rangle \\ & = \sum_\eta \sqrt{\frac{1-\eta}{4}} |\omega_\eta\rangle \otimes \left(\vphantom{\sum} c_{G,j} \langle \phi_\eta |\psi_G\rangle |G_j^s\rangle_{\mathsf{phase}} + c_{\mathcal{E},j} \langle \phi_\eta |\psi_{\mathcal{E}}\rangle |\mathcal{E}_j^s\rangle_{\mathsf{phase}} \right) \otimes |\boldsymbol{0}\rangle_{\mathsf{enc}[H]}\, .
\end{align}
\end{widetext}
What is more, we find 
\begin{align}
\omega_j \langle Q_{G, s};\boldsymbol{0}; \boldsymbol{0}|t_j \rangle = \frac{1}{4}\langle \psi_G| (1 - \hat{F}) \left( \vphantom{\sum}|\psi_G\rangle c_{G,j} \langle \boldsymbol{0} |G^s_j\rangle + |\psi_{\mathcal{E}}\rangle c_{\mathcal{E},j} \langle \boldsymbol{0} |\mathcal{E}^s_j\rangle\right)\, 
\end{align}
which equals $\omega^2_j \langle \boldsymbol{0} |G^s_j\rangle $ in the limit of $c_{G,j}\mapsto 1$, $c_{\mathcal{E},j}\mapsto 0$ and $\omega_j \mapsto \sqrt{(1 - F_G) / 4}$. In the same limit, $\langle Q_{G, s};\boldsymbol{0}; \boldsymbol{0}|p_j\rangle  =  \langle \boldsymbol{0} |G^s_j\rangle$ and so 
\begin{align}
    \langle Q_{G, s};\boldsymbol{0}; \boldsymbol{0}|p_j^\perp\rangle \propto \langle Q_{G, s};\boldsymbol{0}; \boldsymbol{0}|t_j \rangle -\omega_j \langle Q_{G, s};\boldsymbol{0}; \boldsymbol{0}|p_j\rangle = 0 \, .
\end{align}
The contribution of $|p_j^\perp\rangle$ thus vanishes in the limit of the observable error being small. 
 The success probability can, therefore, be lower-bounded with
\begin{align}
    \sum_j \left|\langle Q_{G,s}; \boldsymbol{0}; \boldsymbol{0} | p_j \rangle\right|^2 = \sum_j |c_{G,j}|^2 \cdot |\langle \boldsymbol{0} | G_j^s \rangle|^2\, .
\end{align}
So how big is $|c_{G,j}|$? From the multiplication of the first row of the matrix in Eq.~\eqref{seve46} with the vector $(c_{G,j},\, c_{\mathcal{E},j})^\top$, we learn that
\begin{align}
  \frac{1}{2} \Omega_{j} z\cdot c_{G, j} &= \left(\vphantom{\frac{1}{1}}\frac{1}{2} (1 - F_\mathcal{E})- w^2_{j}\right) c_{\mathcal{E}, j}\, . 
 \end{align}
  such that we find that the ratio of $|c_{G, j}|$ and $|c_{\mathcal{E}, j}|$ is
  \begin{align}
  \left| \frac{c_{G, j}}{c_{\mathcal{E}, j}}\right| \; =& \quad \frac{2}{|\Omega_j z| }\left[ \frac{1}{4}\left| F_G - F_\mathcal{E} \right| \vphantom{\frac{1}{1}}\right.\notag\\&\quad\qquad\left. + \sqrt{\frac{1}{16}\left(F_G - F_\mathcal{E} \right)^2 + \frac{1}{4}|z \Omega_j|^2} \right]\\
   \geq& \quad \frac{1}{|\Omega_j z|}\sqrt{\left(F_G - F_\mathcal{E} \right)^2 + |z \Omega_j|^2}\, . 
\end{align}
In the worst case of $F_G - F_\mathcal{E} = 0$, both coefficients are of the same size such that $|c_{G, j}|^2=1/2$, but in the ideal case where $\Omega_j=0$ we would have $|c_{G, j}|^2=1$.
The worst case success probability is, therefore, 
\begin{align}
\frac{1}{2}\sum_{j=1}^{r} |\langle G^s_j| \boldsymbol{0} \rangle|^2 & = \frac{1}{2}\langle\boldsymbol{0}| \varrho(G,s)| \boldsymbol{0}\rangle\, ,
\end{align}
and the best-case success probability is double that, proving the statement about the success probability. Note that it would be possible to have more refined statements about the success probability when having an idea of the gap $F_G - F_{\mathcal{E}}$.
\section{Numerical framework for iQPE}
\label{sec:iQPE}
In this section, we provide a numerical framework that will allow us to estimate the singular values $\Omega_j$ and overlaps $\langle\boldsymbol{0}|\varrho(G,s) |\boldsymbol{0}\rangle$ depending on the set of functions $P_x(\cdot)$, $Q_x(\cdot)$ for all $x = 1 \dots n$, where $n$ is the number of qubits in the $\mathsf{phase}$ register. For this to be possible, we need a basis for the projectors $\varrho(E,\sigma)$ with respect to a fixed tuple $(E,\sigma)$. Let us call the basis states $|\mathfrak{X}^m_{E,\sigma}\rangle$, where $m$ are the integer labels in $\mathcal{M}$, such that
\begin{align}
    \label{eq:varrhobasis}
    \varrho(E,\sigma)_{\mathsf{phase}} = \sum_{m\in \mathcal{M}} | \mathfrak{X}^m_{E,\sigma}\rangle\!\langle \mathfrak{X}^m_{E,\sigma}|_{\mathsf{phase}}\, .
\end{align}
We would need to be able to make statements about the overlaps $\langle\boldsymbol{0}|\mathfrak{X}^m_{E,\sigma}\rangle$ for us to compute $\langle\boldsymbol{0}|\varrho(G,s) |\boldsymbol{0}\rangle$. To attain $\Omega_j$, however, there is no need for us to expand $|\mathfrak{X}^m_{E,\sigma}\rangle$ into the computational basis. The set of states $|\mathfrak{X}^m_{E,\sigma}\rangle$ itself can function as the basis of dimension $r$: the squared singular values $\Omega_j^2$, can be computed by $\varrho(G,s)\cdot \varrho(\mathcal{E},s) \cdot \varrho(G,s)$ in the basis of $\varrho(G,s)$, if we are given a suitable relation of the overlap of two basis states for different tuples $(E,\sigma)$. This section has, therefore, three goals:
\begin{description}
\item[I] To find the states $|\mathfrak{X}^m_{E,\sigma}\rangle$ and prove that they form a basis by showing that $\langle\mathfrak{X}^j_{E,\sigma} | \mathfrak{X}^k_{E,\sigma} \rangle = \delta_{jk}$;
\item[II] To provide an expression for $\langle\boldsymbol{0}|\varrho(G,s) |\boldsymbol{0}\rangle$ that solely depend on $P_x(\cdot)$ and $Q_x(\cdot)$; 
\item[III] To provide expressions $\langle \mathfrak{X}_{\widetilde{E}, \widetilde{\sigma}}^j| \mathfrak{X}_{\vphantom{\widetilde{E}}E,\sigma}^k \rangle$ for the overlaps of two states from different bases $(E,\sigma)\neq (\widetilde{E}, \widetilde{\sigma})$ in terms of the functions $P_x(\cdot)$ and $Q_x(\cdot)$.
\end{description}
 Before we start, it will be necessary to establish some notation. Throughout this section 
 we will write $b_\ell$ for the binary representation of integers $\ell$ between $0$ and $2^n-1$: 
\begin{align}
    \ell \; \mapsto \; b_\ell = (b_{\ell,1} \, b_{\ell,2} \cdots b_{\ell,n})
\end{align} such that 
\begin{align}
\ell = \sum_{x=1}^n b_{\ell,x} \, 2^{n-x}\, .
\end{align}
and represent $|\ell\rangle$ by
\begin{align}
    |\ell\rangle_{\mathsf{phase}} = |b_{\ell,1}\rangle_{\mathsf{phase}[1]} \otimes |b_{\ell,2}\rangle_{\mathsf{phase}[2]} \otimes \cdots \otimes |b_{\ell,n}\rangle_{\mathsf{phase}[n]}\, ,
\end{align}
where $b_{\ell,1}$ is the most-significant bit and $b_{\ell,n}$ is the least significant qubit. We will access the functions $P_x(\cdot)$, $Q_x(\cdot)$ via the matrix-valued function $\mathrm{M}^{(x)}$ of arbitrary input angles $\vartheta$:
 
 \begin{align}
\label{guseve301}
\left(\begin{matrix}
\mathrm{M}^{(x)}_{00}(\vartheta) & \mathrm{M}^{(x)}_{01}(\vartheta) \\[6pt]
\mathrm{M}^{(x)}_{10}(\vartheta) & \mathrm{M}^{(x)}_{11}(\vartheta) \\
\end{matrix}\right)
 =  
 \left(\begin{matrix}
 P_x(\cos \vartheta) & i Q_x(\cos \vartheta)\\[6pt]
    iQ^*_x(\cos \vartheta) & P^*_x(\cos \vartheta)
\end{matrix} \right) \, .
\end{align}
Let us also introduce a function $\mathfrak{Z}_x(\cdot)$ returning the remainder of an $n$-bit integer after the $x$-th significant bit as a fixed-point number:
\begin{align}
\label{guseve315} \mathfrak{Z}_x(k) = \sum_{y>x} b_{k,y} \, 2^{n-y}\, .
\end{align}
Now we can begin working towards the goals \textbf{I}-\textbf{III}. Let us start with a look at the spectral decomposition of the $\mathsf{iQPE}$ operator: with Eq.~\eqref{guseve301} and 
\begin{widetext}
\begin{align}
\label{guseve309}
\mathcal{V}_x   =\sum_{E,\sigma} |Q_{E,\sigma}\rangle\!\langle Q_{E,\sigma}|\otimes \sum_{\ell=0}^{2^n-1}\exp\left(i 2^{x-1} \pi \left[\vartheta_{E,\sigma} - \frac{\mathfrak{Z}_{x}(\ell)}{2^n} \right] X_{\mathsf{phase}[x]} \right) |\ell\rangle\!\langle\ell|_\mathsf{phase}\, ,
\end{align}
\end{widetext}
we can say
\begin{widetext}
\begin{align}
\label{guseve304}
\mathsf{iQPE} &= \mathcal{W}_1 \mathcal{W}_2 \cdots \mathcal{W}_n \sum_{E,\sigma} |Q_{E,\sigma}\rangle\!\langle Q_{E,\sigma}| \otimes \sum_{\ell = 0}^{2^n-1}|\ell\rangle\!\langle\ell|_\mathsf{phase} \\
& =   \sum_{E,\sigma} |Q_{E,\sigma}\rangle\!\langle Q_{E,\sigma}| \otimes\sum_{k, \, \ell = 0}^{2^n -1} |k\rangle\!\langle\ell|_\mathsf{phase} \prod_{x= 1}^{n} \mathrm{M}_{b_{k,x} b_{\ell, x}}^{(x)}\!\left( 2^{x-1} \pi \left[\vartheta_{E,\sigma} - \frac{\mathfrak{Z}_x(k)}{2^n}\right] \right)\, ,
\end{align}
\end{widetext}
where $\vartheta_{E,\sigma} = \sigma (\arccos E) / (2 \pi) \; \mathrm{mod} \; 1$. Immediately we find 
\begin{align}
\label{guseve305}
&\mathsf{iQPE}^\dagger\; | k \rangle\!\langle k |_\mathsf{phase} \;\, \mathsf{iQPE}   = \notag\\&\qquad\sum_{E,\sigma} |Q_{E,\sigma}\rangle\!\langle Q_{E,\sigma}| \otimes   | \mathfrak{X}_{E,\sigma}^k\rangle\!\langle\mathfrak{X}_{E,\sigma}^k|_\mathsf{phase}
\end{align}
with 
\begin{align}
\label{guseve306}
| \mathfrak{X}_{E,\sigma}^k\rangle_{\mathsf{phase}} = \sum_{\ell=0}^{2^n-1} f_n(\vartheta_{E,\sigma}, k, \ell) |\ell \rangle_{\mathsf{phase}} 
\end{align}
where
\begin{align}
\label{guseve307}
f_n(\vartheta_{E, \sigma}, k, \ell) &= \prod_{x=1}^{n}\mathrm{M}^{(x)*}_{b_{k,x}b_{\ell,x}}\!\left( 2^{x-1} \pi \left[\vartheta_{E,\sigma} - \frac{\mathfrak{Z}_x(k)}{2^n}\right] \right).
\end{align}
The overlap of arbitrary states is now
\begin{widetext}
\begin{align}
\label{guseve310}
\langle \mathfrak{X}_{\widetilde{E}, \widetilde{\sigma}}^j| \mathfrak{X}_{\vphantom{\widetilde{E}}E,\sigma}^k \rangle & =\sum_\ell f_n(\vartheta_{E,\sigma}, k, \ell) \; f^*_n(\vartheta_{\widetilde{E},\widetilde{\sigma}}, j, \ell) \\
\label{guseve311}
& = \prod_{x=1}^n \sum_{b\in\lbrace 0,1 \rbrace} \mathrm{M}^{(x)}_{b_{j,x}\, b}\!\left( 2^{x-1} \pi \left[\vartheta_{\widetilde{E},s} - \frac{\mathfrak{Z}_x(j)}{2^n}\right] \right) \;\, \mathrm{M}^{(x)*}_{b_{k,x}\, b}\!\left( 2^{x-1} \pi \left[\vartheta_{E,s} - \frac{\mathfrak{Z}_x(k)}{2^n}\right] \right)\, .
\end{align}
\end{widetext}
When $(E,\sigma)=(\widetilde{E},\widetilde{\sigma})$ and $j=k$, the complex exponential equals one, and so $\langle \mathfrak{X}^k_{E,\sigma} | \mathfrak{X}_{E,\sigma}^k \rangle = 1$. When $(E,\sigma)=(\widetilde{E},\widetilde{\sigma})$ but $j \neq k$ such that $y$ is the last bit in which $b_k$ and $b_j$ differ, i.e. $b_{k,x} = b_{j,x}$ for $x>y$ and $b_{k,y} = b_{j,y}$ the $y$-th factor in Eq.~\eqref{guseve311} turns into 
\begin{widetext}
\begin{align}
\label{guseve312}
& \sum_b \mathrm{M}^{(x)}_{b_{j,y},\, b}\!\left( \vartheta \right) \cdot \mathrm{M}^{(x)*}_{b_{j,y}+1, \, b}\!\left(\vartheta^\prime \right)   = \left\lbrace \begin{matrix} - i P_x(\cos\vartheta)\, Q^*_x(\cos\vartheta^\prime) + i Q^*_x(\cos\vartheta)\, P_x(\cos\vartheta^\prime)   & \text{if}\; b_{j,y}=0 \\[7pt] \phantom{-}i Q_x(\cos\vartheta)\, P^*_x(\cos\vartheta^\prime) - i P^*_x(\cos\vartheta)\, Q^*_x(\cos\vartheta^\prime) & \text{if}\; b_{j,y}=1
\end{matrix} \right\rbrace \, .
\end{align}
\end{widetext}
Since the $\mathfrak{Z}_y(j) = \mathfrak{Z}_y(k)$ we find $\vartheta =    \vartheta^\prime$ 
which causes the term in Eq.~\eqref{guseve312} to vanish regardless of $b_{j,y}$. It follows that 
\begin{align}
\langle \mathfrak{X}^j_{E,\sigma} | \mathfrak{X}^k_{E,\sigma} \rangle = \delta_{jk} \, ,
\end{align}
and so we achieve goal \textbf{I}. 
With access to the matrix elements $\mathrm{M}_{ab}^{({x})}(\vartheta)$ for arbitrary inputs $\vartheta$, we can calculate the overlaps $\langle \mathfrak{X}^j_{\widetilde{E}, \widetilde{\sigma}}| \mathfrak{X}^k_{\vphantom{\widetilde{E}}E,\sigma} \rangle$ with Eq.~\eqref{guseve311} for the cases where $(E,\sigma)\neq (\widetilde{E}, \widetilde{\sigma})$, achieving goal \textbf{III}.
Also, since
\begin{align}
    \varrho(G,s)_{\mathsf{phase}} = \sum_{m\in \mathcal{M}} \langle Q_{G,s}|\mathsf{iQPE}^\dagger |m\rangle\! \langle m|_{\mathsf{phase}}\, \mathsf{iQPE} |Q_{G,s}\rangle\, ,
\end{align}
is equal to $\varrho(E,\sigma)_{\mathsf{phase}}$ in Eq.~\eqref{eq:varrhobasis}, the maximum success probability can be computed by
\begin{widetext}
\begin{align}
\label{guseve108}
\langle \boldsymbol{0}|\varrho(G,s)|\boldsymbol{0}\rangle & = \sum_m |\langle \mathfrak{X}^m_{G,s} | 0 \rangle|^2\\
& = \sum_{m\in \mathcal{M}} \left|\vphantom{\sum}f_n(\vartheta_{G,s},m,0)\right|^2 \\
& = \sum_{m\in \mathcal{M}} \prod_{x=1}^n \left|\mathrm{M}^{(x)}_{b_{m,x} 0}\!\left( 2^{x-1} \pi \left[\vartheta_{E,s} - \frac{\mathfrak{Z}_x(m)}{2^n}\right] \right)\right|^2\, ,
\end{align}
\end{widetext}
which fulfills the criteria for goal \textbf{II}. Errors and success probabilities for \SEVEplus{} can now be computed when expressions $P_x(z)$ and $Q_x(z)$ can be accessed numerically for arbitrary inputs $z$. This trivially contains \SEVE{} numbers following $P_x(z)=z$ and $Q_x(z)=\sqrt{1 - z^2}$.

\section{Details about resources estimates}\label{sec:resource_details}

In this section we describe the details that went into calculating the resource estimates. We have calculated the resources within a 10-step procedure.
\begin{enumerate}
    \item \textbf{Error budgeting:} The target error of the observable shall be $\varepsilon_{\text{targ}} = \varepsilon / \lambda_F$. Now we have to distribute this error between the systematic error of the expectation value stemming from the discretization error of the inner QPE, and the finite resolution of the outer QPE. We find
\begin{align}
\label{eq:budget0}
\varepsilon_{\text{targ}} = \varepsilon_{\text{in}} + \varepsilon_{\text{out}}\, 
\end{align}
and want to minimize the complexity of the algorithm, which is proportional to the function
\begin{align}
\label{eq:budget1}
\frac{1}{\varepsilon_{\text{out}}} \varkappa({\varepsilon_{\text{in}}})\, 
\end{align}
where the function $\varkappa(\cdot)$ is a multiplier to the complexity of the inner phase estimation routine, in order to compensate the discretization error. Initially defined in Eq.~\eqref{eq:bigcomplexity}, our results show that the function $\varkappa(x) = 1/x$ for \SEVE{} and $\varkappa(x) = \log(1/x)$ for \SEVEplus{}.
Optimizing \eqref{eq:budget1} under the condition \eqref{eq:budget0} using the Lagrange method, we find that $\varepsilon_{\text{in}} = \varepsilon_{\text{out}} = \varepsilon_{\text{targ}}/2$ in the case of \SEVE{}. For \SEVEplus{} we find 
$$
\varepsilon_{\text{in}} = - \frac{\varepsilon_{\text{targ}}}{W_{-1}(-\varepsilon_{\text{targ}} / e) } \, ,
$$
where $W_{-1}(\cdot)$ is the $(-1)$-branch of the Lambert function and $e$ is the Euler number.
\item \textbf{Qubits in the phase register of the outer QPE.} $\varepsilon_{\text{out}}$ is used to set the bit precision $n_{\text{out}}$ of $\mathsf{oQPE}$ by
\begin{align}
n_{\text{out}} = \left\lceil \log \frac{\pi}{\ \varepsilon_{\text{out}}} \right\rceil + 1\, .
\end{align}

\item \textbf{Parameters of the inner QPE:} $\varepsilon_{\text{in}}$ is used to determine either the number of extra qubits $n_{\mathsf{x}}$ in case we are using \SEVE{} or the degree $d$ of the polynomial function \SEVEplus{} from Figure~\ref{fig:seve_comparison}$(a)$. The results in the figure have been precomputed on the basis of a QPE with the spectral gap conditions of \eqref{eq:bin0} and \eqref{eq:bin1}. The expectation values are attained using Eq.~\eqref{eq:error} and the procedure outlined in Appendix \ref{sec:iQPE}.
An important ingredient for the calculation are the values $\arccos G$ and $\arccos \mathcal{E}$ in the ranges based on the ground- and excited state energies, $G$ and $\mathcal{E}$. While $\arccos G$ is found in a closed range between delimiters $\hat{m}$ and $\hat{m} + 1$ in Eq.~\eqref{eq:bin0}, $\arccos \mathcal{E}$ is defined on an open range after $\hat{m}+2$ in Eq.~\eqref{eq:bin1}. Since we only care for the relative distances of $\pm\arccos G$ and $\pm\arccos \mathcal{E}$ to the angle $2^{-n+1}\pi\hat{m}$, where $n$ is the total number of $\mathsf{phase}$ qubits and $\hat{m}$ an integer chosen to be close to $\pm\arccos G$, we can e{.}g{.} set $\hat{m}=0$ and redefine the ranges of $\arccos G$ and $\arccos \mathcal{E}$ accordingly. The results in Figure~\ref{fig:seve_comparison} are attained by maximizing Eq.~\eqref{eq:error} over variations of $\arccos G$ and $\arccos \mathcal{E}$ in their new-defined ranges. 
\item \textbf{Hamiltonian and observable block encoding:} We decompose the Hamiltonian and observable through double factorization \cite{VonBurg2021,Google_THC} to obtain the fermionic bases and low-rank tensors for $\mathcal{B}[\hat{H}]$ and $\mathcal{B}[\hat{F}]$. The vast majority of the gate and auxiliary qubit complexity for implementing the block encoding of the double-factorized Hamiltonian comes from implementing the basis-transforming Givens rotations. Each rotation must be performed to some precision $\beta$, and the angles for each set of Givens rotations per leaf in the rank decomposition must be loaded coherently by a data-loading circuit, or a QROM \cite{low2018trading}. For a system of $N$ orbitals, there will be $N$ many Givens rotations (which must then be uncomputed). For a double-factorized rank $M = O(N^2)$, there will be $M$ different sets of $N$ many Givens rotations. This means that we must load $M$ different sets of $N$ rotations, each with $\beta$ bits of precision, and then we must actually perform the rotations themselves (which we can do via addition into a phase gradient state of size $\beta$). The gate complexity of the aforementioned steps is therefore $O(M/k + N\beta k)$ for the data-loading, and $O(N\beta)$ to implement addition into a phase gradient state, where $k$ is a tunable parameter to trade off between gate and qubit complexity. The qubit complexity for these steps is $O(N\beta) + O(\sqrt{M/\beta})$. The parameter $\beta$ is usually between 10-20 bits. The bits of precision in all Givens rotations can be determined numerically or analytically. Here, we choose the more conservative, analytic estimate given in Eq{.} 76 in \cite{VonBurg2021} which scales logarithmically with the number of spin-orbitals. Practically, one would often truncate the factorization, omitting singular vectors with singular values below a certain threshold. There is no clear consensus on how to assess the impact of an arbitrarily set threshold, and as we want out counts to represent the worst case, we only truncate singular vectors in the single factorization for singular values under $10^{-15}$.

The number of qubits and gates used in these steps is a plurality of all gates and qubits used per iteration of the inner phase estimation, and thus constitutes the majority of the qubit and gate costs for the entire algorithm (since the number of phase qubits for the inner and outer QPEs will be much smaller than $N \beta$).

\item \textbf{Baseline QPE costs:} The cost for $\mathcal{B}[\hat{H}]$ is used to determine the cost $c_{\mathsf{bQPE}}$ of a baseline QPE, running qubitization with a $n_0$-sized phase register big enough to resolve the conditions \eqref{eq:bin0} and \eqref{eq:bin1}, where 
\begin{align}
    n_0 \leq \left\lceil\log\left(\frac{6 \pi \lambda_H}{\delta_H}\right) \right\rceil \, ,
\end{align}
and the complexity $c_{\mathsf{bQPE}}$ roughly $2^{n_0-1}$ times the cost of the Hamiltonian block encoding. 
\item \textbf{Inner QPE costs:} The baseline QPE cost, $c_{\mathsf{bQPE}}$ is modified with the inner QPE parameters from the third step, which is either $n_{\mathsf{x}}$ or $d$, to obtain the cost $c_{\mathsf{iQPE}}$ of the inner QPE, either $c_{\mathsf{iQPE}} \approx 2^{n_{\mathsf{x}}}c_{\mathsf{bQPE}}$ in case of \SEVE{} or $c_{\mathsf{iQPE}} \approx 2 d c_{\mathsf{bQPE}}$ in case of \SEVEplus{}.
\item \textbf{Outer QPE iterate cost:} The cost of $\mathsf{Refl}$ and $\mathcal{B}[\hat{F}]$, where the latter is obtained through double factorization, are combined to obtain the cost $c_{\mathcal{U}}$ of an $\mathsf{oQPE}$ iterate $\mathcal{U}$. We find
\begin{align}
    c_{\mathcal{U}} = 2 c_{\mathsf{iQPE}} + c_{\mathsf{Refl}} + c_{\mathcal{B}[\hat{F}]}\, ,
\end{align}
where $c_{\mathsf{Refl}}$ and $c_{\mathcal{B}[\hat{F}]}$ are the gate costs of the $\mathsf{Refl}$ and the observable block encoding, respectively.
\item \textbf{Outer QPE cost:} The total cost $c_{\mathsf{oQPE}}$ of $\mathsf{oQPE}$ is obtained based on the costs of iterates $\mathcal{U}$ taking into account state-of-the-art tricks like double phase-kickback. The number of times $\mathcal{U}$, or parts thereof are queried has been set by the number of qubits in the $\mathsf{oQPE}$ phase register obtained in step two. A rough estimate would be
\begin{align}
    c_{\mathsf{oQPE}} \approx 2^{n_{\mathsf{out}}-1} c_{\mathcal{U}} \, .
\end{align}
\item \textbf{Ansatz state preparation costs:} The cost of $\mathsf{ASP}$ are estimated by multiple execution of what is essentially a baseline QPE with one additional qubit. The number of repetitions depends on $1/\gamma^2$, the squared inverse overlap amplitude of the Hartree-Fock state $|\mathrm{HF}\rangle$ with the ground state $|\psi_G\rangle$, $\gamma=|\langle \mathrm{HF} | \psi_G\rangle|$, which contributes as a multiplicative constant in this cost of $\mathsf{ASP}$.

When the overlap of the Hartree-Fock state with the ground state is fixed, we can assume that in the worst case, the rest of the state is the first excited state. After having projected the state into one of the qubitization eigenstates $|\mathrm{HF}\rangle \mapsto |\Psi\rangle $, $|\Psi\rangle$ has the form
$$
|\Psi\rangle = a_G |Q_{G,s}\rangle + a_{\mathcal{E}} |Q_{\mathcal{E},s}\rangle\, .
$$ 

for some random $s = \pm 1$, and where $a_G$ and $a_{\mathcal{E}}$ are constants.
Using the notation of Appendix~\ref{sec:errppendix} for the $\mathsf{QPE}$ routine of the ASP, we find
\begin{align}
&\mathsf{QPE} |\Psi; \boldsymbol{0}\rangle   = \notag\\&\qquad\sum_{E=G,\mathcal{E}} a_{E}|Q_{E,s}\rangle \otimes \sum_{k=0}^{2^n-1} f^*\!(\vartheta_{E,s}, k, 0 )|k\rangle_{\mathsf{phase}}
\end{align}
where
$$
|\Psi; \boldsymbol{0}\rangle = |\Psi\rangle \otimes |\boldsymbol{0}\rangle_{\mathsf{phase}} \, .
$$
The ASP circuit shall look a bit like the reflection featuring the inner QPE, but we flip some auxiliary qubit conditionally on $\mathsf{Refl}$ and after the circuit concludes we measure this qubit expecting a flip while also measuring the $\mathsf{phase}$ register expecting $|\boldsymbol{0}\rangle$. The state fidelity $f_{\mathsf{ASP}}$ after a successful projection is then
\begin{align}
f_{\mathsf{ASP}} &=\frac{\left|\langle \Psi; \boldsymbol{0}| \mathsf{Proj} | Q_{G,s}\rangle \otimes |\boldsymbol{0}\rangle_{\mathsf{phase}} \right|^2}{\langle \Psi; \boldsymbol{0}|\mathsf{Proj} |\boldsymbol{0}\rangle\!\langle\boldsymbol{0}|_{\mathsf{phase}} \mathsf{Proj}|\Psi; \boldsymbol{0}\rangle} \\&= \frac{\left|\vphantom{\sum } a_G \cdot \langle \boldsymbol{0} | \varrho(G,s)|\boldsymbol{0}\rangle \right|^2}{|a_G \cdot \langle \boldsymbol{0} | \varrho(G,s)|\boldsymbol{0}\rangle|^2 + 
|a_\mathcal{E}\cdot \langle \boldsymbol{0} | \varrho(\mathcal{E},s)|\boldsymbol{0}\rangle|^2}\, ,
\end{align}
where 
\begin{align}
\mathsf{Proj} &= \mathsf{QPE}^\dagger\cdot(1 - \mathsf{Refl})/2\cdot\mathsf{QPE} \\&= \sum_{E,\sigma} |Q_{E,\sigma}\rangle\! \langle Q_{E,\sigma} | \otimes \varrho(E,\sigma)_{\mathsf{phase}} \, .
\end{align}
The probability to project into the desired state with a single ASP circuit is
\begin{align}
p_{\mathsf{ASP}} &= \langle \Psi;\boldsymbol{0}| \mathsf{Proj} |\boldsymbol{0}\rangle\!\langle\boldsymbol{0}|_{\mathsf{phase}} \mathsf{Proj} |\Psi; \boldsymbol{0}\rangle\\& \geq\frac{1}{2} |a_G \cdot \langle \boldsymbol{0} | \varrho(G,s)|\boldsymbol{0}\rangle|^2 \notag\\&\qquad+ 
\frac{1}{2}|a_\mathcal{E} \cdot \langle \boldsymbol{0} | \varrho(\mathcal{E},s)|\boldsymbol{0}\rangle|^2 \, .
\end{align}
Setting $|a_G| = \gamma$, $|a_{\mathcal{E}}|^2 = 1 - \gamma^2$ we can obtain $p_{\mathsf{ASP}}$ and $f_{\mathsf{ASP}}$ numerically. For $\gamma^2 = 0.01$, we find that $f_{\mathsf{ASP}} \approx 91\%$ and the average gate cost $c_{\mathsf{ASP}}$ of the ASP is $c_{\mathsf{ASP}} = 4 c_{\mathsf{bQPE}}/ \gamma$.
\item \textbf{Total costs.} Costs of $\mathsf{oQPE}$ and $\mathsf{ASP}$ are combined to yield the total gate cost, while smaller cost factors like the creation of phase gradient states are neglected on account of their comparatively small size. 
\end{enumerate}

\onecolumn

\newpage
\section{Molecular geometries}\label{sec:geometries}
\begin{table}[ph!]
\caption{Molecular geometries used for all calculations presented in this work.}\label{tab:molecular_geometries}
\setbox0\hbox{\tabular{@{}l}Ammonia\endtabular}
\setbox1\hbox{\tabular{@{}l}\ce{H2}\endtabular}
\setbox4\hbox{\tabular{@{}l}\ce{Be}\endtabular}
\setbox2\hbox{\tabular{@{}l}Water\endtabular}
\setbox3\hbox{\tabular{@{}l}p-Benzyne\endtabular}
\renewcommand{\arraystretch}{1.3}
\centering
\resizebox{0.5\textwidth}{!}{
\begin{tabular}{p{20mm}rrrr}
\hline\hline
System & Atom & x[\r{A}] & y[\r{A}]& z[\r{A}]\\\hline
\multirow{4}{*}{{\usebox0}} 
&\ce{N}& -1.578718& -0.046611 & 0\\ 
&\ce{H}& -2.158621& 0.136396 &-0.809565\\
&\ce{H}& -2.158621& 0.136396 &0.809565\\
&\ce{H}&-0.849471& 0.658193& 0\\
\hline\multirow{3}{*}{\rotatebox{0}{\usebox2}} 
&\ce{O}& -1.551007& -0.11452 & 0\\ 
&\ce{H}& 1.934259& 0.762503 & 0\\ 
&\ce{H}& -0.599677&0.040712 & 0\\ 
\hline\multirow{2}{*}{\rotatebox{0}{\usebox1}} 
&\ce{H}& 0& 0.435 & 0\\ 
&\ce{H}& 0&-0.435 & 0\\ 
\hline\multirow{1}{*}{\rotatebox{0}{\usebox4}} 
&\ce{Be}& 0& 0 & 0\\
\hline\multirow{10}{*}{\rotatebox{0}{\usebox3}} 
&\ce{C}& 0.73564 & 1.2078 & -2.66526\\ 
&\ce{C}& 1.38154 & -0.02467 & -2.57921\\ 
&\ce{C}& 0.64149 &-1.18586 &-2.3626\\ 
&\ce{C}& -0.74454& -1.11496& -2.2341\\ 
&\ce{C}& -1.39058& 0.11737& -2.3189\\ 
&\ce{C}& -0.65045& 1.27899& -2.53365\\ 
&\ce{H}& 1.30967& 2.10892& -2.83283\\ 
&\ce{H}& 2.45697& -0.07993& -2.67868\\ 
&\ce{H}& -1.31808& -2.01672& -2.06779\\ 
&\ce{H}& -2.46598& 0.17318& -2.21799\\ 
\hline\hline
\end{tabular}}
\end{table}

\newpage
\section{Comprehensive resource data}\label{sec:data_code}

\begin{figure*}[ht!]
    \centering
\includegraphics[scale=0.6]{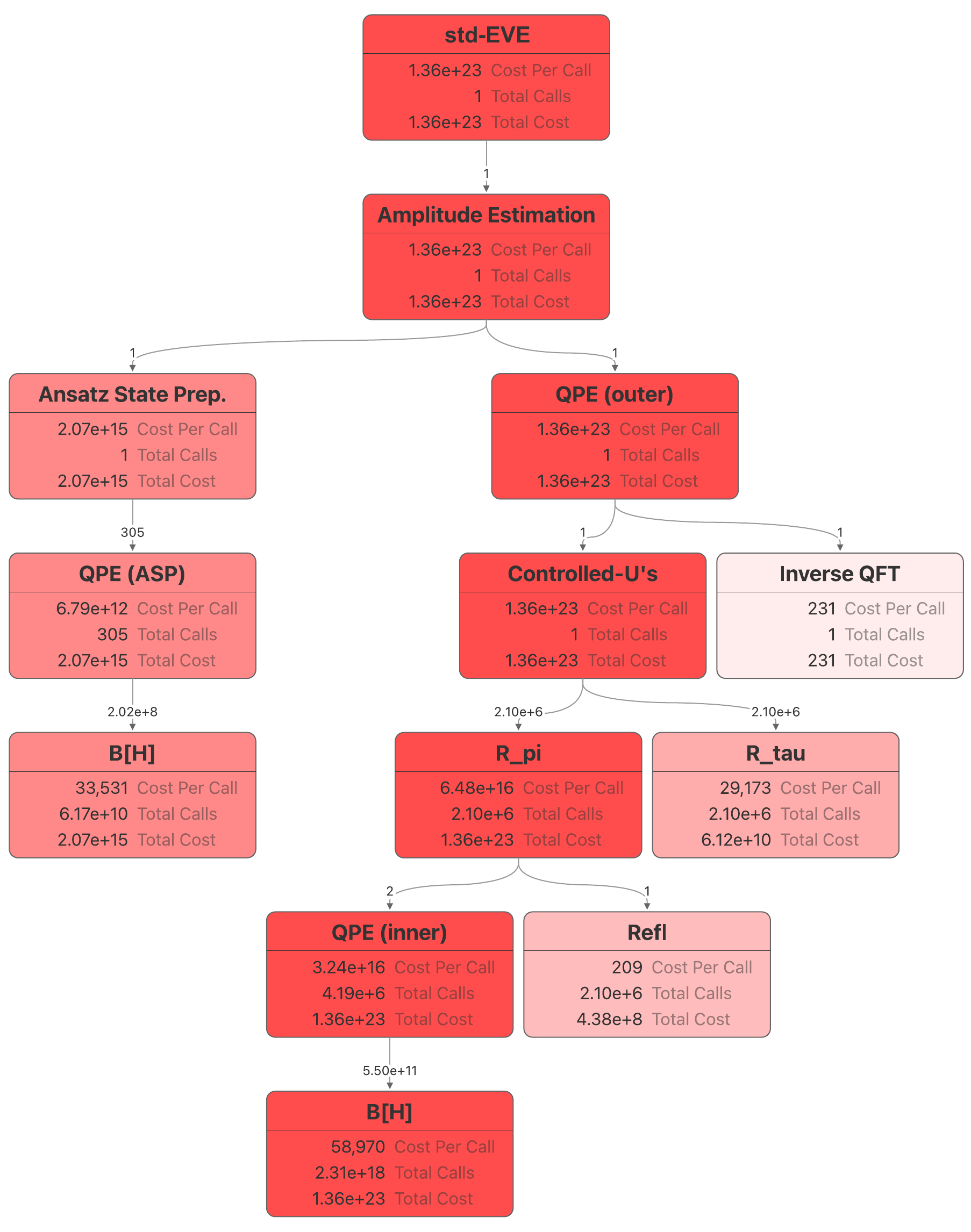}
        \caption{Callgraph depicting the estimated quantum resources required for computing the kinetic energy of a p-Benzyne molecule using \SEVE{}. Routine costs are given in terms of Toffoli gate counts, with each subroutine node depicting the per-call cost, the total number of calls, and the cost when taken over the full algorithm. Note that some routines are deliberately omitted in the count, as they contribute very little. Numbered edges define the number of calls of the target routine within a single call of its parent routine. Darker shading indicates a greater total gate cost for the subroutine. The $\mathsf{ASP}$ comprises several low-precision phase estimation routines in the repeat-until-success circuit. The initial state overlap of the Hartree-Fock state with the Hamiltonian ground state is assumed to be 1\% and is subsequently lifted to above 91\%. Details on the resource estimation of all subroutines can be found in Appendix \ref{sec:resource_details}.}
    \label{fig:callgraph}
\end{figure*}


\begin{sidewaysfigure}
Forces (std-EVE)\\

\resizebox{\textheight}{!}{

\begin{tabular}{|ccc|*6r|*9r|*9r|}
\hline
\multicolumn{3}{|c|}{\bf{Instance}} & \multicolumn{6}{c|}{\bf{Property}} & \multicolumn{9}{c|}{\bf{Gates}} & \multicolumn{9}{c|}{\bf{Qubits}} \\
System & Atom & Axis & N & $\lambda_H$ & $\Delta_H$ & $\lambda_F$ & $\varepsilon$ & $\Lambda_F$ & EVE & ASP & oQPE & $\mathcal{R}_\pi$ & iQPE & $\mathcal{B}[\hat{H}]$ & REFL & $\mathcal{R}_\tau$ & $\mathcal{B}[\hat{F}]$ & EVE & ASP & oQPE & $\mathcal{R}_\pi$ & iQPE & $\mathcal{B}[\hat{H}]$ & REFL & $\mathcal{R}_\tau$ & $\mathcal{B}[\hat{F}]$ \\

\hline
\multirow[c]{9}{*}{Water} & \multirow[c]{3}{*}{0} & x & 24 & 328 & 0.302 & 82 & 0.005 & 16452 & 3.15e+18 & 1.25e+12 & 3.15e+18 & 2.40e+13 & 1.20e+13 & 1.12e+04 & 1.71e+02 & 6.33e+03 & 6.22e+03 & 1.83e+03 & 1.11e+03 & 1.77e+03 & 1.76e+03 & 1.70e+03 & 1.55e+02 & 1.41e+02 & 1.02e+03 & 9.09e+02 \\

 & & y & 24 & 328 & 0.302 & 81 & 0.005 & 16124 & 3.15e+18 & 1.25e+12 & 3.15e+18 & 2.40e+13 & 1.20e+13 & 1.12e+04 & 1.71e+02 & 6.33e+03 & 6.22e+03 & 1.83e+03 & 1.11e+03 & 1.77e+03 & 1.76e+03 & 1.70e+03 & 1.55e+02 & 1.41e+02 & 1.02e+03 & 9.09e+02 \\

 & & z & 24 & 328 & 0.302 & 54 & 0.005 & 10855 & 3.15e+18 & 1.25e+12 & 3.15e+18 & 2.40e+13 & 1.20e+13 & 1.12e+04 & 1.71e+02 & 6.12e+03 & 6.01e+03 & 1.83e+03 & 1.11e+03 & 1.77e+03 & 1.76e+03 & 1.70e+03 & 1.55e+02 & 1.41e+02 & 9.91e+02 & 8.84e+02 \\
 
 & \multirow[c]{3}{*}{1} & x & 24 & 328 & 0.302 & 43 & 0.005 & 8696 & 7.72e+17 & 1.25e+12 & 7.72e+17 & 1.18e+13 & 5.89e+12 & 1.10e+04 & 1.65e+02 & 6.13e+03 & 6.02e+03 & 1.79e+03 & 1.11e+03 & 1.74e+03 & 1.72e+03 & 1.67e+03 & 1.53e+02 & 1.36e+02 & 9.85e+02 & 8.80e+02 \\

 & & y & 24 & 328 & 0.302 & 57 & 0.005 & 11374 & 3.15e+18 & 1.25e+12 & 3.15e+18 & 2.40e+13 & 1.20e+13 & 1.12e+04 & 1.71e+02 & 6.34e+03 & 6.23e+03 & 1.83e+03 & 1.11e+03 & 1.77e+03 & 1.76e+03 & 1.70e+03 & 1.55e+02 & 1.41e+02 & 1.02e+03 & 9.09e+02 \\

 & & z & 24 & 328 & 0.302 & 33 & 0.005 & 6576 & 7.72e+17 & 1.25e+12 & 7.72e+17 & 1.18e+13 & 5.89e+12 & 1.10e+04 & 1.65e+02 & 6.08e+03 & 5.97e+03 & 1.79e+03 & 1.11e+03 & 1.74e+03 & 1.72e+03 & 1.67e+03 & 1.53e+02 & 1.36e+02 & 9.85e+02 & 8.80e+02 \\
 
 & \multirow[c]{3}{*}{2} & x & 24 & 328 & 0.302 & 61 & 0.005 & 12110 & 3.15e+18 & 1.25e+12 & 3.15e+18 & 2.40e+13 & 1.20e+13 & 1.12e+04 & 1.71e+02 & 6.33e+03 & 6.23e+03 & 1.83e+03 & 1.11e+03 & 1.77e+03 & 1.76e+03 & 1.70e+03 & 1.55e+02 & 1.41e+02 & 1.02e+03 & 9.09e+02 \\

 & & y & 24 & 328 & 0.302 & 35 & 0.005 & 6940 & 7.72e+17 & 1.25e+12 & 7.72e+17 & 1.18e+13 & 5.89e+12 & 1.10e+04 & 1.65e+02 & 6.12e+03 & 6.02e+03 & 1.79e+03 & 1.11e+03 & 1.74e+03 & 1.72e+03 & 1.67e+03 & 1.53e+02 & 1.36e+02 & 9.85e+02 & 8.80e+02 \\

 & & z & 24 & 328 & 0.302 & 32 & 0.005 & 6451 & 7.72e+17 & 1.25e+12 & 7.72e+17 & 1.18e+13 & 5.89e+12 & 1.10e+04 & 1.65e+02 & 6.09e+03 & 5.98e+03 & 1.79e+03 & 1.11e+03 & 1.74e+03 & 1.72e+03 & 1.67e+03 & 1.53e+02 & 1.36e+02 & 9.85e+02 & 8.80e+02 \\
\hline  
\multirow[c]{12}{*}{Ammonia} & \multirow[c]{3}{*}{0} & x & 29 & 434 & 0.280 & 108 & 0.005 & 21559 & 1.56e+19 & 2.90e+12 & 1.56e+19 & 5.95e+13 & 2.98e+13 & 1.39e+04 & 1.78e+02 & 7.81e+03 & 7.70e+03 & 2.17e+03 & 1.33e+03 & 2.12e+03 & 2.10e+03 & 2.04e+03 & 1.67e+02 & 1.48e+02 & 1.21e+03 & 1.10e+03 \\

 & & y & 29 & 434 & 0.280 & 99 & 0.005 & 19891 & 7.81e+18 & 2.90e+12 & 7.81e+18 & 5.95e+13 & 2.98e+13 & 1.39e+04 & 1.78e+02 & 7.79e+03 & 7.68e+03 & 2.17e+03 & 1.33e+03 & 2.12e+03 & 2.10e+03 & 2.04e+03 & 1.67e+02 & 1.48e+02 & 1.21e+03 & 1.10e+03 \\

 & & z & 29 & 434 & 0.280 & 113 & 0.005 & 22587 & 1.56e+19 & 2.90e+12 & 1.56e+19 & 5.95e+13 & 2.98e+13 & 1.39e+04 & 1.78e+02 & 7.79e+03 & 7.68e+03 & 2.17e+03 & 1.33e+03 & 2.12e+03 & 2.10e+03 & 2.04e+03 & 1.67e+02 & 1.48e+02 & 1.21e+03 & 1.10e+03 \\
 
 & \multirow[c]{3}{*}{1} & x & 29 & 434 & 0.280 & 53 & 0.005 & 10604 & 3.83e+18 & 2.90e+12 & 3.83e+18 & 2.93e+13 & 1.46e+13 & 1.36e+04 & 1.72e+02 & 7.55e+03 & 7.44e+03 & 2.13e+03 & 1.33e+03 & 2.08e+03 & 2.06e+03 & 2.00e+03 & 1.65e+02 & 1.43e+02 & 1.17e+03 & 1.07e+03 \\

 & & y & 29 & 434 & 0.280 & 39 & 0.005 & 7786 & 1.92e+18 & 2.90e+12 & 1.92e+18 & 2.93e+13 & 1.46e+13 & 1.36e+04 & 1.70e+02 & 7.29e+03 & 7.18e+03 & 2.13e+03 & 1.33e+03 & 2.08e+03 & 2.06e+03 & 2.00e+03 & 1.65e+02 & 1.39e+02 & 1.14e+03 & 1.03e+03 \\

 & & z & 29 & 434 & 0.280 & 63 & 0.005 & 12603 & 3.83e+18 & 2.90e+12 & 3.83e+18 & 2.93e+13 & 1.46e+13 & 1.36e+04 & 1.72e+02 & 7.55e+03 & 7.44e+03 & 2.13e+03 & 1.33e+03 & 2.08e+03 & 2.06e+03 & 2.00e+03 & 1.65e+02 & 1.43e+02 & 1.17e+03 & 1.07e+03 \\
 
 & \multirow[c]{3}{*}{2} & x & 29 & 434 & 0.280 & 53 & 0.005 & 10604 & 3.83e+18 & 2.90e+12 & 3.83e+18 & 2.93e+13 & 1.46e+13 & 1.36e+04 & 1.72e+02 & 7.55e+03 & 7.44e+03 & 2.13e+03 & 1.33e+03 & 2.08e+03 & 2.06e+03 & 2.00e+03 & 1.65e+02 & 1.43e+02 & 1.17e+03 & 1.07e+03 \\

 & & y & 29 & 434 & 0.280 & 39 & 0.005 & 7786 & 1.92e+18 & 2.90e+12 & 1.92e+18 & 2.93e+13 & 1.46e+13 & 1.36e+04 & 1.70e+02 & 7.30e+03 & 7.20e+03 & 2.13e+03 & 1.33e+03 & 2.08e+03 & 2.06e+03 & 2.00e+03 & 1.65e+02 & 1.39e+02 & 1.14e+03 & 1.03e+03 \\

 & & z & 29 & 434 & 0.280 & 58 & 0.005 & 11563 & 3.83e+18 & 2.90e+12 & 3.83e+18 & 2.93e+13 & 1.46e+13 & 1.36e+04 & 1.72e+02 & 7.54e+03 & 7.43e+03 & 2.13e+03 & 1.33e+03 & 2.08e+03 & 2.06e+03 & 2.00e+03 & 1.65e+02 & 1.43e+02 & 1.17e+03 & 1.07e+03 \\
 
 & \multirow[c]{3}{*}{3} & x & 29 & 434 & 0.280 & 55 & 0.005 & 10912 & 3.83e+18 & 2.90e+12 & 3.83e+18 & 2.93e+13 & 1.46e+13 & 1.36e+04 & 1.72e+02 & 7.54e+03 & 7.43e+03 & 2.13e+03 & 1.33e+03 & 2.08e+03 & 2.06e+03 & 2.00e+03 & 1.65e+02 & 1.43e+02 & 1.17e+03 & 1.07e+03 \\

 & & y & 29 & 434 & 0.280 & 50 & 0.005 & 10062 & 1.92e+18 & 2.90e+12 & 1.92e+18 & 2.93e+13 & 1.46e+13 & 1.36e+04 & 1.72e+02 & 7.54e+03 & 7.43e+03 & 2.13e+03 & 1.33e+03 & 2.08e+03 & 2.06e+03 & 2.00e+03 & 1.65e+02 & 1.43e+02 & 1.17e+03 & 1.07e+03 \\

 & & z & 29 & 434 & 0.280 & 41 & 0.005 & 8122 & 1.92e+18 & 2.90e+12 & 1.92e+18 & 2.93e+13 & 1.46e+13 & 1.36e+04 & 1.69e+02 & 7.26e+03 & 7.16e+03 & 2.13e+03 & 1.33e+03 & 2.08e+03 & 2.06e+03 & 2.00e+03 & 1.65e+02 & 1.37e+02 & 1.14e+03 & 1.03e+03 \\
\hline  
\multirow[c]{30}{*}{p-Benzyne} & \multirow[c]{3}{*}{0} & x & 104 & 3833 & 0.114 & 1074 & 0.005 & 214796 & 1.36e+23 & 2.38e+15 & 1.36e+23 & 6.48e+16 & 3.24e+16 & 5.90e+04 & 2.22e+02 & 3.09e+04 & 3.08e+04 & 7.93e+03 & 5.12e+03 & 7.86e+03 & 7.84e+03 & 7.76e+03 & 3.45e+02 & 1.80e+02 & 4.19e+03 & 4.05e+03 \\

 & & y & 104 & 3833 & 0.114 & 1050 & 0.005 & 210035 & 1.36e+23 & 2.38e+15 & 1.36e+23 & 6.48e+16 & 3.24e+16 & 5.90e+04 & 2.22e+02 & 3.09e+04 & 3.08e+04 & 7.93e+03 & 5.12e+03 & 7.86e+03 & 7.84e+03 & 7.76e+03 & 3.45e+02 & 1.80e+02 & 4.19e+03 & 4.05e+03 \\

 & & z & 104 & 3833 & 0.114 & 705 & 0.005 & 140947 & 3.35e+22 & 2.38e+15 & 3.35e+22 & 3.20e+16 & 1.60e+16 & 5.81e+04 & 2.16e+02 & 3.01e+04 & 3.00e+04 & 7.81e+03 & 5.12e+03 & 7.74e+03 & 7.72e+03 & 7.65e+03 & 3.43e+02 & 1.75e+02 & 4.08e+03 & 3.94e+03 \\
 
 & \multirow[c]{3}{*}{1} & x & 104 & 3833 & 0.114 & 1051 & 0.005 & 210185 & 1.36e+23 & 2.38e+15 & 1.36e+23 & 6.48e+16 & 3.24e+16 & 5.90e+04 & 2.22e+02 & 3.09e+04 & 3.08e+04 & 7.93e+03 & 5.12e+03 & 7.86e+03 & 7.84e+03 & 7.76e+03 & 3.45e+02 & 1.80e+02 & 4.19e+03 & 4.05e+03 \\

 & & y & 104 & 3833 & 0.114 & 1093 & 0.005 & 218623 & 1.36e+23 & 2.38e+15 & 1.36e+23 & 6.48e+16 & 3.24e+16 & 5.90e+04 & 2.22e+02 & 3.09e+04 & 3.08e+04 & 7.93e+03 & 5.12e+03 & 7.86e+03 & 7.84e+03 & 7.76e+03 & 3.45e+02 & 1.80e+02 & 4.19e+03 & 4.05e+03 \\

 & & z & 104 & 3833 & 0.114 & 704 & 0.005 & 140782 & 3.35e+22 & 2.38e+15 & 3.35e+22 & 3.20e+16 & 1.60e+16 & 5.81e+04 & 2.16e+02 & 3.01e+04 & 3.00e+04 & 7.81e+03 & 5.12e+03 & 7.74e+03 & 7.72e+03 & 7.65e+03 & 3.43e+02 & 1.75e+02 & 4.08e+03 & 3.94e+03 \\
 
 & \multirow[c]{3}{*}{2} & x & 104 & 3833 & 0.114 & 1003 & 0.005 & 200618 & 1.36e+23 & 2.38e+15 & 1.36e+23 & 6.48e+16 & 3.24e+16 & 5.90e+04 & 2.22e+02 & 3.09e+04 & 3.08e+04 & 7.93e+03 & 5.12e+03 & 7.86e+03 & 7.84e+03 & 7.76e+03 & 3.45e+02 & 1.80e+02 & 4.19e+03 & 4.05e+03 \\

 & & y & 104 & 3833 & 0.114 & 970 & 0.005 & 193941 & 1.36e+23 & 2.38e+15 & 1.36e+23 & 6.48e+16 & 3.24e+16 & 5.90e+04 & 2.22e+02 & 3.09e+04 & 3.08e+04 & 7.93e+03 & 5.12e+03 & 7.86e+03 & 7.84e+03 & 7.76e+03 & 3.45e+02 & 1.80e+02 & 4.19e+03 & 4.05e+03 \\

 & & z & 104 & 3833 & 0.114 & 662 & 0.005 & 132418 & 3.35e+22 & 2.38e+15 & 3.35e+22 & 3.20e+16 & 1.60e+16 & 5.81e+04 & 2.16e+02 & 3.01e+04 & 3.00e+04 & 7.81e+03 & 5.12e+03 & 7.74e+03 & 7.72e+03 & 7.65e+03 & 3.43e+02 & 1.75e+02 & 4.08e+03 & 3.94e+03 \\
 
 & \multirow[c]{3}{*}{3} & x & 104 & 3833 & 0.114 & 1084 & 0.005 & 216890 & 1.36e+23 & 2.38e+15 & 1.36e+23 & 6.48e+16 & 3.24e+16 & 5.90e+04 & 2.22e+02 & 3.09e+04 & 3.08e+04 & 7.93e+03 & 5.12e+03 & 7.86e+03 & 7.84e+03 & 7.76e+03 & 3.45e+02 & 1.80e+02 & 4.19e+03 & 4.05e+03 \\

 & & y & 104 & 3833 & 0.114 & 1071 & 0.005 & 214123 & 1.36e+23 & 2.38e+15 & 1.36e+23 & 6.48e+16 & 3.24e+16 & 5.90e+04 & 2.22e+02 & 3.09e+04 & 3.08e+04 & 7.93e+03 & 5.12e+03 & 7.86e+03 & 7.84e+03 & 7.76e+03 & 3.45e+02 & 1.80e+02 & 4.19e+03 & 4.05e+03 \\

 & & z & 104 & 3833 & 0.114 & 712 & 0.005 & 142300 & 3.35e+22 & 2.38e+15 & 3.35e+22 & 3.20e+16 & 1.60e+16 & 5.81e+04 & 2.16e+02 & 3.01e+04 & 3.00e+04 & 7.81e+03 & 5.12e+03 & 7.74e+03 & 7.72e+03 & 7.65e+03 & 3.43e+02 & 1.75e+02 & 4.08e+03 & 3.94e+03 \\
 
 & \multirow[c]{3}{*}{4} & x & 104 & 3833 & 0.114 & 1065 & 0.005 & 213018 & 1.36e+23 & 2.38e+15 & 1.36e+23 & 6.48e+16 & 3.24e+16 & 5.90e+04 & 2.22e+02 & 3.09e+04 & 3.08e+04 & 7.93e+03 & 5.12e+03 & 7.86e+03 & 7.84e+03 & 7.76e+03 & 3.45e+02 & 1.80e+02 & 4.19e+03 & 4.05e+03 \\

 & & y & 104 & 3833 & 0.114 & 1082 & 0.005 & 216434 & 1.36e+23 & 2.38e+15 & 1.36e+23 & 6.48e+16 & 3.24e+16 & 5.90e+04 & 2.22e+02 & 3.09e+04 & 3.08e+04 & 7.93e+03 & 5.12e+03 & 7.86e+03 & 7.84e+03 & 7.76e+03 & 3.45e+02 & 1.80e+02 & 4.19e+03 & 4.05e+03 \\

 & & z & 104 & 3833 & 0.114 & 712 & 0.005 & 142498 & 3.35e+22 & 2.38e+15 & 3.35e+22 & 3.20e+16 & 1.60e+16 & 5.81e+04 & 2.16e+02 & 3.01e+04 & 3.00e+04 & 7.81e+03 & 5.12e+03 & 7.74e+03 & 7.72e+03 & 7.65e+03 & 3.43e+02 & 1.75e+02 & 4.08e+03 & 3.94e+03 \\
 
 & \multirow[c]{3}{*}{5} & x & 104 & 3833 & 0.114 & 1020 & 0.005 & 203926 & 1.36e+23 & 2.38e+15 & 1.36e+23 & 6.48e+16 & 3.24e+16 & 5.90e+04 & 2.22e+02 & 3.09e+04 & 3.08e+04 & 7.93e+03 & 5.12e+03 & 7.86e+03 & 7.84e+03 & 7.76e+03 & 3.45e+02 & 1.80e+02 & 4.19e+03 & 4.05e+03 \\

 & & y & 104 & 3833 & 0.114 & 945 & 0.005 & 189006 & 1.36e+23 & 2.38e+15 & 1.36e+23 & 6.48e+16 & 3.24e+16 & 5.90e+04 & 2.22e+02 & 3.09e+04 & 3.08e+04 & 7.93e+03 & 5.12e+03 & 7.86e+03 & 7.84e+03 & 7.76e+03 & 3.45e+02 & 1.80e+02 & 4.19e+03 & 4.05e+03 \\

 & & z & 104 & 3833 & 0.114 & 659 & 0.005 & 131877 & 3.35e+22 & 2.38e+15 & 3.35e+22 & 3.20e+16 & 1.60e+16 & 5.81e+04 & 2.16e+02 & 3.01e+04 & 3.00e+04 & 7.81e+03 & 5.12e+03 & 7.74e+03 & 7.72e+03 & 7.65e+03 & 3.43e+02 & 1.75e+02 & 4.08e+03 & 3.94e+03 \\
 
 & \multirow[c]{3}{*}{6} & x & 104 & 3833 & 0.114 & 175 & 0.005 & 35063 & 2.03e+21 & 2.38e+15 & 2.03e+21 & 7.76e+15 & 3.88e+15 & 5.64e+04 & 2.03e+02 & 2.80e+04 & 2.79e+04 & 7.58e+03 & 5.12e+03 & 7.51e+03 & 7.49e+03 & 7.43e+03 & 3.39e+02 & 1.63e+02 & 3.86e+03 & 3.72e+03 \\

 & & y & 104 & 3833 & 0.114 & 196 & 0.005 & 39294 & 2.03e+21 & 2.38e+15 & 2.03e+21 & 7.76e+15 & 3.88e+15 & 5.64e+04 & 2.05e+02 & 2.80e+04 & 2.79e+04 & 7.58e+03 & 5.12e+03 & 7.51e+03 & 7.50e+03 & 7.43e+03 & 3.39e+02 & 1.67e+02 & 3.86e+03 & 3.73e+03 \\

 & & z & 104 & 3833 & 0.114 & 130 & 0.005 & 26064 & 2.03e+21 & 2.38e+15 & 2.03e+21 & 7.76e+15 & 3.88e+15 & 5.64e+04 & 2.03e+02 & 2.81e+04 & 2.79e+04 & 7.58e+03 & 5.12e+03 & 7.51e+03 & 7.49e+03 & 7.43e+03 & 3.39e+02 & 1.63e+02 & 3.86e+03 & 3.72e+03 \\
 
 & \multirow[c]{3}{*}{7} & x & 104 & 3833 & 0.114 & 206 & 0.005 & 41130 & 4.13e+21 & 2.38e+15 & 4.13e+21 & 1.57e+16 & 7.87e+15 & 5.73e+04 & 2.09e+02 & 2.89e+04 & 2.88e+04 & 7.69e+03 & 5.12e+03 & 7.63e+03 & 7.61e+03 & 7.54e+03 & 3.41e+02 & 1.68e+02 & 3.97e+03 & 3.83e+03 \\

 & & y & 104 & 3833 & 0.114 & 162 & 0.005 & 32361 & 2.03e+21 & 2.38e+15 & 2.03e+21 & 7.76e+15 & 3.88e+15 & 5.64e+04 & 2.03e+02 & 2.80e+04 & 2.79e+04 & 7.58e+03 & 5.12e+03 & 7.51e+03 & 7.49e+03 & 7.43e+03 & 3.39e+02 & 1.63e+02 & 3.86e+03 & 3.72e+03 \\

 & & z & 104 & 3833 & 0.114 & 129 & 0.005 & 25742 & 2.03e+21 & 2.38e+15 & 2.03e+21 & 7.76e+15 & 3.88e+15 & 5.64e+04 & 2.03e+02 & 2.80e+04 & 2.79e+04 & 7.58e+03 & 5.12e+03 & 7.51e+03 & 7.49e+03 & 7.43e+03 & 3.39e+02 & 1.63e+02 & 3.86e+03 & 3.72e+03 \\
 
 & \multirow[c]{3}{*}{8} & x & 104 & 3833 & 0.114 & 182 & 0.005 & 36481 & 2.03e+21 & 2.38e+15 & 2.03e+21 & 7.76e+15 & 3.88e+15 & 5.64e+04 & 2.03e+02 & 2.80e+04 & 2.79e+04 & 7.58e+03 & 5.12e+03 & 7.51e+03 & 7.49e+03 & 7.43e+03 & 3.39e+02 & 1.63e+02 & 3.86e+03 & 3.72e+03 \\

 & & y & 104 & 3833 & 0.114 & 207 & 0.005 & 41305 & 4.13e+21 & 2.38e+15 & 4.13e+21 & 1.57e+16 & 7.87e+15 & 5.73e+04 & 2.09e+02 & 2.89e+04 & 2.87e+04 & 7.69e+03 & 5.12e+03 & 7.63e+03 & 7.61e+03 & 7.54e+03 & 3.41e+02 & 1.68e+02 & 3.97e+03 & 3.83e+03 \\

 & & z & 104 & 3833 & 0.114 & 131 & 0.005 & 26152 & 2.03e+21 & 2.38e+15 & 2.03e+21 & 7.76e+15 & 3.88e+15 & 5.64e+04 & 2.03e+02 & 2.80e+04 & 2.79e+04 & 7.58e+03 & 5.12e+03 & 7.51e+03 & 7.49e+03 & 7.43e+03 & 3.39e+02 & 1.63e+02 & 3.86e+03 & 3.72e+03 \\
 
 & \multirow[c]{3}{*}{9} & x & 104 & 3833 & 0.114 & 217 & 0.005 & 43306 & 8.25e+21 & 2.38e+15 & 8.25e+21 & 1.57e+16 & 7.87e+15 & 5.73e+04 & 2.09e+02 & 2.89e+04 & 2.87e+04 & 7.69e+03 & 5.12e+03 & 7.63e+03 & 7.61e+03 & 7.54e+03 & 3.41e+02 & 1.68e+02 & 3.97e+03 & 3.83e+03 \\

 & & y & 104 & 3833 & 0.114 & 161 & 0.005 & 32212 & 2.03e+21 & 2.38e+15 & 2.03e+21 & 7.76e+15 & 3.88e+15 & 5.64e+04 & 2.03e+02 & 2.80e+04 & 2.79e+04 & 7.58e+03 & 5.12e+03 & 7.51e+03 & 7.49e+03 & 7.43e+03 & 3.39e+02 & 1.63e+02 & 3.86e+03 & 3.72e+03 \\

 & & z & 104 & 3833 & 0.114 & 129 & 0.005 & 25870 & 2.03e+21 & 2.38e+15 & 2.03e+21 & 7.76e+15 & 3.88e+15 & 5.64e+04 & 2.03e+02 & 2.80e+04 & 2.79e+04 & 7.58e+03 & 5.12e+03 & 7.51e+03 & 7.49e+03 & 7.43e+03 & 3.39e+02 & 1.63e+02 & 3.86e+03 & 3.72e+03 \\
\hline  

\end{tabular}

}
\vspace{10pt}
\end{sidewaysfigure}


\begin{sidewaysfigure}
Forces (qsp-EVE)\\

\resizebox{\textheight}{!}{

\begin{tabular}{|ccc|*6r|*9r|*9r|}
\hline
\multicolumn{3}{|c|}{\bf{Instance}} & \multicolumn{6}{c|}{\bf{Property}} & \multicolumn{9}{c|}{\bf{Gates}} & \multicolumn{9}{c|}{\bf{Qubits}} \\
System & Atom & Axis & N & $\lambda_H$ & $\Delta_H$ & $\lambda_F$ & $\varepsilon$ & $\Lambda_F$ & EVE & ASP & oQPE & $\mathcal{R}_\pi$ & iQPE & $\mathcal{B}[\hat{H}]$ & REFL & $\mathcal{R}_\tau$ & $\mathcal{B}[\hat{F}]$ & EVE & ASP & oQPE & $\mathcal{R}_\pi$ & iQPE & $\mathcal{B}[\hat{H}]$ & REFL & $\mathcal{R}_\tau$ & $\mathcal{B}[\hat{F}]$ \\

\hline
\multirow[c]{9}{*}{Water} & \multirow[c]{3}{*}{0} & x & 24 & 328 & 0.302 & 82 & 0.005 & 16452 & 8.84e+15 & 1.25e+12 & 8.83e+15 & 1.35e+11 & 6.74e+10 & 8.36e+03 & 1.15e+02 & 6.30e+03 & 6.22e+03 & 1.36e+03 & 1.11e+03 & 1.32e+03 & 1.31e+03 & 1.25e+03 & 1.27e+02 & 1.27e+02 & 9.88e+02 & 9.09e+02 \\

 & & y & 24 & 328 & 0.302 & 81 & 0.005 & 16124 & 8.84e+15 & 1.25e+12 & 8.83e+15 & 1.35e+11 & 6.74e+10 & 8.36e+03 & 1.15e+02 & 6.30e+03 & 6.22e+03 & 1.36e+03 & 1.11e+03 & 1.32e+03 & 1.31e+03 & 1.25e+03 & 1.27e+02 & 1.27e+02 & 9.88e+02 & 9.09e+02 \\

 & & z & 24 & 328 & 0.302 & 54 & 0.005 & 10855 & 8.84e+15 & 1.25e+12 & 8.83e+15 & 1.35e+11 & 6.74e+10 & 8.36e+03 & 1.15e+02 & 6.09e+03 & 6.01e+03 & 1.36e+03 & 1.11e+03 & 1.32e+03 & 1.31e+03 & 1.25e+03 & 1.27e+02 & 1.27e+02 & 9.63e+02 & 8.84e+02 \\
 
 & \multirow[c]{3}{*}{1} & x & 24 & 328 & 0.302 & 43 & 0.005 & 8696 & 4.42e+15 & 1.25e+12 & 4.42e+15 & 1.35e+11 & 6.74e+10 & 8.36e+03 & 1.13e+02 & 6.10e+03 & 6.02e+03 & 1.36e+03 & 1.11e+03 & 1.32e+03 & 1.30e+03 & 1.25e+03 & 1.27e+02 & 1.23e+02 & 9.59e+02 & 8.80e+02 \\

 & & y & 24 & 328 & 0.302 & 57 & 0.005 & 11374 & 8.84e+15 & 1.25e+12 & 8.83e+15 & 1.35e+11 & 6.74e+10 & 8.36e+03 & 1.15e+02 & 6.31e+03 & 6.23e+03 & 1.36e+03 & 1.11e+03 & 1.32e+03 & 1.31e+03 & 1.25e+03 & 1.27e+02 & 1.27e+02 & 9.88e+02 & 9.09e+02 \\

 & & z & 24 & 328 & 0.302 & 33 & 0.005 & 6576 & 4.42e+15 & 1.25e+12 & 4.42e+15 & 1.35e+11 & 6.74e+10 & 8.36e+03 & 1.13e+02 & 6.05e+03 & 5.97e+03 & 1.36e+03 & 1.11e+03 & 1.32e+03 & 1.30e+03 & 1.25e+03 & 1.27e+02 & 1.23e+02 & 9.59e+02 & 8.80e+02 \\
 
 & \multirow[c]{3}{*}{2} & x & 24 & 328 & 0.302 & 61 & 0.005 & 12110 & 8.84e+15 & 1.25e+12 & 8.83e+15 & 1.35e+11 & 6.74e+10 & 8.36e+03 & 1.15e+02 & 6.31e+03 & 6.23e+03 & 1.36e+03 & 1.11e+03 & 1.32e+03 & 1.31e+03 & 1.25e+03 & 1.27e+02 & 1.27e+02 & 9.88e+02 & 9.09e+02 \\

 & & y & 24 & 328 & 0.302 & 35 & 0.005 & 6940 & 4.42e+15 & 1.25e+12 & 4.42e+15 & 1.35e+11 & 6.74e+10 & 8.36e+03 & 1.13e+02 & 6.10e+03 & 6.02e+03 & 1.36e+03 & 1.11e+03 & 1.32e+03 & 1.30e+03 & 1.25e+03 & 1.27e+02 & 1.23e+02 & 9.59e+02 & 8.80e+02 \\

 & & z & 24 & 328 & 0.302 & 32 & 0.005 & 6451 & 4.42e+15 & 1.25e+12 & 4.42e+15 & 1.35e+11 & 6.74e+10 & 8.36e+03 & 1.13e+02 & 6.06e+03 & 5.98e+03 & 1.36e+03 & 1.11e+03 & 1.32e+03 & 1.30e+03 & 1.25e+03 & 1.27e+02 & 1.23e+02 & 9.59e+02 & 8.80e+02 \\
\hline  
\multirow[c]{12}{*}{Ammonia} & \multirow[c]{3}{*}{0} & x & 29 & 434 & 0.280 & 108 & 0.005 & 21559 & 2.20e+16 & 2.90e+12 & 2.19e+16 & 1.67e+11 & 8.37e+10 & 1.02e+04 & 1.18e+02 & 7.78e+03 & 7.70e+03 & 1.60e+03 & 1.33e+03 & 1.56e+03 & 1.54e+03 & 1.49e+03 & 1.37e+02 & 1.33e+02 & 1.18e+03 & 1.10e+03 \\

 & & y & 29 & 434 & 0.280 & 99 & 0.005 & 19891 & 2.20e+16 & 2.90e+12 & 2.19e+16 & 1.67e+11 & 8.37e+10 & 1.02e+04 & 1.18e+02 & 7.76e+03 & 7.68e+03 & 1.60e+03 & 1.33e+03 & 1.56e+03 & 1.54e+03 & 1.49e+03 & 1.37e+02 & 1.33e+02 & 1.18e+03 & 1.10e+03 \\

 & & z & 29 & 434 & 0.280 & 113 & 0.005 & 22587 & 2.20e+16 & 2.90e+12 & 2.19e+16 & 1.67e+11 & 8.37e+10 & 1.02e+04 & 1.18e+02 & 7.76e+03 & 7.68e+03 & 1.60e+03 & 1.33e+03 & 1.56e+03 & 1.54e+03 & 1.49e+03 & 1.37e+02 & 1.33e+02 & 1.18e+03 & 1.10e+03 \\
 
 & \multirow[c]{3}{*}{1} & x & 29 & 434 & 0.280 & 53 & 0.005 & 10604 & 1.10e+16 & 2.90e+12 & 1.10e+16 & 1.67e+11 & 8.37e+10 & 1.02e+04 & 1.16e+02 & 7.52e+03 & 7.44e+03 & 1.60e+03 & 1.33e+03 & 1.56e+03 & 1.54e+03 & 1.49e+03 & 1.37e+02 & 1.29e+02 & 1.14e+03 & 1.07e+03 \\

 & & y & 29 & 434 & 0.280 & 39 & 0.005 & 7786 & 5.49e+15 & 2.90e+12 & 5.49e+15 & 1.67e+11 & 8.37e+10 & 1.02e+04 & 1.14e+02 & 7.26e+03 & 7.18e+03 & 1.60e+03 & 1.33e+03 & 1.56e+03 & 1.54e+03 & 1.49e+03 & 1.37e+02 & 1.25e+02 & 1.11e+03 & 1.03e+03 \\

 & & z & 29 & 434 & 0.280 & 63 & 0.005 & 12603 & 1.10e+16 & 2.90e+12 & 1.10e+16 & 1.67e+11 & 8.37e+10 & 1.02e+04 & 1.16e+02 & 7.52e+03 & 7.44e+03 & 1.60e+03 & 1.33e+03 & 1.56e+03 & 1.54e+03 & 1.49e+03 & 1.37e+02 & 1.29e+02 & 1.14e+03 & 1.07e+03 \\
 
 & \multirow[c]{3}{*}{2} & x & 29 & 434 & 0.280 & 53 & 0.005 & 10604 & 1.10e+16 & 2.90e+12 & 1.10e+16 & 1.67e+11 & 8.37e+10 & 1.02e+04 & 1.16e+02 & 7.52e+03 & 7.44e+03 & 1.60e+03 & 1.33e+03 & 1.56e+03 & 1.54e+03 & 1.49e+03 & 1.37e+02 & 1.29e+02 & 1.14e+03 & 1.07e+03 \\

 & & y & 29 & 434 & 0.280 & 39 & 0.005 & 7786 & 5.49e+15 & 2.90e+12 & 5.49e+15 & 1.67e+11 & 8.37e+10 & 1.02e+04 & 1.14e+02 & 7.27e+03 & 7.20e+03 & 1.60e+03 & 1.33e+03 & 1.56e+03 & 1.54e+03 & 1.49e+03 & 1.37e+02 & 1.25e+02 & 1.11e+03 & 1.03e+03 \\

 & & z & 29 & 434 & 0.280 & 58 & 0.005 & 11563 & 1.10e+16 & 2.90e+12 & 1.10e+16 & 1.67e+11 & 8.37e+10 & 1.02e+04 & 1.16e+02 & 7.51e+03 & 7.43e+03 & 1.60e+03 & 1.33e+03 & 1.56e+03 & 1.54e+03 & 1.49e+03 & 1.37e+02 & 1.29e+02 & 1.14e+03 & 1.07e+03 \\
 
 & \multirow[c]{3}{*}{3} & x & 29 & 434 & 0.280 & 55 & 0.005 & 10912 & 1.10e+16 & 2.90e+12 & 1.10e+16 & 1.67e+11 & 8.37e+10 & 1.02e+04 & 1.16e+02 & 7.51e+03 & 7.43e+03 & 1.60e+03 & 1.33e+03 & 1.56e+03 & 1.54e+03 & 1.49e+03 & 1.37e+02 & 1.29e+02 & 1.14e+03 & 1.07e+03 \\

 & & y & 29 & 434 & 0.280 & 50 & 0.005 & 10062 & 1.10e+16 & 2.90e+12 & 1.10e+16 & 1.67e+11 & 8.37e+10 & 1.02e+04 & 1.16e+02 & 7.51e+03 & 7.43e+03 & 1.60e+03 & 1.33e+03 & 1.56e+03 & 1.54e+03 & 1.49e+03 & 1.37e+02 & 1.29e+02 & 1.14e+03 & 1.07e+03 \\

 & & z & 29 & 434 & 0.280 & 41 & 0.005 & 8122 & 5.49e+15 & 2.90e+12 & 5.49e+15 & 1.67e+11 & 8.37e+10 & 1.02e+04 & 1.13e+02 & 7.23e+03 & 7.16e+03 & 1.60e+03 & 1.33e+03 & 1.56e+03 & 1.54e+03 & 1.49e+03 & 1.37e+02 & 1.23e+02 & 1.11e+03 & 1.03e+03 \\
\hline  
\multirow[c]{30}{*}{p-Benzyne} & \multirow[c]{3}{*}{0} & x & 104 & 3833 & 0.114 & 1074 & 0.005 & 214796 & 2.79e+19 & 2.38e+15 & 2.79e+19 & 2.66e+13 & 1.33e+13 & 4.37e+04 & 1.50e+02 & 3.09e+04 & 3.08e+04 & 5.89e+03 & 5.12e+03 & 5.84e+03 & 5.82e+03 & 5.75e+03 & 3.09e+02 & 1.62e+02 & 4.15e+03 & 4.05e+03 \\

 & & y & 104 & 3833 & 0.114 & 1050 & 0.005 & 210035 & 2.79e+19 & 2.38e+15 & 2.79e+19 & 2.66e+13 & 1.33e+13 & 4.37e+04 & 1.50e+02 & 3.09e+04 & 3.08e+04 & 5.89e+03 & 5.12e+03 & 5.84e+03 & 5.82e+03 & 5.75e+03 & 3.09e+02 & 1.62e+02 & 4.15e+03 & 4.05e+03 \\

 & & z & 104 & 3833 & 0.114 & 705 & 0.005 & 140947 & 1.39e+19 & 2.38e+15 & 1.39e+19 & 2.66e+13 & 1.33e+13 & 4.37e+04 & 1.48e+02 & 3.01e+04 & 3.00e+04 & 5.89e+03 & 5.12e+03 & 5.84e+03 & 5.82e+03 & 5.75e+03 & 3.09e+02 & 1.58e+02 & 4.04e+03 & 3.94e+03 \\
 
 & \multirow[c]{3}{*}{1} & x & 104 & 3833 & 0.114 & 1051 & 0.005 & 210185 & 2.79e+19 & 2.38e+15 & 2.79e+19 & 2.66e+13 & 1.33e+13 & 4.37e+04 & 1.50e+02 & 3.09e+04 & 3.08e+04 & 5.89e+03 & 5.12e+03 & 5.84e+03 & 5.82e+03 & 5.75e+03 & 3.09e+02 & 1.62e+02 & 4.15e+03 & 4.05e+03 \\

 & & y & 104 & 3833 & 0.114 & 1093 & 0.005 & 218623 & 2.79e+19 & 2.38e+15 & 2.79e+19 & 2.66e+13 & 1.33e+13 & 4.37e+04 & 1.50e+02 & 3.09e+04 & 3.08e+04 & 5.89e+03 & 5.12e+03 & 5.84e+03 & 5.82e+03 & 5.75e+03 & 3.09e+02 & 1.62e+02 & 4.15e+03 & 4.05e+03 \\

 & & z & 104 & 3833 & 0.114 & 704 & 0.005 & 140782 & 1.39e+19 & 2.38e+15 & 1.39e+19 & 2.66e+13 & 1.33e+13 & 4.37e+04 & 1.48e+02 & 3.01e+04 & 3.00e+04 & 5.89e+03 & 5.12e+03 & 5.84e+03 & 5.82e+03 & 5.75e+03 & 3.09e+02 & 1.58e+02 & 4.04e+03 & 3.94e+03 \\
 
 & \multirow[c]{3}{*}{2} & x & 104 & 3833 & 0.114 & 1003 & 0.005 & 200618 & 2.79e+19 & 2.38e+15 & 2.79e+19 & 2.66e+13 & 1.33e+13 & 4.37e+04 & 1.50e+02 & 3.09e+04 & 3.08e+04 & 5.89e+03 & 5.12e+03 & 5.84e+03 & 5.82e+03 & 5.75e+03 & 3.09e+02 & 1.62e+02 & 4.15e+03 & 4.05e+03 \\

 & & y & 104 & 3833 & 0.114 & 970 & 0.005 & 193941 & 2.79e+19 & 2.38e+15 & 2.79e+19 & 2.66e+13 & 1.33e+13 & 4.37e+04 & 1.50e+02 & 3.09e+04 & 3.08e+04 & 5.89e+03 & 5.12e+03 & 5.84e+03 & 5.82e+03 & 5.75e+03 & 3.09e+02 & 1.62e+02 & 4.15e+03 & 4.05e+03 \\

 & & z & 104 & 3833 & 0.114 & 662 & 0.005 & 132418 & 1.39e+19 & 2.38e+15 & 1.39e+19 & 2.66e+13 & 1.33e+13 & 4.37e+04 & 1.48e+02 & 3.01e+04 & 3.00e+04 & 5.89e+03 & 5.12e+03 & 5.84e+03 & 5.82e+03 & 5.75e+03 & 3.09e+02 & 1.58e+02 & 4.04e+03 & 3.94e+03 \\
 
 & \multirow[c]{3}{*}{3} & x & 104 & 3833 & 0.114 & 1084 & 0.005 & 216890 & 2.79e+19 & 2.38e+15 & 2.79e+19 & 2.66e+13 & 1.33e+13 & 4.37e+04 & 1.50e+02 & 3.09e+04 & 3.08e+04 & 5.89e+03 & 5.12e+03 & 5.84e+03 & 5.82e+03 & 5.75e+03 & 3.09e+02 & 1.62e+02 & 4.15e+03 & 4.05e+03 \\

 & & y & 104 & 3833 & 0.114 & 1071 & 0.005 & 214123 & 2.79e+19 & 2.38e+15 & 2.79e+19 & 2.66e+13 & 1.33e+13 & 4.37e+04 & 1.50e+02 & 3.09e+04 & 3.08e+04 & 5.89e+03 & 5.12e+03 & 5.84e+03 & 5.82e+03 & 5.75e+03 & 3.09e+02 & 1.62e+02 & 4.15e+03 & 4.05e+03 \\

 & & z & 104 & 3833 & 0.114 & 712 & 0.005 & 142300 & 1.39e+19 & 2.38e+15 & 1.39e+19 & 2.66e+13 & 1.33e+13 & 4.37e+04 & 1.48e+02 & 3.01e+04 & 3.00e+04 & 5.89e+03 & 5.12e+03 & 5.84e+03 & 5.82e+03 & 5.75e+03 & 3.09e+02 & 1.58e+02 & 4.04e+03 & 3.94e+03 \\
 
 & \multirow[c]{3}{*}{4} & x & 104 & 3833 & 0.114 & 1065 & 0.005 & 213018 & 2.79e+19 & 2.38e+15 & 2.79e+19 & 2.66e+13 & 1.33e+13 & 4.37e+04 & 1.50e+02 & 3.09e+04 & 3.08e+04 & 5.89e+03 & 5.12e+03 & 5.84e+03 & 5.82e+03 & 5.75e+03 & 3.09e+02 & 1.62e+02 & 4.15e+03 & 4.05e+03 \\

 & & y & 104 & 3833 & 0.114 & 1082 & 0.005 & 216434 & 2.79e+19 & 2.38e+15 & 2.79e+19 & 2.66e+13 & 1.33e+13 & 4.37e+04 & 1.50e+02 & 3.09e+04 & 3.08e+04 & 5.89e+03 & 5.12e+03 & 5.84e+03 & 5.82e+03 & 5.75e+03 & 3.09e+02 & 1.62e+02 & 4.15e+03 & 4.05e+03 \\

 & & z & 104 & 3833 & 0.114 & 712 & 0.005 & 142498 & 1.39e+19 & 2.38e+15 & 1.39e+19 & 2.66e+13 & 1.33e+13 & 4.37e+04 & 1.48e+02 & 3.01e+04 & 3.00e+04 & 5.89e+03 & 5.12e+03 & 5.84e+03 & 5.82e+03 & 5.75e+03 & 3.09e+02 & 1.58e+02 & 4.04e+03 & 3.94e+03 \\
 
 & \multirow[c]{3}{*}{5} & x & 104 & 3833 & 0.114 & 1020 & 0.005 & 203926 & 2.79e+19 & 2.38e+15 & 2.79e+19 & 2.66e+13 & 1.33e+13 & 4.37e+04 & 1.50e+02 & 3.09e+04 & 3.08e+04 & 5.89e+03 & 5.12e+03 & 5.84e+03 & 5.82e+03 & 5.75e+03 & 3.09e+02 & 1.62e+02 & 4.15e+03 & 4.05e+03 \\

 & & y & 104 & 3833 & 0.114 & 945 & 0.005 & 189006 & 2.79e+19 & 2.38e+15 & 2.79e+19 & 2.66e+13 & 1.33e+13 & 4.37e+04 & 1.50e+02 & 3.09e+04 & 3.08e+04 & 5.89e+03 & 5.12e+03 & 5.84e+03 & 5.82e+03 & 5.75e+03 & 3.09e+02 & 1.62e+02 & 4.15e+03 & 4.05e+03 \\

 & & z & 104 & 3833 & 0.114 & 659 & 0.005 & 131877 & 1.39e+19 & 2.38e+15 & 1.39e+19 & 2.66e+13 & 1.33e+13 & 4.37e+04 & 1.48e+02 & 3.01e+04 & 3.00e+04 & 5.89e+03 & 5.12e+03 & 5.84e+03 & 5.82e+03 & 5.75e+03 & 3.09e+02 & 1.58e+02 & 4.04e+03 & 3.94e+03 \\
 
 & \multirow[c]{3}{*}{6} & x & 104 & 3833 & 0.114 & 175 & 0.005 & 35063 & 3.49e+18 & 2.38e+15 & 3.48e+18 & 2.66e+13 & 1.33e+13 & 4.37e+04 & 1.43e+02 & 2.80e+04 & 2.79e+04 & 5.88e+03 & 5.12e+03 & 5.83e+03 & 5.81e+03 & 5.75e+03 & 3.09e+02 & 1.48e+02 & 3.82e+03 & 3.72e+03 \\

 & & y & 104 & 3833 & 0.114 & 196 & 0.005 & 39294 & 6.97e+18 & 2.38e+15 & 6.97e+18 & 2.66e+13 & 1.33e+13 & 4.37e+04 & 1.45e+02 & 2.80e+04 & 2.79e+04 & 5.88e+03 & 5.12e+03 & 5.83e+03 & 5.82e+03 & 5.75e+03 & 3.09e+02 & 1.52e+02 & 3.83e+03 & 3.73e+03 \\

 & & z & 104 & 3833 & 0.114 & 130 & 0.005 & 26064 & 3.49e+18 & 2.38e+15 & 3.48e+18 & 2.66e+13 & 1.33e+13 & 4.37e+04 & 1.43e+02 & 2.80e+04 & 2.79e+04 & 5.88e+03 & 5.12e+03 & 5.83e+03 & 5.81e+03 & 5.75e+03 & 3.09e+02 & 1.48e+02 & 3.82e+03 & 3.72e+03 \\
 
 & \multirow[c]{3}{*}{7} & x & 104 & 3833 & 0.114 & 206 & 0.005 & 41130 & 6.97e+18 & 2.38e+15 & 6.97e+18 & 2.66e+13 & 1.33e+13 & 4.37e+04 & 1.45e+02 & 2.89e+04 & 2.88e+04 & 5.88e+03 & 5.12e+03 & 5.83e+03 & 5.82e+03 & 5.75e+03 & 3.09e+02 & 1.52e+02 & 3.93e+03 & 3.83e+03 \\

 & & y & 104 & 3833 & 0.114 & 162 & 0.005 & 32361 & 3.49e+18 & 2.38e+15 & 3.48e+18 & 2.66e+13 & 1.33e+13 & 4.37e+04 & 1.43e+02 & 2.80e+04 & 2.79e+04 & 5.88e+03 & 5.12e+03 & 5.83e+03 & 5.81e+03 & 5.75e+03 & 3.09e+02 & 1.48e+02 & 3.82e+03 & 3.72e+03 \\

 & & z & 104 & 3833 & 0.114 & 129 & 0.005 & 25742 & 3.49e+18 & 2.38e+15 & 3.48e+18 & 2.66e+13 & 1.33e+13 & 4.37e+04 & 1.43e+02 & 2.80e+04 & 2.79e+04 & 5.88e+03 & 5.12e+03 & 5.83e+03 & 5.81e+03 & 5.75e+03 & 3.09e+02 & 1.48e+02 & 3.82e+03 & 3.72e+03 \\
 
 & \multirow[c]{3}{*}{8} & x & 104 & 3833 & 0.114 & 182 & 0.005 & 36481 & 3.49e+18 & 2.38e+15 & 3.48e+18 & 2.66e+13 & 1.33e+13 & 4.37e+04 & 1.43e+02 & 2.80e+04 & 2.79e+04 & 5.88e+03 & 5.12e+03 & 5.83e+03 & 5.81e+03 & 5.75e+03 & 3.09e+02 & 1.48e+02 & 3.82e+03 & 3.72e+03 \\

 & & y & 104 & 3833 & 0.114 & 207 & 0.005 & 41305 & 6.97e+18 & 2.38e+15 & 6.97e+18 & 2.66e+13 & 1.33e+13 & 4.37e+04 & 1.45e+02 & 2.88e+04 & 2.87e+04 & 5.88e+03 & 5.12e+03 & 5.83e+03 & 5.82e+03 & 5.75e+03 & 3.09e+02 & 1.52e+02 & 3.93e+03 & 3.83e+03 \\

 & & z & 104 & 3833 & 0.114 & 131 & 0.005 & 26152 & 3.49e+18 & 2.38e+15 & 3.48e+18 & 2.66e+13 & 1.33e+13 & 4.37e+04 & 1.43e+02 & 2.80e+04 & 2.79e+04 & 5.88e+03 & 5.12e+03 & 5.83e+03 & 5.81e+03 & 5.75e+03 & 3.09e+02 & 1.48e+02 & 3.82e+03 & 3.72e+03 \\
 
 & \multirow[c]{3}{*}{9} & x & 104 & 3833 & 0.114 & 217 & 0.005 & 43306 & 6.97e+18 & 2.38e+15 & 6.97e+18 & 2.66e+13 & 1.33e+13 & 4.37e+04 & 1.45e+02 & 2.88e+04 & 2.87e+04 & 5.88e+03 & 5.12e+03 & 5.83e+03 & 5.82e+03 & 5.75e+03 & 3.09e+02 & 1.52e+02 & 3.93e+03 & 3.83e+03 \\

 & & y & 104 & 3833 & 0.114 & 161 & 0.005 & 32212 & 3.49e+18 & 2.38e+15 & 3.48e+18 & 2.66e+13 & 1.33e+13 & 4.37e+04 & 1.43e+02 & 2.80e+04 & 2.79e+04 & 5.88e+03 & 5.12e+03 & 5.83e+03 & 5.81e+03 & 5.75e+03 & 3.09e+02 & 1.48e+02 & 3.82e+03 & 3.72e+03 \\

 & & z & 104 & 3833 & 0.114 & 129 & 0.005 & 25870 & 3.49e+18 & 2.38e+15 & 3.48e+18 & 2.66e+13 & 1.33e+13 & 4.37e+04 & 1.43e+02 & 2.80e+04 & 2.79e+04 & 5.88e+03 & 5.12e+03 & 5.83e+03 & 5.81e+03 & 5.75e+03 & 3.09e+02 & 1.48e+02 & 3.82e+03 & 3.72e+03 \\
\hline  

\end{tabular}
}

\vspace{10pt}
\end{sidewaysfigure}


\begin{sidewaysfigure}
Dipole (std-EVE)\\

\resizebox{\textheight}{!}{

\begin{tabular}{|cc|*6r|*9r|*9r|}
\hline
\multicolumn{2}{|c|}{\bf{Instance}} & \multicolumn{6}{c|}{\bf{Property}} & \multicolumn{9}{c|}{\bf{Gates}} & \multicolumn{9}{c|}{\bf{Qubits}} \\
System & Axis & N & $\lambda_H$ & $\Delta_H$ & $\lambda_F$ & $\varepsilon$ & $\Lambda_F$ & EVE & ASP & oQPE & $\mathcal{R}_\pi$ & iQPE & $\mathcal{B}[\hat{H}]$ & REFL & $\mathcal{R}_\tau$ & $\mathcal{B}[\hat{F}]$ & EVE & ASP & oQPE & $\mathcal{R}_\pi$ & iQPE & $\mathcal{B}[\hat{H}]$ & REFL & $\mathcal{R}_\tau$ & $\mathcal{B}[\hat{F}]$ \\

\hline
\multirow[c]{3}{*}{Water} & x & 24 & 328 & 0.302 & 65 & 0.004 & 16264 & 3.15e+18 & 1.25e+12 & 3.15e+18 & 2.40e+13 & 1.20e+13 & 1.12e+04 & 1.63e+02 & 5.99e+03 & 5.89e+03 & 1.82e+03 & 1.11e+03 & 1.76e+03 & 1.75e+03 & 1.70e+03 & 1.55e+02 & 1.25e+02 & 1.00e+03 & 8.93e+02 \\

 & y & 24 & 328 & 0.302 & 17 & 0.004 & 4342 & 1.90e+17 & 1.25e+12 & 1.90e+17 & 5.78e+12 & 2.89e+12 & 1.08e+04 & 1.51e+02 & 5.59e+03 & 5.49e+03 & 1.75e+03 & 1.11e+03 & 1.70e+03 & 1.68e+03 & 1.64e+03 & 1.51e+02 & 1.15e+02 & 9.38e+02 & 8.35e+02 \\
\hline 
\multirow[c]{3}{*}{Ammonia} & x & 29 & 434 & 0.280 & 91 & 0.004 & 22654 & 1.56e+19 & 2.90e+12 & 1.56e+19 & 5.95e+13 & 2.98e+13 & 1.39e+04 & 1.69e+02 & 7.37e+03 & 7.26e+03 & 2.16e+03 & 1.33e+03 & 2.11e+03 & 2.09e+03 & 2.04e+03 & 1.67e+02 & 1.30e+02 & 1.19e+03 & 1.08e+03 \\

 & y & 29 & 434 & 0.280 & 20 & 0.004 & 5010 & 4.71e+17 & 2.90e+12 & 4.71e+17 & 1.44e+13 & 7.18e+12 & 1.34e+04 & 1.57e+02 & 6.89e+03 & 6.78e+03 & 2.08e+03 & 1.33e+03 & 2.03e+03 & 2.01e+03 & 1.97e+03 & 1.63e+02 & 1.20e+02 & 1.12e+03 & 1.01e+03 \\

 & z & 29 & 434 & 0.280 & 26 & 0.004 & 6522 & 9.41e+17 & 2.90e+12 & 9.41e+17 & 1.44e+13 & 7.18e+12 & 1.34e+04 & 1.57e+02 & 6.89e+03 & 6.78e+03 & 2.08e+03 & 1.33e+03 & 2.03e+03 & 2.01e+03 & 1.97e+03 & 1.63e+02 & 1.20e+02 & 1.12e+03 & 1.01e+03 \\
\hline 
\multirow[c]{3}{*}{p-Benzyne} & x & 104 & 3833 & 0.114 & 239 & 0.004 & 59649 & 8.25e+21 & 2.38e+15 & 8.25e+21 & 1.57e+16 & 7.87e+15 & 5.73e+04 & 1.97e+02 & 2.75e+04 & 2.74e+04 & 7.68e+03 & 5.12e+03 & 7.62e+03 & 7.60e+03 & 7.54e+03 & 3.41e+02 & 1.44e+02 & 3.94e+03 & 3.81e+03 \\

 & y & 104 & 3833 & 0.114 & 205 & 0.004 & 51147 & 8.25e+21 & 2.38e+15 & 8.25e+21 & 1.57e+16 & 7.87e+15 & 5.73e+04 & 1.97e+02 & 2.75e+04 & 2.74e+04 & 7.68e+03 & 5.12e+03 & 7.62e+03 & 7.60e+03 & 7.54e+03 & 3.41e+02 & 1.44e+02 & 3.94e+03 & 3.81e+03 \\

 & z & 104 & 3833 & 0.114 & 481 & 0.004 & 120328 & 3.35e+22 & 2.38e+15 & 3.35e+22 & 3.20e+16 & 1.60e+16 & 5.81e+04 & 2.03e+02 & 2.83e+04 & 2.82e+04 & 7.80e+03 & 5.12e+03 & 7.73e+03 & 7.71e+03 & 7.65e+03 & 3.43e+02 & 1.49e+02 & 4.05e+03 & 3.92e+03 \\
\hline 

\end{tabular}

}
\vspace{10pt}


Dipole (qsp-EVE)\\

\resizebox{\textheight}{!}{

\begin{tabular}{|cc|*6r|*9r|*9r|}
\hline
\multicolumn{2}{|c|}{\bf{Instance}} & \multicolumn{6}{c|}{\bf{Property}} & \multicolumn{9}{c|}{\bf{Gates}} & \multicolumn{9}{c|}{\bf{Qubits}} \\
System & Axis & N & $\lambda_H$ & $\Delta_H$ & $\lambda_F$ & $\varepsilon$ & $\Lambda_F$ & EVE & ASP & oQPE & $\mathcal{R}_\pi$ & iQPE & $\mathcal{B}[\hat{H}]$ & REFL & $\mathcal{R}_\tau$ & $\mathcal{B}[\hat{F}]$ & EVE & ASP & oQPE & $\mathcal{R}_\pi$ & iQPE & $\mathcal{B}[\hat{H}]$ & REFL & $\mathcal{R}_\tau$ & $\mathcal{B}[\hat{F}]$ \\

\hline
\multirow[c]{3}{*}{Water} & x & 24 & 328 & 0.302 & 65 & 0.004 & 16264 & 8.98e+15 & 1.25e+12 & 8.98e+15 & 1.37e+11 & 6.85e+10 & 8.36e+03 & 1.07e+02 & 5.96e+03 & 5.89e+03 & 1.36e+03 & 1.11e+03 & 1.32e+03 & 1.30e+03 & 1.25e+03 & 1.27e+02 & 1.11e+02 & 9.72e+02 & 8.93e+02 \\

 & y & 24 & 328 & 0.302 & 17 & 0.004 & 4342 & 2.25e+15 & 1.25e+12 & 2.24e+15 & 1.37e+11 & 6.85e+10 & 8.36e+03 & 1.03e+02 & 5.56e+03 & 5.49e+03 & 1.35e+03 & 1.11e+03 & 1.31e+03 & 1.30e+03 & 1.25e+03 & 1.27e+02 & 1.03e+02 & 9.14e+02 & 8.35e+02 \\
\hline 
\multirow[c]{3}{*}{Ammonia} & x & 29 & 434 & 0.280 & 91 & 0.004 & 22654 & 2.23e+16 & 2.90e+12 & 2.23e+16 & 1.70e+11 & 8.51e+10 & 1.02e+04 & 1.09e+02 & 7.34e+03 & 7.26e+03 & 1.59e+03 & 1.33e+03 & 1.55e+03 & 1.54e+03 & 1.49e+03 & 1.37e+02 & 1.15e+02 & 1.16e+03 & 1.08e+03 \\

 & y & 29 & 434 & 0.280 & 20 & 0.004 & 5010 & 5.58e+15 & 2.90e+12 & 5.57e+15 & 1.70e+11 & 8.51e+10 & 1.02e+04 & 1.05e+02 & 6.86e+03 & 6.78e+03 & 1.59e+03 & 1.33e+03 & 1.55e+03 & 1.53e+03 & 1.49e+03 & 1.37e+02 & 1.07e+02 & 1.09e+03 & 1.01e+03 \\

 & z & 29 & 434 & 0.280 & 26 & 0.004 & 6522 & 5.58e+15 & 2.90e+12 & 5.57e+15 & 1.70e+11 & 8.51e+10 & 1.02e+04 & 1.05e+02 & 6.86e+03 & 6.78e+03 & 1.59e+03 & 1.33e+03 & 1.55e+03 & 1.53e+03 & 1.49e+03 & 1.37e+02 & 1.07e+02 & 1.09e+03 & 1.01e+03 \\
\hline 
\multirow[c]{3}{*}{p-Benzyne} & x & 104 & 3833 & 0.114 & 239 & 0.004 & 59649 & 7.07e+18 & 2.38e+15 & 7.06e+18 & 2.69e+13 & 1.35e+13 & 4.37e+04 & 1.33e+02 & 2.75e+04 & 2.74e+04 & 5.87e+03 & 5.12e+03 & 5.82e+03 & 5.80e+03 & 5.75e+03 & 3.09e+02 & 1.28e+02 & 3.91e+03 & 3.81e+03 \\

 & y & 104 & 3833 & 0.114 & 205 & 0.004 & 51147 & 7.07e+18 & 2.38e+15 & 7.06e+18 & 2.69e+13 & 1.35e+13 & 4.37e+04 & 1.33e+02 & 2.75e+04 & 2.74e+04 & 5.87e+03 & 5.12e+03 & 5.82e+03 & 5.80e+03 & 5.75e+03 & 3.09e+02 & 1.28e+02 & 3.91e+03 & 3.81e+03 \\

 & z & 104 & 3833 & 0.114 & 481 & 0.004 & 120328 & 1.41e+19 & 2.38e+15 & 1.41e+19 & 2.69e+13 & 1.35e+13 & 4.37e+04 & 1.35e+02 & 2.83e+04 & 2.82e+04 & 5.88e+03 & 5.12e+03 & 5.82e+03 & 5.80e+03 & 5.75e+03 & 3.09e+02 & 1.32e+02 & 4.02e+03 & 3.92e+03 \\
\hline 

\end{tabular}

}
\vspace{10pt}


Kinetic Energy (std-EVE)\\

\resizebox{\textheight}{!}{

\begin{tabular}{|c|*6r|*9r|*9r|}
\hline
\bf{Instance} & \multicolumn{6}{c|}{\bf{Property}} & \multicolumn{9}{c|}{\bf{Gates}} & \multicolumn{9}{c|}{\bf{Qubits}} \\
System & N & $\lambda_H$ & $\Delta_H$ & $\lambda_F$ & $\varepsilon$ & $\Lambda_F$ & EVE & ASP & oQPE & $\mathcal{R}_\pi$ & iQPE & $\mathcal{B}[\hat{H}]$ & REFL & $\mathcal{R}_\tau$ & $\mathcal{B}[\hat{F}]$ & EVE & ASP & oQPE & $\mathcal{R}_\pi$ & iQPE & $\mathcal{B}[\hat{H}]$ & REFL & $\mathcal{R}_\tau$ & $\mathcal{B}[\hat{F}]$ \\

\hline
\ce{H2} & 10 & 71 & 0.472 & 17 & 0.002 & 10697 & 1.56e+17 & 1.35e+10 & 1.56e+17 & 1.19e+12 & 5.94e+11 & 4.42e+03 & 1.50e+02 & 2.60e+03 & 2.50e+03 & 9.32e+02 & 4.88e+02 & 8.85e+02 & 8.67e+02 & 8.20e+02 & 1.11e+02 & 1.20e+02 & 5.24e+02 & 4.33e+02 \\
\hline
Be & 14 & 65 & 0.207 & 15 & 0.002 & 9534 & 2.15e+17 & 3.70e+10 & 2.15e+17 & 3.28e+12 & 1.64e+12 & 6.12e+03 & 1.54e+02 & 3.56e+03 & 3.46e+03 & 1.15e+03 & 6.40e+02 & 1.10e+03 & 1.09e+03 & 1.04e+03 & 1.22e+02 & 1.21e+02 & 6.58e+02 & 5.64e+02 \\
\hline
Water & 24 & 328 & 0.302 & 99 & 0.002 & 61792 & 5.21e+19 & 1.25e+12 & 5.21e+19 & 9.94e+13 & 4.97e+13 & 1.16e+04 & 1.75e+02 & 6.40e+03 & 6.29e+03 & 1.89e+03 & 1.11e+03 & 1.84e+03 & 1.82e+03 & 1.76e+03 & 1.59e+02 & 1.35e+02 & 1.06e+03 & 9.51e+02 \\
\hline
Ammonia & 29 & 434 & 0.280 & 91 & 0.002 & 56635 & 1.29e+20 & 2.90e+12 & 1.29e+20 & 2.47e+14 & 1.23e+14 & 1.43e+04 & 1.79e+02 & 7.62e+03 & 7.50e+03 & 2.24e+03 & 1.33e+03 & 2.19e+03 & 2.17e+03 & 2.12e+03 & 1.71e+02 & 1.36e+02 & 1.23e+03 & 1.12e+03 \\
\hline
p-Benzyne & 104 & 3833 & 0.114 & 316 & 0.002 & 197448 & 1.36e+23 & 2.38e+15 & 1.36e+23 & 6.48e+16 & 3.24e+16 & 5.90e+04 & 2.09e+02 & 2.92e+04 & 2.90e+04 & 7.91e+03 & 5.12e+03 & 7.84e+03 & 7.82e+03 & 7.76e+03 & 3.45e+02 & 1.54e+02 & 4.16e+03 & 4.03e+03 \\
\hline

\end{tabular}

}
\vspace{10pt}


Kinetic Energy (qsp-EVE)\\

\resizebox{\textheight}{!}{

\begin{tabular}{|c|*6r|*9r|*9r|}
\hline
\bf{Instance} & \multicolumn{6}{c|}{\bf{Property}} & \multicolumn{9}{c|}{\bf{Gates}} & \multicolumn{9}{c|}{\bf{Qubits}} \\
System & N & $\lambda_H$ & $\Delta_H$ & $\lambda_F$ & $\varepsilon$ & $\Lambda_F$ & EVE & ASP & oQPE & $\mathcal{R}_\pi$ & iQPE & $\mathcal{B}[\hat{H}]$ & REFL & $\mathcal{R}_\tau$ & $\mathcal{B}[\hat{F}]$ & EVE & ASP & oQPE & $\mathcal{R}_\pi$ & iQPE & $\mathcal{B}[\hat{H}]$ & REFL & $\mathcal{R}_\tau$ & $\mathcal{B}[\hat{F}]$ \\

\hline
\ce{H2} & 10 & 71 & 0.472 & 17 & 0.002 & 10697 & 4.08e+14 & 1.35e+10 & 4.08e+14 & 6.22e+09 & 3.11e+09 & 3.19e+03 & 9.40e+01 & 2.57e+03 & 2.50e+03 & 6.65e+02 & 4.88e+02 & 6.32e+02 & 6.15e+02 & 5.68e+02 & 8.30e+01 & 1.06e+02 & 4.96e+02 & 4.33e+02 \\
\hline
Be & 14 & 65 & 0.207 & 15 & 0.002 & 9534 & 5.65e+14 & 3.70e+10 & 5.65e+14 & 1.72e+10 & 8.62e+09 & 4.42e+03 & 9.80e+01 & 3.53e+03 & 3.46e+03 & 8.29e+02 & 6.40e+02 & 7.95e+02 & 7.79e+02 & 7.32e+02 & 9.40e+01 & 1.07e+02 & 6.30e+02 & 5.64e+02 \\
\hline
Water & 24 & 328 & 0.302 & 99 & 0.002 & 61792 & 3.82e+16 & 1.25e+12 & 3.82e+16 & 1.46e+11 & 7.29e+10 & 8.36e+03 & 1.11e+02 & 6.36e+03 & 6.29e+03 & 1.36e+03 & 1.11e+03 & 1.32e+03 & 1.30e+03 & 1.25e+03 & 1.27e+02 & 1.19e+02 & 1.03e+03 & 9.51e+02 \\
\hline
Ammonia & 29 & 434 & 0.280 & 91 & 0.002 & 56635 & 4.74e+16 & 2.90e+12 & 4.74e+16 & 1.81e+11 & 9.04e+10 & 1.02e+04 & 1.11e+02 & 7.58e+03 & 7.50e+03 & 1.60e+03 & 1.33e+03 & 1.56e+03 & 1.54e+03 & 1.49e+03 & 1.37e+02 & 1.19e+02 & 1.20e+03 & 1.12e+03 \\
\hline
p-Benzyne & 104 & 3833 & 0.114 & 316 & 0.002 & 197448 & 2.98e+19 & 2.38e+15 & 2.98e+19 & 2.84e+13 & 1.42e+13 & 4.37e+04 & 1.37e+02 & 2.91e+04 & 2.90e+04 & 5.88e+03 & 5.12e+03 & 5.83e+03 & 5.81e+03 & 5.75e+03 & 3.09e+02 & 1.36e+02 & 4.13e+03 & 4.03e+03 \\
\hline

\end{tabular}

}
\vspace{10pt}
	
\end{sidewaysfigure}
\end{document}